%% file: JRSSB_Final.tex
\documentclass{statsoc}
\usepackage{epsfig}
\usepackage{amsmath,amsfonts,amssymb,graphicx,algorithmic,algorithm}
\usepackage{bm, url}
\usepackage{times}
\usepackage{natbib}
\usepackage[mathscr]{euscript}
\input{commands}

\newcommand{\Prob}{\mathscr{P}}
\newcommand{\WW}{\mathscr{W}}
\newcommand{\LL}{\mathscr{L}}
\newcommand{\MM}{\mathscr{M}}
\newcommand{\dt}{{\rm d}t}
\newcommand{\ds}{{\rm d}s}

\title[Generic solution of the  heterogeneity-induced competing risk problem in survival analysis]
{Generic solution of the heterogeneity-induced\\ competing risk problem in survival analysis}
\author[J van Baardewijk et al.]
{J van Baardewijk$^{1}$, H Garmo$^{2}$, 
M Van Hemelrijck$^2$, L Holmberg$^2$, and ACC Coolen$^{1,3}$}

\address{\footnotesize  $1$  King's College London, Institute for Mathematical and Molecular Biomedicine, 
 Hodgkin Building,
London SE1 1UL, UK
\\[1mm]
$2$ King's College London, School of Medicine, Cancer Epidemiology Group,
Guy's Hospital, London SE1 9RT, UK.
\\[1mm]
$3$
London Institute for Mathematical Sciences, 35a South St, Mayfair, London W1K 2XF, UK
\\[2mm]
\hspace*{\fill}{\small\em Final version of June 30th 2013}\\
\hspace*{\fill}{\small\em corresponding author: ton.coolen@kcl.ac.uk}}

\begin{document}

\begin{abstract}
Most papers implicitly assume competing risks to be induced by residual cohort heterogeneity, i.e. heterogeneity that is not captured by the recorded covariates. Based on this observation 
we develop a generic statistical description of competing risks that unifies the main schools of thought.  Assuming heterogeneity-induced competing risks is much weaker than assuming risk independence.  However, we show that it still imposes sufficient constraints to solve the competing risk problem, and 
derive exact formulae for decontaminated primary risk hazard rates and cause-specific survival functions.  
The canonical description  is in terms of a cohort's covariate-constrained functional distribution of individual hazard rates of all risks. Assuming proportional hazards at the level of individuals leads to a natural parametrisation of this distribution,  from which Cox regression, frailty and random effects models, and latent class models can all be recovered in special limits, and which also generates parametrised cumulative incidence functions (the language of Fine and Gray). 
 We demonstrate with synthetic data how the generic method can uncover and map a cohort's substructure, if such substructure exists, and remove heterogeneity-induced false protectivity and false exposure effects. Application to real survival data from the ULSAM study, with prostate cancer as the primary risk, is found to give plausible alternative explanations for previous counter-intuitive inferences.
\end{abstract}

\section{Introduction}

For general introductions to the survival analysis literature we refer to the excellent textbooks 
\citep{Hougaard,KleinBook,Ibrahim,Crowder}.
The competing risk problem in survival analysis is the question of how to handle the possible contamination   of   those characteristics of the primary risk that can be extracted from survival data, such as its hazard rate or its cumulative incidence function, by informative censoring. 
Decontaminating primary risk characteristics means finding what their values would have been in the hypothetical situation where all non-primary risks were disabled. This is nontrivial, because disabling non-primary risks will not only set their cause-specific hazard rates to zero, but it will generally affect also the hazard rate of the primary risk.
If all risks  have statistically independent event times censoring is not informative, so there is no problem and many simple methods are available for analysis and regression, such as the survival function estimators of \citep{KM} or the proportional hazards method \citep{Cox}.
Unfortunately, one usually cannot know beforehand whether the risks in a study  are uncorrelated, and  there are many cases where the  independence assumption is clearly incorrect. Tsiatis' nonidentifiability theorem \citep{Tsiatis} shows that without additional assumptions it is not possible to infer presence or absence of risk correlations unambiguously from survival data.  Unaccounted for risk correlations invalidate the standard interpretations of methods such as \citep{KM,Cox}, and can lead to `false protectivity' effects \citep{DiSerio}  and incorrect inferences \citep{AndersenReview,Dignam}. 

Risk correlations are often fingerprints of residual heterogeneity in cohorts, i.e. of variability in individuals and diseases that is not captured by the covariates.  For instance, a primary and a secondary risk (or disease) may share common molecular pathways or be jointly influenced by common environmental or lifestyle factors that were not measured. We would then find that those individuals most likely to be censored by the secondary risk are not random, but would be the ones most likely to report a primary risk event (or vice versa), even if they are indistinguishable in covariate terms. Or, that which we presently regard as a single disease could in fact be a spectrum of distinct diseases, each with their own specific associations with covariates. 
Many authors have therefore tried to model residual cohort heterogeneity, usually by postulating individual cause-specific hazard rates $h_r^i(t)$ of the Cox type, but with additional individualised risk multipliers:
\begin{eqnarray}
h_r^i(t)=\lambda_r(t)\rme^{\sum_{\mu=1}^p\beta^\mu_r z^\mu_i+\xi_r^i}
\end{eqnarray}
Here $r$ and $i$ label, respectively, the different risks and the individuals in our cohort, $\lambda_r(t)$ is a common time-dependent base hazard rate of risk $r$, 
$(\beta_r^1,\ldots,\beta_r^p)$ is a vector of regression coefficents for risk $r$, and $(z_i^1,\ldots,z_i^p)$ is a vector of covariates of individual $i$. 
 The`frailty factors'  $\xi_r^i$ are assumed to be sampled from a given parametrised distribution, whose parameters must be estimated from the data. If the frailty factors $\xi_r^i$  do not depend on the individual's covariates we would speak of `frailty models', e.g. \citep{Vaupel,Zahl,Yashin,Gorfine}. Frailty models are often regarded as representing the impact of unobserved covariates, see e.g. the discussion in \citep{KeidingAndersenKlein}. Models in which the frailty factors depend on the observed covariates, e.g. $\xi_r^i=\sum_{\mu=1}^p\gamma^\mu_{ir} z^\mu_i $,  are called 
`random effects models', e.g. \citep{Vaida,DiSerio} or the more recent application to breast cancer sub-types \citep{Rosner}. See also the textbooks \citep{Wienke,Duchateau}. 
If the distribution of frailty factors takes the form of discrete clusters (latent classes), an idea that goes back to \citep{Lazarsfeld},  we obtain the so-called latent class models. See e.g. \citep{Muhten}
which combines frailty and random effects with covariate-dependent class sizes as in  \citep{Reboussin}.
Further variations include e.g. frailty factors that are allowed to evolve over time, and models in which the cluster membership of individuals (which represents the heterogeneity) is assumed known. 
Most frailty and random effects studies, however, quantify only the hazard rates of the primary risk. For instance, in the references above the only exceptions are \citep{Zahl,DiSerio,Gorfine}. Although one may capture many consequences of cohort heterogeneity (such as time-dependence of regression parameters caused by cohort filtering), without modelling also the non-primary risks it is not possible to deal with the competing risk problem. Moreover, none of the above studies derive explicit formulae for decontaminated primary risk measures such as cause-specific hazard rates or  survival functions.

The line of work initiated by \citep{FineGray} does not try to deal with the decontamination question. Instead it focuses on finding parametrisations of the covariate-conditioned cumulative incidence function $F_1(t|\bz)$ of the primary risk, for which \citep{FineGray}  propose\footnote{The authors of \citep{FineGray}  were critcised for interpreting the quantity $\lambda(t)\rme^{\sum_{\mu=1}^p\beta^\mu z^\mu}$, with $\lambda(t)=\rmd\Lambda(t)/\dt$, as a hazard rate, since this requires unnatural risk sets. 
We fail to see the need for this interpretation and regard (\ref{eq:FG}) simply as a parametrisation for the cumulative incidence function. In addition, \citep{FineGray} 
also includes time-dependent covariates, but this is not their main point.} 
\begin{eqnarray}
F_1(t|\bz)=\Phi\big(\Lambda(t)\rme^{\sum_{\mu=1}^p\beta^\mu z^\mu}\big),~~~~~~~~\Phi(x)=1-\rme^{-x}
\label{eq:FG}
\end{eqnarray}
This is conceptually similar to \citep{Cox}, one just parametrises a diffent quantity: in \citep{Cox} the covariate-conditioned hazard rate of the primary risk, in \citep{FineGray} the cumulative incidence function. In particular, \citep{Cox} and 
\citep{FineGray} both model the primary risk profile in the presence of all other risks. 
The approach of  \citep{FineGray} thus studies risks that compete, but does not address the competing risk problem. 
The hope is that by parametrising $F_1(t|\bz)$ directly, one may capture more heterogeneity-induced effects.
The quantity $F_1(t|\bz)$ appears more informative than the primary risk hazard rate; it is directly measurable and involves also the crude hazard rates of non-primary risks. However, the latter could have been estimated with Cox regression too.  The price paid in \citep{FineGray} for the advantages of the cumulative incidence function is in parameter estimation: expressing the data likelihood in terms of cumulative incidence functions is much more cumbersome  than expressing it in terms of hazard rates. Further developments  involve e.g. alternative choices for the function $\Phi(x)$ \citep{Fine,KleinAndersen}, application to the cumulative incidence of non-primary risks \citep{JeongFine}, and the inclusion of frailty factors \citep{Katsahian}. 

So we face the unsatisfactory situation of having multiple distinct and diverging approaches to the modeling of heterogeneity and competing risks.  Only few actually address the competing risk problem, which requires modelling the hazard rates of {\em all} risks and their correlations (not just that of the primary risk), and none lead to explicit formulae for decontamined primary risk measures. In this paper we try to construct a generic statistical description of competing risks and a resolution of the competing risk problem that unifies the various schools of thought described above. Our work is based on the observation that virtually all papers implicitly assume that correlations between competing risks are induced by residual cohort heterogeneity. We show how this simple and transparent assumption leads in a natural way to a formalism with exact formulae for decontaminated primary risk measures, in which Cox regression, 
frailty models, random effect models, and latent class models are all included as special cases, and which produces transparent parametrisations of the cumulative incidence function (which is the language of  Fine and Gray). 

This paper is organised as follows. In section 2 we define the relevant survival analysis quantities and their relations, and state the competing risk problem in mathematical terms. We then inspect in section 3 the relation between cohort level and individual level statistical descriptions, classify the different levels of risk complexity from the competing risk perspective, and define what we mean by heterogeneity-induced competing risks. We derive the implications of having heterogeneity-induced competing risks, and show how that the canonical description for solving the heterogeneity-induced competing risk problem is in terms of the  covariate-conditioned functional distribution $\WW[h_0,\ldots,h_R|\bz]$ of the individual hazard rates of all risks over the cohort. In section 4 we obtain a generic parametrisation of  this  distribution, which reduces the mathematical description to a joint covariate-conditioned (functional) distribution $\MM(\bbeta_0,\ldots,\bbeta_R;\lambda_0,\ldots,\lambda_R|\bz)$ over the cohort of individual regression parameters and base hazard rates for all risks, which often simplifies to a covariate-independent form $\MM(\bbeta_0,\ldots,\bbeta_R;\lambda_0,\ldots,\lambda_R)$. We work out the theory in more detail for a natural family of parametrisations $\MM(\bbeta_0,\ldots,\bbeta_R;\lambda_0,\ldots,\lambda_R)$. 
We show how the conventional methods (Cox regression, frailty models, random effect models, latent class models) are recovered in special limits, and derive the parametrised cumulative incidence functions for all risks. 
The remaining sections are devoted to applications of the formalism to synthetic data, as well as real survival data from the ULSAM longitudinal study \citep{ULSAM,ULSAMpaper}, focusing on prostate cancer as the primary risk. The latter application is found to result in appealing and transparent novel explanations for previously counter-intuitive inferences. We end with a summary of our findings.

\section{Definitions and general identities}

In this section we recall the basic definitions and objectives of survival analysis, and define the competing risk problem in mathematical terms. In doing so we will try to stay as close as possible to the notation conventions and terminology of \citep{KleinBook}. 

\subsection{Survival probability and crude cause specific hazard rates}

We assume having a cohort of individuals who are subject to $R$ true risks, labelled by $r=1\ldots R$. We use $r=0$ to indicate the end-of-trial censoring event, since for the mathematical structure of the theory there is no difference between censoring due to alternative true risks and censoring due to termination of the trial. Most of the mathematical relations of survival analysis can be derived directly from the joint distribution $\Prob(t_0,\ldots,t_R)$ of event times $(t_0,\ldots,t_R)$, where $t_r\geq 0$ is the time at which risk $r$ triggers a failure event.\footnote{This starting point is not fully general, since  it assumes that all risks will ultimately lead to failure. One can include the possibility that events have a finite probability of not happening at {\em any} time, by adding for each risk $r$ a binary random variable $\tau_r$ to indicate whether or not the calamity button of risk $r$ is pressed at time $t_r$.}
From this distribution follow the survival function $S(t)$, 
i.e. the probability that all events happen later than time $t$, 
\begin{eqnarray}
S(t)&=&\int_0^\infty\!\!\!\!\ldots\!\int_0^\infty\!\dt_0\ldots \dt_R~\Prob(t_0,\ldots,t_R)\prod_{r=0}^R\theta(t_r-t)
\label{eq:survival}
\end{eqnarray}
and the crude cause-specific hazard rates $h_r(t)$, i.e. the probability per unit time that failure $r$ occurs at time $t$ if until that time none of the possible events has yet occurred:
\begin{eqnarray}
h_r(t)&=&\frac{1}{S(t)}\int_0^\infty\!\!\!\!\ldots\!\int_0^\infty\!\dt_0\ldots \dt_R~\Prob(t_0,\ldots,t_R)\delta(t-t_r)\prod_{r^\prime\neq r}^R\theta(t_{r^\prime}-t)
\label{eq:hazardrates}
\end{eqnarray}
Here we used the delta-distribution $\delta(x)$, defined by the identity $\int_{-\infty}^\infty\!\rmd x~\delta(x)f(x)=f(0)$, and the step function, defined by  $\theta(x>0)=1$ and $\theta(x<0)=0$. It is easy to show that the survival function can be written in terms of the crude hazard rates  as
\begin{eqnarray}
S(t)&=&\rme^{-\sum_{r=0}^R \int_0^t\ds~h_r(s)},
\label{eq:S_inrates}
\end{eqnarray}
We assume that we can only observe the timing and the risk label of the {\em earliest} event. The crude cause-specific hazard rates provide the link between theory and observations, since the probability density $P(t,r)$ to find the earliest event occurring at time $t$ and corresponding to risk $r$, is given by
\begin{eqnarray}
P(t,r)&=& h_r(t)\rme^{-\sum_{r^\prime=0}^R \int_0^t\ds~h_{r^\prime}(s)},
\label{eq:dataprob_in_cruderates}
\end{eqnarray}
%In fact, we can also go from the functions $P(t,r)$ to the crude hazard rates, via 
%\begin{eqnarray}
%h_r(t)&=&\frac{P(t,r)}{\sum_{r^\prime=0}^R \int_t^\infty\!\ds~P(s,r^\prime)}
%\end{eqnarray}
The above relations hold irrespective of whether  $\Prob(t_0,\ldots,t_R)$ describes a large or a small cohort, or even a single individual, although obviously the values of $\Prob(t_0,\ldots,t_R)$ would be different. However, at some point in our paper we will work simultaneously with both cohort level and individual level descriptions, and it will become necessary to specify with further indices to which we refer. 

Conditioning on covariate information is trivial in the above picture. For simplicity we assume the covariates to be discrete; in the case of continuous covariates one usually finds the same formulae as those which we derive here, but with integrals instead of sums. Knowing the values $\bz\in\R^p$ of $p$ covariates means starting from the  distribution $\Prob(t_0,\ldots,t_R|\bz)$ which gives the event time statistics of the sub-cohort  of those individuals $i$ that have covariate vector $\bz_i=\bz$. It is related to the previous distribution of the full cohort via  $\Prob(t_0,\ldots,t_R)=\sum_{\bz}\Prob(t_0,\ldots,t_R|\bz)P(\bz)$, where $P(\bz)$ gives the fraction of the cohort that have covariates $\bz$. We then obtain the following {\em covariate-conditioned} survival functions and crude cause-specific hazard rates:
\begin{eqnarray}
S(t|\bz)&=&\int_0^\infty\!\!\!\!\ldots\!\int_0^\infty\!\dt_0\ldots \dt_R~\Prob(t_0,\ldots,t_R|\bz)\prod_{r=0}^R\theta(t_r-t)
\label{eq:covariates_S}
\\
h_r(t|\bz)&=&\frac{1}{S(t|\bz)}\int_0^\infty\!\!\!\!\ldots\!\int_0^\infty\!\dt_0\ldots \dt_R~\Prob(t_0,\ldots,t_R|\bz)\delta(t-t_r)\prod_{r^\prime\neq r}^R\theta(t_{r^\prime}-t)
\label{eq:covariates_h}
\end{eqnarray}
with the usual relation between survival and crude hazard rates, and the usual link to observations:
\begin{eqnarray}
S(t|\bz)&=&\rme^{-\sum_{r^\prime=0}^R\int_0^t\ds~h_{r^\prime}(s|\bz)}
\\
P(t,r|\bz)&=&h_r(t|\bz)\rme^{-\sum_{r^\prime=0}^R\int_0^t\ds~h_{r^\prime}(s|\bz)}
\label{eq:conditioned_dataprob}
\end{eqnarray}

If we study a cohort of $N$ individuals, with coresponding $N$ covariate vectors $\{\bz_1,\ldots,\bz_N\}$, the  survival data $D$ usually consist of 
$N$ samples of event time and event type pairs $(t,r)$:
\begin{eqnarray}
D=\{(t_1,r_1),\ldots,(t_N,r_N)\}
\end{eqnarray}
Since the probability density for an individual with covariate vector $\bz$ to report the pair $(t,r)$ is given by (\ref{eq:conditioned_dataprob}), 
the data likelihood $P(D)= \prod_{i=1}^N P(t_i,r_i|\bz_i)$  obeys
\begin{eqnarray}
\log P(D)&=& \sum_{i=1}^N \log \Big\{h_{r_i}(t_i|\bz_i)\rme^{-\sum_{r=0}^R\int_0^{t_i}\dt~h_{r}(t|\bz_i)}\Big\}
\nonumber
\\
&=& \sum_{r=0}^R\Big\{ \sum_{i=1}^N\delta_{r,r_i}\log h_{r}(t_i|\bz_i)
-\sum_{i=1}^N\int_0^{t_i}\!\dt~h_{r}(t|\bz_i)\Big\}
\end{eqnarray}

\subsection{Decontaminated cause-specific risk measures -- the competing risk problem}

The aim of survival analysis is to extract statistical patterns from survival data, 
that allow us to make risk predictions  for new individuals, usually conditioned on knowledge of their covariates.  We are often interested in one specific {\em primary} risk. 
Many relevant risk-specific quantities can be calculated once we know the crude hazard rates. 
 For instance, the cause-specific cumulative incidence function $F_r(t)$, i.e. the probability that event  $r$ has been {\em observed} at any time prior to time $t$,  is given by
\begin{eqnarray}
F_r(t)&=& \int_0^t\!\dt^\prime~S(t^\prime)h_r(t^\prime)
\end{eqnarray}
Although $F_r(t)$ refers to risk $r$ specifically, it can be heavily influenced by the other risks. For instance, if $F_r(t)$ is small, this may be because event $r$ is intrinsically unlikely, or because it has the habit of being preceeded by alternative events $r^\prime\neq r$. One cannot tell. The problem lies in the difference between events having been {\em observed}  and events having {\em happened} prior to a given time. 

To obtain decontaminated information on an individual primary risk $r$ one must consider the 
hypothetical situation where all other risks $r^\prime\!\neq r$ are disabled. This means  replacing\footnote{A valid critical note that has been made  at this step \citep{Prentice_etal} is that one cannot be sure that this hypothesis is appropriate; it may be that correlated risks share biochemical pathways such that they can never be deactivated independently.}
\begin{eqnarray}
\Prob(t_0,\ldots,t_R)~\to~ \Prob(t_r)\lim_{\Lambda\to \infty}\prod_{r^\prime\neq r}^R \delta(t_{r^\prime}-\Lambda)
\label{eq:hypothetical}
\end{eqnarray}
with the the marginal  event time distribution $\Prob(t_r)=\int\!\ldots\!\int[\prod_{s\neq r}\dt_s] \Prob(t_0,\ldots,t_R)$ of risk $r$.
 Insertion of (\ref{eq:hypothetical}) into (\ref{eq:survival},\ref{eq:hazardrates}) gives, as expected, 
zero values for all non-primary crude hazard rates, but it also affects the value of the primary risk hazard rate.  One now finds the following formulae for the decontaminated cause-specific survival function and hazard rate for risk $r$, indicated with tildes to distinguish them from their crude counterparts:
\begin{eqnarray}
&&
\tilde{S}_r(t)=\int_t^\infty\!\!\dt_r~\Prob(t_r),~~~~~~~~
\tilde{h}_r(t)=-\frac{\rmd}{\dt}\log \tilde{S}_r(t)
\label{eq:decontaminated1}
\end{eqnarray}
or, in the case of covariate conditioning, 
\begin{eqnarray}
&&
\tilde{S}_r(t|\bz)=\int_t^\infty\!\!\dt_r~\Prob(t_r|\bz),~~~~~~~~
\tilde{h}_r(t|\bz)=-\frac{\rmd}{\dt}\log \tilde{S}_r(t|\bz)
\label{eq:decontaminated2}
\end{eqnarray}
In general one will indeed find that $\tilde{h}_r(t)\neq h_r(t)$ and $\tilde{h}_r(t|\bz)\neq h_r(t|\bz)$. 

Equations (\ref{eq:decontaminated1},\ref{eq:decontaminated2}) tell us that to determine the decontaminated risk measures for the primary risk $r$ we must estimate the marginal distributions  $\Prob(t_r)$ or $\Prob(t_r|\bz)$ from survival data. Tsiatis showed \citep{Tsiatis}  that this is impossible without further assumptions. 
His identifiability theorem states that 
for {\em every} $\Prob(t_0,\ldots,t_R)$ there is an alternative  distribution $\overline\Prob(t_0,\ldots,t_R)$ 
that describes {\em independent} times, but such that $\Prob$ and $\overline\Prob$ both generate identical cause-specific hazard rates for all risks:
\begin{eqnarray}
\overline\Prob(t_0,\ldots,t_R)=\prod_{r=0}^R\Big(h_r(t)e^{-\int_0^t\ds~h_r(s)}\Big)
\end{eqnarray}
in which $\{h_r(t)\}$ are the cause-specific hazard rates of $\Prob(t_0,\ldots,t_R)$.
Hence the only information that can be estimated from survival data alone are the (covariate-conditioned) crude cause-specicifc hazard rates. One cannot calculate the distributions $\Prob(t_0,\ldots,t_R)$ or $\Prob(t_0,\ldots,t_R|\bz)$ and their marginals. Without further information or assumptions there is therefore no way to disentangle the different risks and identify the decontaminated cause-specific hazard rates and survival functions. This, in a nutshell, is the competing risk problem. 

One obvious and simple way out is to assume that all risks are statistically independent, i.e. that $\Prob(t_0,\ldots,t_R|\bz)=\prod_{r=0}^R\Prob(t_r|\bz)$. This solves trivially the competing risk problem, since now one finds immediately that $h_r(t|\bz)=\tilde{h}_r(t|\bz)$ for all $r$, and 
\begin{eqnarray}
\tilde{S}_r(t|\bz)&=& \rme^{-\int_0^t\ds~h_r(s|\bz)}
\label{eq:indepS}
\end{eqnarray}
The independence assumption underlies the clinical use of e.g. Cox's proportional hazards regression \citep{Cox} and of Kaplan-Meier estimators of the cause-specific survival function \citep{KM}, which would otherwise be unjustified tools for quantifying cause-specific  survival  regularities. Assuming risk independence may be  acceptable in specific cases. For diseases that share molecular pathways, however, the event times  will be strongly correlated. This can lead to false protectivity effects, and the independence assumption may generate  nonsensical claims.

\section{Heterogeneity-induced competing risks}
\label{sec:HICR}

We introduce a different assumption on the nature of the event time correlations of different risks:
we assume these to be caused by residual cohort heterogeneity. We will call this {\em heterogeneity-induced} competing risks.
The assumption is transparent and much weaker than assuming risk independence, but still imposes sufficient constraints to allow us to solve the computing risk problem. 

\subsection{Connection between cohort level and individual level descriptions}

To give a precise definition of heterogeneity-induced competing risks we first need to describe the connection between cohort-level and individual level risk descriptions. The standard survival analysis formalism is built solely on the starting point of a joint event time distributions; it can therefore also be applied directly to risk at the level of individuals. Let $N$ be the number of individuals  in the cohort to which the distribution $\Prob(t_0,\ldots,t_R)$ refers, labelled by $i=1\ldots N$.  We  write the joint event time distribution of individual $i$ in this cohort as $\Prob_i(t_0,\ldots,t_R)$, and the crude cause-specific hazard rates of individual $i$ as $h_r^i(t)$. It then follows directly from the general theory that 
\begin{eqnarray}
S_i(t)&=&\int_0^\infty\!\!\!\!\ldots\!\int_0^\infty\!\dt_0\ldots \dt_R~\Prob_i(t_0,\ldots,t_R)\prod_{r=0}^R\theta(t_r-t)
\\
h^i_r(t)&=&\frac{1}{S_i(t)}\int_0^\infty\!\!\!\!\ldots\!\int_0^\infty\!\dt_0\ldots \dt_R~\Prob_i(t_0,\ldots,t_R)\delta(t-t_r)\prod_{r^\prime\neq r}^R\theta(t_{r^\prime}-t)
\label{eq:rates_fori}
\end{eqnarray}
and
\begin{eqnarray}
P_i(t,r)&=& h^i_r(t)\rme^{-\sum_{r^\prime=0}^R \int_0^t\ds~h^i_{r^\prime}(s)},
\label{eq:indiv_dataprob}
\end{eqnarray}
Here $S_i(t)=\exp[-\sum_{r=0}^R \int_0^t\ds~h^i_r(s)]$ 
is the survival function of individual $i$, and $P_i(t,r)$ is the probability that the first event for individual $i$ occurs at time $t$ and corresponds to risk $r$. 
When collecting survival data in a cohort, we have the added uncertainty of not knowing which individuals  were picked from the population, so 
the connection between the two levels is simply given by
\begin{eqnarray}
\Prob(t_0,\ldots,t_R)&=&\frac{1}{N}\sum_{i=1}^{N}\Prob_i(t_0,\ldots,t_R)
\label{eq:link_P}
\\
\Prob(t_0,\ldots,t_R|\bz)&=&\frac{\sum_{i,~\bz_i=\bz}\Prob_i(t_0,\ldots,t_R)}
{\sum_{i,~\bz_i=\bz}1}
\label{eq:link_PZ}
\end{eqnarray}
For quantities that depend linearly on the joint event time distribution, the link between cohort level and individual level is a simple averaging over the label $i$, possibly conditioned on covariates, e.g.
\begin{eqnarray}
&&
S(t)=\frac{1}{N}\sum_{i=1}^N S_i(t),~~~~~~~~~~~~~~~~~P(t,r)=\frac{1}{N}\sum_{i=1}^NP_i(t,r)
\\
&&
S(t|\bz)=\frac{\sum_{i,~\bz_i=\bz}S_i(t)}{\sum_{i,~\bz_i=\bz}1},~~~~~~~~P(t,r|\bz)=\frac{\sum_{i,~\bz_i=\bz}P_i(t,r)}{\sum_{i,~\bz_i=\bz}1}
\label{eq:link_dataprob}
\end{eqnarray}
However, quantities such as the crude cause-specific hazard rates depend  in a more complicated way on $\Prob(t_0,\ldots,t_R)$, via their conditioning on survival. As a consequence, cohort level cause-specific hazard rates are {\em not} direct averages over their individual level counterparts. Instead one finds (see appendix \ref{app:link_for_rates} for details):
\begin{eqnarray}
h_r(t)&=&\frac{\sum_{i=1}^N h^i_r(t)\rme^{-\sum_{r^\prime=0}^R\int_0^t\ds~h^i_{r^\prime}(s)}}
{\sum_{i=1}^N\rme^{-\sum_{r^\prime=0}^R\int_0^t\ds~h^i_{r^\prime}(s)}},
\label{eq:link1}
\\
h_r(t|\bz)&=&\frac{\sum_{i,\bz^i=\bz} h^i_r(t)\rme^{-\sum_{r^\prime=0}^R\int_0^t\ds~h^i_{r^\prime}(s)}}
{\sum_{i,\bz^i=\bz} \rme^{-\sum_{r^\prime=0}^R\int_0^t\ds~h^i_{r^\prime}(s)}}
\label{eq:link2}
\end{eqnarray}

\subsection{Heterogeneous cohorts and the different levels of risk complexity}

We always assume our cohorts to be heterogeneous at the level of covariates. The heterogeneity of concern here is in the relation between covariates and risks. 
A homogeneous cohort is one in which the relation between covariates and risks is uniform, so that the
distribution $\Prob_i(t_0,\ldots,t_R)$ can depend on $i$ only via 
$\bz_i$. Put differently, there exists a function $\overline{\Prob}(t_0,\ldots,t_R|\bz)$ such that 
\begin{eqnarray}
\Prob_i(t_0,\ldots,t_R)=\overline{\Prob}(t_0,\ldots,t_R|\bz_i)~~~{\it for~all}~i
\label{eq:homogeneous}
\end{eqnarray}
The same is then true for the
 cause-specific hazard rates:
$h_r^i(t)=\overline{h}_r(t|\bz_i)$ for all $i$, in which $\overline{h}_r(t|\bz)$ is related to 
$\overline{\Prob}(t_0,\ldots,t_R|\bz)$ via equations (\ref{eq:covariates_S},\ref{eq:covariates_h}). 
It also follows directly from (\ref{eq:link2}) that at cohort level the covariate-conditioned event time distribution is  ${\Prob}(t_0,\ldots,t_R|\bz)=\overline{\Prob}(t_0,\ldots,t_R|\bz)$, and 
the covariate-constrained 
crude hazard rates are $h_r(t|\bz)=\overline{h}_r(t|\bz)$, as expected. A special property of homogeneous cohorts is that uncorrelated  individual level risks, i.e. 
${\Prob}_i(t_0,\ldots,t_R)= \prod_{r=0}^R{\Prob}_{i}(t_r)$, imply uncorrelated  covariate-conditioned cohort level risks. This follows from (\ref{eq:link_PZ}):
\begin{eqnarray}
{\Prob}(t_0,\ldots,t_R|\bz)&=& \frac{\sum_{i,~\bz_i=\bz}\prod_{r=0}^R{\Prob}_{i}(t_r)}{\sum_{i,~\bz_i=\bz}1}
~=~\frac{\sum_{i,~\bz_i=\bz}\prod_{r=0}^R\overline{\Prob}(t_r|\bz_i)}{\sum_{i,~\bz_i=\bz}1}
\nonumber
\\
&=& \prod_{r=0}^R\overline{\Prob}(t_r|\bz)~=~\prod_{r=0}^R\Prob(t_r|\bz)
\end{eqnarray}

Heterogeneous cohorts, in contrast, are those where (\ref{eq:homogeneous}) does not hold, i.e. our individuals have further `hidden' features, not captured by the covariates, that impact upon their risks. In such cohorts one will observe a gradual `filtering': high-risk individuals will drop out early, causing time dependencies at cohort level that have no counterpart at the level of individuals. For instance, in the simplest case where all individuals have stationary hazard rates, viz. $h_r^i(t)=h_r^i$, one would according to (\ref{eq:link1},\ref{eq:link2}) still find time dependent crude hazard rates at cohort level. In heterogeneous cohorts it is no longer true that having uncorrelated  individual level risks implies having uncorrelated  covariate-conditioned cohort level risks. 
It is a trivial exercise to devise examples where ${\Prob}_i(t_0,\ldots,t_R)= \prod_{r=0}^R{\Prob}_{i}(t_r)$, but still ${\Prob}(t_0,\ldots,t_R|\bz)\neq \prod_{r=0}^R\Prob(t_r|\bz)$. 

It is clear that risk correlations can be generated at different levels, and that there is a natural hierarchy of cohorts in terms of risk complexity, with implications for the applicability of methods:
\begin{itemize}
\item {\sc Level 1: homogeneous cohort, no competing risks}~\\[2mm]
\hspace*{5mm} individual: ~$\Prob_i(t_0,\ldots,t_R)=\prod_{r=0}^R\overline{\Prob}(t_r|\bz_i)$\\[1mm]
\hspace*{5mm} cohort:~~~~~ ${\Prob}(t_0,\ldots,t_R|\bz)=\prod_{r=0}^R\Prob(t_r|\bz)$
\\[3mm]
Although the members of the cohort will be different 
in their recorded covariates, they are homogeneous in terms of the link between covariates and risk. For each individual, the event times  of all risks are statistically independent, and their probabilities are determined fully by the covariates alone. Since there is no residual heterogeneity, there is no competing risk problem; crude and true cause-specific hazard rates and survival functions are identical.

\item {\sc Level 2: heterogeneous cohort, no competing risks}~\\[2mm]
\hspace*{5mm} individual:~ $\Prob_i(t_0,\ldots,t_R)=\prod_{r=0}^R\Prob_{i}(t_r)$\\[1mm]
\hspace*{5mm} cohort:~~~~~ ${\Prob}(t_0,\ldots,t_R|\bz)=\prod_{r=0}^R\Prob(t_r|\bz)$
\\[3mm]
Here for each individual the event times  of all risks are still statistically independent, but their susceptibilities are no longer determined by their recorded covariates alone (reflecting e.g. disease sub-groups or the impact of further unobserved covariates).  However, this residual heterogeneity does not manifest itself in  risk correlations at cohort level. 
One will therefore observe heterogeneity-induced effects, such as `cohort filtering', but  no competing risks.  

\item {\sc Level 3: heterogeneity-induced competing risks}~\\[2mm]
\hspace*{5mm} individual:~ $\Prob_i(t_0,\ldots,t_R)=\prod_{r=0}^R\Prob_{i}(t_r)$\\[1mm]
\hspace*{5mm} cohort:~~~~~ ${\Prob}(t_0,\ldots,t_R|\bz)\neq \prod_{r=0}^R\Prob(t_r|\bz)$
\\[3mm]
Here for each individual the event times  of all risks are statistically independent, but their susceptibilities are not determined by their recorded covariates alone, similar to level 2.  However, now this residual cohort heterogeneity leads to  risk correlations at cohort level, reflecting e.g. common unobserved risk factors or co-morbitities, and thereby to informative censoring.
One  will now observe competing risks phenomena, such as false protectivity and false exposure.

\item {\sc Level 4:  individual and cohort level competing risks}~ \\[2mm]
\hspace*{5mm} individual:~ $\Prob_i(t_0,\ldots,t_R)\neq \prod_{r=0}^R\Prob_{i}(t_r)$\\[1mm]
\hspace*{5mm} cohort:~~~~~ ${\Prob}(t_0,\ldots,t_R|\bz)\neq \prod_{r=0}^R\Prob(t_r|\bz)$
\\[3mm] 
This is the most complex situation from a modelling point of view, where both at the level of individuals and at cohort level the event times of different risks are correlated. We will again observe competing risk phenomena, but can no longer say where these are generated. 
\end{itemize}
In fact, having correlations amongst non-primary risks is harmless in the context of 
decontaminating primary risk measures. The only issue is whether there are correlations between primary and non-primary risks. So we could in principle make a further distinction between having 
${\Prob}(t_0,\ldots,t_R|\bz)=\prod_{r=0}^R\Prob(t_r|\bz)$ and ${\Prob}(t_0,\ldots,t_R|\bz)=\Prob(t_r|\bz)
\Prob(t_0,\ldots,t_{r-1},t_{r+1},\ldots,t_R|\bz)$; the latter property being weaker but still sufficient. 
Here we will not persue this distinction; it is obvious how the theory should be adapted to accommodate non-primary risk correlations. 

Levels 1 and 2 are those where the assumption of statistically independent risks, underlying the clinical use of e.g. Cox regression and Kaplan-Meier estimators, is valid.
At level 2 there is still no competing risk problem, but the heterogeneity demands parametrisations of crude primary hazard rates at cohort level that are more complex than those used in Cox regression, which is the rationale behind the development of frailty models and random effects models, as well as the latent class models of \citep{Muhten}.
All these methodologies still only model the cause-specific hazard rate of the {\em primary} risk, and therefore cannot handle cohorts beyond complexity level 2. 

 Level 4, which includes homogeneous cohorts with individual level competing risks, represents the most complex scenario, which we will not deal with in this paper. 
Our focus is on level 3: that of cohorts with 
 heterogeneity-induced competing risks.  Here the correlations between cohort level event times  have their origin strictly in {\em correlations between disease susceptibilities of individuals}, e.g. someone with a high hazard rate for a disease A may also be likely to have a high hazard rate for B, for reasons not explained by the recorded covariates. 
Most papers on frailty, random effects and latent class models assume implicitly that competing risks are induced by such residual heterogeneity.
We now show that the assumption of heterogeneity-induced competing risks leads to a transparent resolution of the competing risk problem. 

\subsection{Implications of having heterogeneity-induced competing risks}

In the case of heterogeneity-induced competing risks we have independent event times at the level of individuals, hence for each individual $i$ we can be sure that
\begin{eqnarray}
\Prob_{i}(t_r)&=&h_r^i(t)\rme^{-\int_0^t\ds~h_r^i(s)}
\end{eqnarray}
The covariate-conditioned cohort level event time marginals are therefore
\begin{eqnarray}
\Prob_{r}(t_r|\bz)&=&\frac{\sum_{i,\bz_i=\bz}h_r^i(t)\rme^{-\int_0^t\ds~h_r^i(s)}}
{\sum_{i,\bz_i=\bz}1}
\end{eqnarray}
and via (\ref{eq:decontaminated2})
we can write the decontaminated cause-specific survival function and hazard rate as
\begin{eqnarray}
\tilde{S}_r(t|\bz)&=&
\frac{\sum_{i,\bz_i=\bz}\rme^{-\int_0^{t}\ds~h_r^i(s)}}
{\sum_{i,\bz_i=\bz}1}
\label{eq:trueStilde}
\\
\tilde{h}_r(t|\bz)&=&
\frac{\sum_{i,\bz_i=\bz}h_r^i(t)\rme^{-\int_0^{t}\ds~h_r^i(s)}}{\sum_{i,\bz_i=\bz}\rme^{-\int_0^{t}\ds~h_r^i(s)}}
\label{eq:truehtilde}
\end{eqnarray}
Here we used $\int_0^\infty\!\ds~h_r^i(s)=\infty$ for all $(i,r)$, which follows from the assumed normalisation of 
$\Prob_i(t_0,\ldots,t_R)$. Expressions (\ref{eq:trueStilde},\ref{eq:truehtilde}) are similar but not identical to the
formulae (\ref{eq:indepS},\ref{eq:link2}) for the decontaminated cause-specific survival function and the  crude covariate-conditioned cause-specific
hazard rates which we would have taken to be correct  had we assumed all risks to be independent:
 \begin{eqnarray}
S_r(t|\bz)&=& \rme^{-\int_0^t\ds~h_r(s|\bz)}
\label{eq:falseS}
\\
h_r(t|\bz)&=&\frac{\sum_{i,\bz^i=\bz} h^i_r(t)\rme^{-\sum_{r^\prime=0}^R\int_0^t\ds~h^i_{r^\prime}(s)}}
{\sum_{i,\bz^i=\bz} \rme^{-\sum_{r^\prime=0}^R\int_0^t\ds~h^i_{r^\prime}(s)}}
\label{eq:falseh}
\end{eqnarray}
All this is easily interpreted. 
In formula (\ref{eq:truehtilde}) the probability that individual $i$ survives until time $t$ 
 is correctly given by the factor $\exp[-\int_0^{t}\ds~h_r^i(s)]$ (which causes the `cohort filtering'), since no risk other than $r$ is active.  In contrast, in (\ref{eq:falseh}) {\em all} risks contribute to cohort filtering. The formulae (\ref{eq:truehtilde}) and (\ref{eq:falseh}) will therefore be non-identical,  unless we have risk independence, which in (\ref{eq:falseh}) would give rise to an identical factor in numerator and denominator that would drop out. 
Indeed, the differences between (\ref{eq:trueStilde},\ref{eq:truehtilde}) and  (\ref{eq:falseS},\ref{eq:falseh}) quantify the severity of the competing risk problem in the cohort at hand. We also see that in the case of a homogenous cohort, where $h_r^i(t)=\overline{h}_r(t|\bz_i)$ for all $(r,i)$, one indeed recovers $\tilde{S}_r(t|\bz)=S_r(t|\bz)$  and $\tilde{h}_r(t|\bz)=h_r(t|\bz)$.  

Similarly, we can work out the formula that provides the link between the theory and survival data. 
Inserting (\ref{eq:indiv_dataprob}) into (\ref{eq:link_dataprob}) immediately leads us to 
\begin{eqnarray}
P(t,r|\bz)=\frac{\sum_{i,~\bz_i=\bz} h^i_r(t)\rme^{-\sum_{r^\prime=0}^R \int_0^t\ds~h^i_{r^\prime}(s)}}{\sum_{i,~\bz_i=\bz}1}
\label{eq:truedataprob}
\end{eqnarray}
We conclude that the assumption that competing risks (if present) are of the heterogeneity-induced type allows one to derive relatively simple formulae both for the decontaminated cause-specific quantities of interest
and for the likelihood of observing individual survival data. What remains is to identify the {\em minimal} level of description required for evaluating these formulae, and to determine how the minimum required information can in practice be estimated from survival data.

\subsection{Canonical level of description for resolving heterogeneity-induced competing risks}

The canonical level of description is the minimal set of observables in terms of which we can write both the decontaminated risk-specific quantities (\ref{eq:trueStilde},\ref{eq:truehtilde}) (so that we can calculate what we are interested in),  and the data likelihood 
(\ref{eq:truedataprob}) (so it can be estimated from survival data).  In (\ref{eq:trueStilde},\ref{eq:truehtilde}) we need the covariate-constrained  distribution of individual hazard rates for the primary risk. In (\ref{eq:truedataprob}) we need in addition the covariate-constrained  distribution of the cumulative rates of non-primary risks. In combination we see that the minimal description would be the functional distribution
\begin{eqnarray} 
\WW[h_r,h_{\!/r}|\bz]&=& \frac{\sum_{i,\bz_i=\bz}\delta_{\rm F}\big[h_r\!-\!h_r^i\big]\delta_{\rm F}\big[h_{\!/r}\!-\!\sum_{r^\prime\neq r}h^i_{r^\prime}\big]}{\sum_{i,\bz_i=\bz}1}
\label{eq:minimal}
\end{eqnarray}
Here $\delta_{\rm F}$ denotes the functional $\delta$-distribution, defined by the functional integral identity 
\begin{eqnarray}
\int\{\rmd f\}\delta_{\rm F}[f]G[f]=\left.G[f]\right|_{f(t)=0~\forall t\geq 0}
\end{eqnarray}
 $\WW[h_r,h_{\!/r}|\bz]$ tells us for each possible choice of the function pair $\{h_r(t),h_{\!/r}(t)\}$: which fraction of those individuals in our cohort that have covariates $\bz$ also have the individual primary
hazard rates $h_r^i(t)=h_r(t)$ and the cumulative non-primary hazard  rates $\sum_{r^\prime\neq r}h_r^i(t)=h_{\!/r}(t)$. 

In practice it will often be advantageous to relax slightly our requirement of a {\em minimal} description. The non-primary risks will usually be mutually very different in their characteristics, so finding an efficient parametrisation of the dependence  on $\sum_{r^\prime\neq r}h_{r^\prime}^i(t)$ in $\WW[h_r,h_{\!/r}|\bz]$ will be awkward. 
A slightly redundant alternative choice, but one that is more easily parametrised,  would be 
\begin{eqnarray} 
\WW[h_0,\ldots,h_R|\bz]&=& \frac{\sum_{i,\bz_i=\bz}\prod_{r=0}^R\delta_{\rm F}\big[h_r\!-\!h_r^i\big]}{\sum_{i,\bz_i=\bz}1}
\label{eq:practical}
\end{eqnarray}
 It gives the joint functional distribution over the cohort of all $R+1$ individual cause-specific hazard rates at all times. The distribution (\ref{eq:minimal}) follows from (\ref{eq:practical})  via
\begin{eqnarray}
\WW[h_r,h_{\!/r}|\bz]&=&\int\!\{\rmd h_0^\prime\ldots \rmd h_R^\prime\}~\WW[h^\prime_0,\ldots,h^\prime_R|\bz]~\delta_{\rm F}\big[h_r\!-\!h_r^\prime\big]\delta_{\rm F}\big[h_{\!/r}\!-\!\sum_{r^\prime\neq r}h^\prime_{r^\prime}\big]
\end{eqnarray}
For independent risks one would simply find the factorised form $\WW[h_0,\ldots,h_R|\bz]=\prod_{r=0}^R \WW[h_r|\bz]$.
If we know (\ref{eq:practical})  we can write the decontaminated risk-specific quantities (\ref{eq:trueStilde},\ref{eq:truehtilde}) as 
\begin{eqnarray}
\tilde{S}_r(t|\bz)&=& \int\!\{\rmd h_0\ldots \rmd h_R\}~\WW[h_0,\ldots,h_R|\bz]~\rme^{-\int_0^{t}\ds~h_r(s)}
\label{eq:trueStilde_inW}
\\
\tilde{h}_r(t|\bz)&=&\frac{ \int\!\{\rmd h_0\ldots \rmd h_R\}~\WW[h_0,\ldots,h_R|\bz]~h_r(t)\rme^{-\int_0^{t}\ds~h_r(s)}}
{ \int\!\{\rmd h_0\ldots \rmd h_R\}~\WW[h_0,\ldots,h_R|\bz]~\rme^{-\int_0^{t}\ds~h_r(s)}}
\label{eq:truehtilde_inW}
\end{eqnarray}
whereas their `crude' counterparts, which  would be reported upon assuming independent risks, are
\begin{eqnarray}
S_r(t|\bz)&=& \rme^{-\int_0^t\ds~h_r(s|\bz)}
\label{eq:falseS_inW}
\\
h_r(t|\bz)&=& \frac{\int\!\{\rmd h_0\ldots \rmd h_R\}~\WW[h_0,\ldots,h_R|\bz]~
h_r(t)\rme^{-\sum_{r^\prime=0}^R\int_0^t\ds~h_{r^\prime}(s)}}{\int\!\{\rmd h_0\ldots \rmd h_R\}~\WW[h_0,\ldots,h_R|\bz]~
\rme^{-\sum_{r^\prime=0}^R\int_0^t\ds~h_{r^\prime}(s)}}
\label{eq:falseh_inW}
\end{eqnarray}
Having formulae for the latter is useful for quantifying  the impact of competing risks in the cohort, via comparison to 
(\ref{eq:trueStilde_inW},\ref{eq:truehtilde_inW}). 
One can easily confirm that if the primary risk $r$ is not correlated with the non-primary risks (i.e. if $\WW[h_0,\ldots,h_R|\bz]=\WW[h_1|\bz]\WW[h_0,h_2,\ldots,h_R|\bz]$), or if there is just one risk, the formulae (\ref{eq:trueStilde}) and (\ref{eq:falseS_inW})
as well as (\ref{eq:truehtilde_inW}) and (\ref{eq:falseh_inW}) become pairwise identical, as expected.

The data likelihood (\ref{eq:truedataprob}) acquires the form 
\begin{eqnarray}
P(t,r|\bz)=\int\!\{\rmd h_0\ldots \rmd h_R\}~\WW[h_0,\ldots,h_R|\bz] ~h_r(t)\rme^{-\sum_{r^\prime=0}^R \int_0^t\ds~h_{r^\prime}(s)}
\label{eq:dataprob_inW}
\end{eqnarray}
An alternative formula for $P(t,r|\bz)$ follows upon combining 
(\ref{eq:link2}) with (\ref{eq:falseh_inW}). In appendix \ref{app:twoformulas} we show that the two formulae are indeed identical, as they should.
Finally,  the covariate-conditioned cause-specific cumulative incidence functions $F_r(t|\bz)=\int_0^t \!\ds ~P(s,r|\bz)$ can be written as
\begin{eqnarray}
F_r(t|\bz)=\int\!\{\rmd h_0\ldots \rmd h_R\}~\WW[h_0,\ldots,h_R|\bz] \int_0^t \!\rmd t^\prime ~h_r(t^\prime)\rme^{-\sum_{r^\prime=0}^R \int_0^{t^\prime}\ds~h_{r^\prime}(s)}
\label{eq:CIF_inW}
\end{eqnarray}

The level of description (\ref{eq:practical}) is both sufficient and necessary for handling heterogeneity-induced competing risks, apart from the trivial option to combine the non-primary risks $r^\prime\neq r$ into a single non-primary risk, which would lead to (\ref{eq:minimal}). More specifically, one cannot work with the crude cohort-level covariate-conditioned hazard rates alone: whereas the latter  can all be calculated from $\WW[h_0,\ldots,h_R|\bz]$ via (\ref{eq:falseh_inW}), the converse is not true. 
 In fact it is easy to show that for any $\WW[h_0,\ldots,h_R|\bz]$ there exists an alternative distribution $\overline{\WW}[h_0,\ldots,h_R|\bz]$ that describes a {\em homogeneous} cohort, such that $\WW$ and $\overline{\WW}$ give identical crude cohort-level covariate-conditioned cause-specific hazard rates, namely
\begin{eqnarray}
 \overline{\WW}[h_0,\ldots,h_R|\bz]=\prod_{r=0}^R\delta_{\rm F}[h_r\!-\!h_r(\bz)]
\end{eqnarray}
in which $h_r(\bz)$ is the function of time given in (\ref{eq:falseh_inW}).

\subsection{Estimation of $\WW[h_0,\ldots,h_R|\bz]$ from survival data}
\label{sub:estimation}

When there is a limited supply of survival data one must determine 
the relevant quantities in parametrised form, to avoid overfitting, and estimate the parameters from the data. 
It is still true that the data likelihood can be expressed in terms of the crude cohort-level covariate-conditioned cause-specific hazard rates, so one cannot extract information on $\WW[h_0,\ldots,h_R|\bz]$ from survival data that is  not contained in $\{h_t(t|\bz)\}$. 
However, even relatively simple and natural parametrisations of $\WW[h_0,\ldots,h_R|\bz]$ will via (\ref{eq:falseh_inW}) correspond to nontrivial crude conditioned hazard rates (with time dependencies caused by cohort filtering), that one would have been very unlikely to propose when parametrising directly at the level of the crude hazard rates. This situation mirrors that of using frailty or latent class models for chorts at complexity level 2.

We thus assume $\WW[h_0,\ldots,h_R|\bz]$ to be a member of a parametrised family of conditioned distributions $\WW[h_0,\ldots,h_R|\bz,\btheta]$, in which $\btheta\in\Omega$  denotes the vector of parameters and $\Omega$ is its value domain. For our cohort of $N$ individuals, with covariate vectors $\{\bz_1,\ldots,\bz_N\}$, the  available survival data consist of the 
$N$ samples of event time and event type pairs,  
$D=\{(t_1,r_1),\ldots,(t_N,r_N)\}$. 
Since the probability density for an individual with covariate vector $\bz$ to report the pair $(t,r)$ is given by (\ref{eq:dataprob_inW}), 
the data likelihood $P(D|\btheta)= \prod_{i=1}^N P(t_i,r_i|\bz_i)$ given the parameters $\btheta$ is
\begin{eqnarray}
P(D|\btheta)&=& \prod_{i=1}^N \int\!\{\rmd h_0\ldots \rmd h_R\}~\WW[h_0,\ldots,h_R|\bz_i,\btheta] ~h_{r_i}(t_i)\rme^{-\sum_{r^\prime=0}^R \int_0^{t_i}\ds~h_{r^\prime}(s)}
\end{eqnarray}
If we concentrate all the survival data in two empirical distributions,
\begin{eqnarray}
\hat{P}(t,r|\bz)=\frac{\sum_{i,~\bz_i=\bz}\delta(t-t_i)\delta_{r,r_i}}{\sum_{i,~\bz_i=\bz}1},~~~~~~~~\hat{P}(\bz)=\frac{1}{N}\sum_{i=1}^N \delta_{\bz,\bz_i}
\end{eqnarray}
(with the Kronecker delta-symbol, $\delta_{ab}=1$ if $a=b$ and $\delta_{ab}=0$ otherwise)
we can write the log-likelihood $\LL(\btheta)=\log P(D|\btheta)$ of the observed data as
\begin{eqnarray}
\LL(\btheta)&=&N\sum_{\bz} \hat{P}(\bz)\sum_{r=0}^R \int\!\dt~\hat{P}(t,r|\bz)
\log\int\!\{\rmd h_0 \ldots \rmd h_R\}~\WW[h_0,\ldots,h_R|\bz,\btheta] 
\nonumber
\\
&&\hspace*{30mm}\times~h_{r}(t)\rme^{-\sum_{r^\prime=0}^R \int_0^{t}\ds~h_{r^\prime}(s)}
\label{eq:general_loglikelihood}
\end{eqnarray}
This log-likelihood can be interpreted in terms of the dissimilarity of the  empirical function 
$\hat{P}(t,r|\bz)$ and the  model prediction $\hat{P}(t,r|\bz,\btheta)$ , i.e. the result of substituting $\WW[h_0,\ldots,h_R|\bz,\btheta] $ into (\ref{eq:dataprob_inW}):
\begin{eqnarray}
\frac{\LL(\btheta)}{N}=
\sum_{\bz} \hat{P}(\bz)\Big\{\sum_{r=0}^R \int\!\!\dt~\hat{P}(t,r|\bz)
\log  \hat{P}(t,r|\bz)
-\!
\sum_{r=0}^R \int\!\!\dt~\hat{P}(t,r|\bz)
\log  \Big(\frac{\hat{P}(t,r|\bz)}{P(t,r|\bz,\btheta)}\Big)
\Big\}
\end{eqnarray}
The first (entropic) term is independent of $\btheta$ and the second term is minus the Kullback-Leibler distance $D(\hat{P}||P)$  \citep{InfoTheory}
between $\hat{P}$ and $P$,  hence finding the most probable parameters $\btheta$ is equivalent to minimizing $D(\hat{P}||P)$.

From this starting point one can follow different routes for estimating $\btheta$, each with specific advantages and limitations, and each with different computational costs. For instance, 
in maximum likelihood (ML) estimation one simply uses the value $\hat{\btheta}$ for which the data are most likely, 
\begin{eqnarray}
\hat{\btheta}_{\rm ML}&=& {\rm argmax}_{\btheta\in\Omega}~\LL(\btheta)
\label{eq:MP}
\end{eqnarray}
In the Bayesian formalism 
 one does not commit oneself to one choice for $\btheta$, but one uses the full Bayesian posterior parameter probability $P(\btheta|D)$. 
Given a parameter prior $P(\btheta)$ this would give
\begin{eqnarray}
\log P(\btheta|D)&=&  \LL(\btheta)+\log P(\btheta)- \log \int_{\Omega}\!\rmd\btheta^\prime~P(\btheta^\prime)e^{\LL(\btheta^\prime)}
\label{eq:Bayes}
\end{eqnarray}
Finally, in maximum a posteriori 
probability (MAP) estimation one uses the value $\hat{\btheta}$ for which the Bayesian posterior parameter probability is maximal,
\begin{eqnarray}
\hat{\btheta}_{\rm MAP}&=& {\rm argmax}_{\btheta\in\Omega}~\big[ \LL(\btheta)+\log P(\btheta)\big]
\label{eq:MAP}
\end{eqnarray}
For sufficiently large data sets the above three estimation methods would all become equivalent, i.e. $\lim_{N\to\infty}\hat{\btheta}_{\rm MAP}=\lim_{N\to\infty}\hat{\btheta}_{\rm ML}$ and $\lim_{N\to\infty} P(\btheta|D)=\delta(\btheta-\btheta_{\rm ML})$. This follows from  
the property $\lim_{N\to\infty} \LL(\btheta)/N=\lim_{N\to\infty}[ \LL(\btheta)+\log P(\btheta)]/N=\lim_{N\to\infty}P(\btheta|D)/N$.

There are obviously multiple and more advanced variations  on the above parameter estimation protocols. For instance, one could reduce the overfitting danger in the ML method by including Aikake's Information Criterion (AIC) or the Bayesian Information Criterion (BIC). Alternative  Bayesian routes involve e.g. hyperparameter estimation, or variational approximations of the posterior parameter distribution to reduce computation costs, or model selection to select good parametrisations $\WW[h_0,\ldots,h_R|\bz,\btheta]$. See e.g. \citep{MacKay}.

\section{Parametrisation of $\WW[h_0,\ldots,h_R|\bz]$}

A transparent class of parametrisations for $\WW[h_0,\ldots,h_R|\bz,\btheta]$ is obtained by assuming that the proportional hazards assumption of Cox holds at the level of individuals. Here we work out the relevant equations, and show how the resulting theory includes the conventional methods (e.g. Cox regression, frailty models, random effect models, latent class analysis)   as special simplified cases.

\subsection{Generic parametrisation}

We note that for each individual $i$ in our cohort we can always write the individual cause-specific hazard rates in the form 
$h_r^i(t)=\lambda^i_r(t)\exp(\beta^{0i}_r+\sum_{\mu=1}^p\beta_r^{\mu i}z^i_\mu)$. 
The time-dependence for each risk is concentrated in $\lambda^i_r(t)$. The  parameters $\beta^{0i}_{r}$ represent individual risk-specific frailties, which have to be normalised in such a way as to remove the redundancy of the parametrisation, i.e. to eliminate the invariance of the individual hazard rates under the transformation $\{\lambda_r^i(t),\beta_r^{0i}\}\to \{\lambda_r^i(t)\rme^{-\zeta_r^i},\beta_r^{0i}+\zeta_r^i\}$.
According to (\ref{eq:practical}), we can then write 
$\WW[h_0,\ldots,h_R|\bz]$ as
\begin{eqnarray}
\WW[h_0,\ldots,h_R|\bz,\MM]&=& \int\!\rmd\bbeta_0\ldots\rmd\bbeta_R\int\!\{\rmd\lambda_0\ldots\rmd\lambda_R\}~\MM(\bbeta_0,\ldots,\bbeta_R;\lambda_0,\ldots,\lambda_R|\bz)
\nonumber
\\
&&\times
\prod_{r=0}^R\delta_{\rm F}\big[h_r\!-\!\lambda_r\rme^{\beta_r^0+\sum_{\mu=1}^p\beta_r^\mu z_\mu}\big]
\label{eq:parametrisation}
\end{eqnarray}
with the short-hand $\bbeta_r=(\beta_r^0,\ldots,\beta_r^p)$, and with
\begin{eqnarray}
\MM(\bbeta_0,\ldots,\bbeta_R;\lambda_0,\ldots,\lambda_R|\bz)&=& \frac{\sum_{i,~\bz_i=\bz}\prod_{r=0}^R \big\{
\delta_{\rm F}[\lambda_r\!-\!\lambda_r^i] \delta(\bbeta_r\!-\!\bbeta_r^{i})\big\}}
{\sum_{i,~\bz_i=\bz}1}
\end{eqnarray}
 In the language of the previous subsection we thus have a parametrisation in which $\btheta=\MM$. Note that (\ref{eq:parametrisation}) is still completely general. In particular, it does not yet imply a proportional hazards assumption at the level of individuals unless $\MM(\bbeta_0,\ldots,\bbeta_R;\lambda_0,\ldots,\lambda_R|\bz)$ is independent of $\bz$. However, it is a useful representation only if $\MM(\bbeta_0,\ldots,\bbeta_R;\lambda_0,\ldots,\lambda_R|\bz)$ depends in  a relatively simple way on the 
parameters $\{\bbeta_0,\ldots,\bbeta_R\}$ and the functions $\{\lambda_0,\ldots,\lambda_R\}$. 
To compactify our notation further we introduce the short-hands $\bbeta\cdot\bz=\beta^0+\sum_{\mu=1}^p\beta^\mu z^\mu$ and $\Lambda_t(t)=\int_0^{t}\ds~\lambda_{r}(s)$. 
Inserting (\ref{eq:parametrisation}) into (\ref{eq:general_loglikelihood}) then gives the data  log-likelihood $\LL(\MM)$
corresponding to (\ref{eq:parametrisation}):
\begin{eqnarray}
\LL(\MM)&=&
N\sum_{\bz} \hat{P}(\bz)\sum_{r=0}^R \int\!\dt~\hat{P}(t,r|\bz)
\log\int\!\rmd\bbeta_0\ldots\rmd\bbeta_R\int\!\{\rmd\lambda_0,\ldots,\lambda_R\}
\nonumber
\\
&&~\times ~\MM(\bbeta_0,\ldots,\bbeta_R;\lambda_0,\ldots,\lambda_R|\bz) ~
 \lambda_r(t)~\rme^{\bbeta_r\cdot\bz-\sum_{r^\prime=0}^R
\Lambda_{r^\prime}(t)\exp(\bbeta_{r^\prime}\cdot\bz)
}
\end{eqnarray}
This is equivalent to
\begin{eqnarray}
\LL(\MM)&=&
\sum_{i=1}^N 
\log\int\!\rmd\bbeta_0\ldots\rmd\bbeta_R\int\!\{\rmd\lambda_0,\ldots,\lambda_R\}~\MM(\bbeta_0,\ldots,\bbeta_R;\lambda_0,\ldots,\lambda_R|\bz_i) 
\nonumber
\\
&&
\hspace*{30mm}
\times \lambda_{r_i}(t_i)~\rme^{\bbeta_{r_i}\cdot\bz_i-\sum_{r^\prime=0}^R
\Lambda_{r^\prime}(t_i)\exp(\bbeta_{r^\prime}\cdot\bz_i)
}
\label{eq:generic_loglikelihood}
\end{eqnarray}
The individual cause-specific hazard rates of all individuals are written in a form reminiscent of \citep{Cox}, but with time-dependent factors and time-independent regression and frailty parameters for the $R+1$ risks that are not uniform over the cohort, but distributed according to the distribution $\MM(\bbeta_0,\ldots,\bbeta_R;\lambda_0,\ldots,\lambda_R|\bz)$, in the spirit of fraily and random effects models. However, here this is done for {\em all risks} simultaneously, so the complexities of competing risks and false protectivities are 
captured by the correlation structure of the joint distribution $\MM(\bbeta_0,\ldots,\bbeta_R;\lambda_0,\ldots,\lambda_R|\bz)$.

All applications and examples in  the remainder of this paper are  based on the generic parametrisation (\ref{eq:parametrisation}).  
Given (\ref{eq:parametrisation}) one obtains formulae for the various decontaminated and `crude' cause-specific quantities of interest,  which are fully exact as long as $\MM(\bbeta_0,\ldots,\bbeta_R;\lambda_0,\ldots,\lambda_R|\bz)$ is kept general. 
We write the single-risk marginals of $\MM(\bbeta_0,\ldots,\bbeta_R;\lambda_0,\ldots,\lambda_R|\bz)$  as
\begin{eqnarray}
\MM(\bbeta_r;\lambda_r|\bz)= \int\!\Big( \prod_{r^\prime\neq r}\rmd\bbeta_{r^\prime}\{\rmd\lambda_{r^\prime}\}\Big)\MM(\bbeta_0,\ldots,\bbeta_R;\lambda_0,\ldots,\lambda_R|\bz)
\end{eqnarray}
For the decontaminated cause-specific survival functions and hazard rates we then get
\begin{eqnarray}
\tilde{S}_r(t|\bz)&=&  \int\!\rmd\bbeta_r\{\rmd\lambda_r\}~\MM(\bbeta_r;\lambda_r|\bz) ~\rme^{- \exp(\bbeta_{r}\cdot \bz)\Lambda_{r}(t)}
\label{eq:Stilde_generic}
\\
\tilde{h}_r(t|\bz)&=&  \frac{ \int\!\rmd\bbeta_r\{\rmd\lambda_r\}~\MM(\bbeta_r;\lambda_r|\bz) ~\lambda_r(t)\rme^{\bbeta_r\cdot \bz- \exp(\bbeta_{r}\cdot \bz)\Lambda_{r}(t)}}
{  \int\!\rmd\bbeta_r\{\rmd\lambda_r\}~\MM(\bbeta_r;\lambda_r|\bz) ~\rme^{- \exp(\bbeta_{r}\cdot \bz)\Lambda_{r}(t)}}
\label{eq:htilde_generic}
\end{eqnarray}
The crude hazard rates and the data probability become
\begin{eqnarray}
h_r(t|\bz)&=& \label{eq:h_generic}
\\
&&\hspace*{-20mm} \frac{\int\!\rmd\bbeta_0\ldots\rmd\bbeta_R\int\!\{\rmd\lambda_0\ldots\rmd\lambda_R\}~\MM(\bbeta_0,\ldots,\bbeta_R;\lambda_0,\ldots,\lambda_R|\bz) ~\lambda_r(t)
\rme^{\bbeta_r\cdot \bz-\sum_{r^\prime=0}^R \exp(\bbeta_{r^\prime}\cdot \bz)\Lambda_{r^\prime}(t)}}{\int\!\rmd\bbeta_0\ldots\rmd\bbeta_R\int\!\{\rmd\lambda_0\ldots\rmd\lambda_R\}~\MM(\bbeta_0,\ldots,\bbeta_R;\lambda_0,\ldots,\lambda_R|\bz) ~
\rme^{-\sum_{r^\prime=0}^R \exp(\bbeta_{r^\prime}\cdot \bz)\Lambda_{r^\prime}(t)}}
\nonumber
\\[1.5mm]
P(t,r|\bz)&=&
\\
&&\hspace*{-20mm}
\int\!\!\rmd\bbeta_0\ldots\rmd\bbeta_R\!\int\!\{\rmd\lambda_0,\ldots,\lambda_R\}~\MM(\bbeta_0,\ldots,\bbeta_R;\lambda_0,\ldots,\lambda_R|\bz) \lambda_r(t)\rme^{\bbeta_r\cdot \bz-\sum_{r^\prime=0}^R \exp(\bbeta_{r^\prime}\cdot \bz)\Lambda_{r^\prime}(t)}
\nonumber
\end{eqnarray}
and, finally, the covariate-conditioned cumulative cause-specific incidence functions are
\begin{eqnarray}
F_r(t|\bz)&\!\!=\!\!& \!\int\!\rmd\bbeta_0\ldots\rmd\bbeta_R\int\!\{\rmd\lambda_0\ldots\rmd\lambda_R\}~\MM(\bbeta_0,\ldots,\bbeta_R;\lambda_0,\ldots,\lambda_R|\bz)
\nonumber
\\
&&\times\int_0^t\!\!\rmd t^\prime~\lambda_r(t^\prime)\rme^{\bbeta_r\cdot \bz-\sum_{r^\prime=0}^R \exp(\bbeta_{r^\prime}\cdot \bz)\Lambda_{r^\prime}(t^\prime)}
\end{eqnarray}

\subsection{Connection with conventional regression methods}

Since the parametrisation (\ref{eq:parametrisation}) is generic, {\em all} existing regression methods that are compatible with the assumption of heterogeneity-induced competing risks will 
in principle correspond to specific choices for the covariate-conditioned distribution $\MM(\bbeta_0,\ldots,\bbeta_R;\lambda_0,\ldots,\lambda_R|\bz)$. We label the primary risk as $r=1$. 
All methods that assume primary and non-primary risks to be independent would have 
$\MM(\bbeta_0,\ldots,\bbeta_R;\lambda_0,\ldots,\lambda_R|\bz)= \MM(\bbeta_1,\lambda_1|\bz)\MM(\bbeta_0,\bbeta_2,\ldots,\bbeta_R;\lambda_0,\lambda_2,\ldots,\lambda_R|\bz)$ with some specific choice for the form of $\MM(\bbeta_1,\lambda_1|\bz)$. Examples from this group are
\begin{itemize}
\item
{\em Cox's proportional hazards regression  \citep{Cox}}
\\[2mm]
Here one assumes that there is no variability in the parameters $(\bbeta_1,\lambda_1)$ of the primary risk. Elimination of parameter  redundancy then means that $\beta_1^0$ is absorbed into $\lambda_1(t)$, and we find
\begin{eqnarray}
\MM(\bbeta_1;\lambda_1|\bz)&=& \delta_{\rm F}[\lambda_1\!-\!\hat{\lambda}]~\delta(\beta_1^0)\prod_{\mu=1}^p\delta(\beta_1^\mu\!-\!\hat{\beta}^\mu)
\label{eq:Cox}
\end{eqnarray}
Via the maximum likelihood method one can express the base hazard rate $\hat{\lambda}(t)$ 
in terms of the regression coefficients $\{\hat{\beta}^\mu\}$ (giving Breslow's formula), substitution of which then leads 
directly to Cox's equations   \citep{Cox}. See appendix \ref{app:links} for details. 

\item{\em Simple frailty models}
\\[2mm]
In simple frailty models, such as \citep{Vaupel,Yashin},  the frailty parameters of different risks are assumed to be statistically independent,
so the heterogeneity of the cohort that impacts upon the primary risk is concentrated in the random parameter $\beta_1^0$:
\begin{eqnarray}
\MM(\bbeta_1;\lambda_1|\bz)&=& \delta_{\rm F}[\lambda_1\!-\!\hat{\lambda}]~g(\beta_1^0)\prod_{\mu=1}^p\delta(\beta_1^\mu\!-\!\hat{\beta}^\mu)
\label{eq:frailty}
\end{eqnarray}
One usually chooses the frailty distribution $g(\beta_1^0)$ to be of a specific  parametrised form
that allows one to do various relevant integrals over $\beta_1^0$  analytically.  See appendix \ref{app:links} for details.

\item{\em Simple random effects models}
\\[2mm]
In simple random effects models, such as  \citep{Vaida},  one still assumes the parameters of the primary risk to be independent of the  non-primary ones, but now the regression coeficients that couple to the covariates are non-uniform:
\begin{eqnarray}
\MM(\bbeta_1;\lambda_1|\bz)&=& \delta_{\rm F}[\lambda_1\!-\!\hat{\lambda}]~W(\bbeta_1)
\label{eq:random_effect}
\end{eqnarray}
One then assumes a specific parametrized form for the distribution $W(\bbeta_1)$ and estimates its parameters from the data. 

 \item {\em Latent class models}
\\[2mm]
The latent class models of \citep{Muhten} are recovered upon assuming the cohort to consists of a finite number of discrete  sub-cohorts. Each  is of the  type (\ref{eq:Cox}), but with a distinct base hazard rate and distinc regression coefficients. The probabilities $w_\ell$ for individuals to belong to each sub-cohort $\ell$ are allowed to depend on their covariates $\bz$, as in  \citep{Reboussin}:
\begin{eqnarray}
\MM(\bbeta_1;\lambda_1|\bz)&=& \sum_{\ell=1}^L w(\ell|\bz)~\delta_{\rm F}[\lambda_1\!-\!\hat{\lambda}^\ell]~\delta(\beta_1^0)\prod_{\mu=1}^p\delta(\beta_1^\mu\!-\!\hat{\beta}^{\ell\mu})
\label{eq:latentclass}
\\
w(\ell|\bz)&=& \frac{\rme^{\alpha_0^\ell+\sum_{\mu=1}^p \alpha_\mu^\ell z^\mu}}
{\sum_{\ell^\prime=1}^L\rme^{\alpha_0^{\ell^\prime}+\sum_{\mu=1}^p \alpha_\mu^{\ell^\prime} z^\mu}}
\end{eqnarray}
\end{itemize}
The above models all focus on the parameters of the primary risk only, and thereby lose the ability to deal with the competing risk problem. Only few papers try to characterise all risk and their possible parameter interactions simultaneously, such as \cite{Zahl} or \citep{DiSerio}, but they do not yet develop their ideas into full systematic regression and/or decontamination protocols. 
Of course there are multiple variations on the above models. These include versions with time-dependent covariates, and models with non-latent classes in the sense that for each individual $i$ one knows the class label $\ell(i)\in\{1,\ldots,L\}$. It is easy to see how they would fit into the generic formulation.

\subsection{A simple  latent class parametrisation for heterogeneity-induced competing risks}

\begin{figure}[t]
%\unitlength=0.13mm
\unitlength=0.122mm
{\small
%\hspace*{30mm}
\hspace*{32mm}
\begin{picture}(400,270)
\put(300,120){\oval(1070,260)}

\put(-25,120){\oval(380,220)}
\put(-20,195){\here{\underline{\em Latent class 1}}}
\put(-20,140){\here{\em fraction: $w_1$}}
\put(0,90){\here{\em for all $r\!>\!0$:\hspace*{30mm}}}
\put(-20,55){\here{$h_r^i(t)=\hat{\lambda}_r(t)\rme^{\hat{\beta}^{10}_r+\sum_{\mu=1}^p \hat{\beta}^{1\mu}_r z^\mu_i}$}}

\put(300,120){\here{$\cdots\cdots\cdots\cdots\cdots$}}

\put(625,120){\oval(380,220)}
\put(630,195){\here{\underline{\em Latent class $L$}}}
\put(630,140){\here{\em fraction: $w_L$}}
\put(650,90){\here{\em for all $r\!>\!0$:\hspace*{30mm}}}
\put(630,55){\here{$h_r^i(t)=\hat{\lambda}_r(t)\rme^{\hat{\beta}^{L 0}_r+\sum_{\mu=1}^p \hat{\beta}^{L\mu}_r z^\mu_i}$}}

\end{picture}
}
\vsp
\caption{Illustration of the parametrisation (\ref{eq:our_model}). 
All individuals $i$ in the cohort are assumed to have personalised cause-specific hazard rates $h_r^i(t)$ which for all risks $r=1\ldots R$ are of the proportional hazards form. 
The cohort is allowed to be heterogeneous in that it may consist of $L$ sub-cohorts (or `latent classes'), labelled by $\ell =1\ldots L$. Each latent class $\ell$ contains individuals with risk-specific frailties $\hat{\beta}_r^{\ell 0}$ and with risk-specific regression parameters $\hat{\beta}_r^{\ell \mu}$ to capture the impact of covariates. 
The base hazard rates $\hat{\lambda}_r(t)$ of the risks are assumed not to vary between individuals.  The class membership of the individuals in our data set is not known a priori, but can be inferred {\em a posteriori}.}
\label{fig:heterogeneity}
\end{figure}
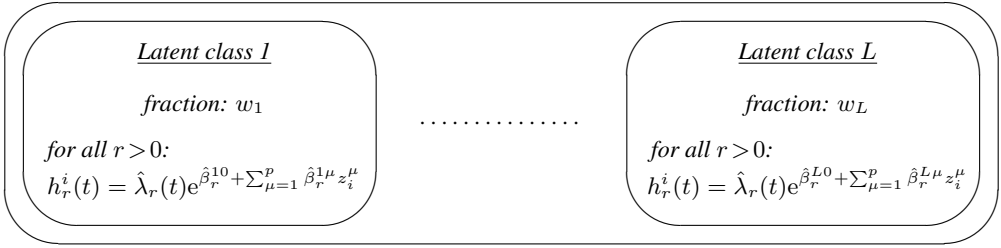

Any description that includes {\em all} risks and their correlations, a prerequisite for decontaminating primary risk measures, will have significantly more parameters than those limited to the primary risk. In view of the overfitting danger it is then vital that one limits the complexity of the chosen parametrisation. 
 The difference between frailty and random effects models is only in whether the risk variability relates to known or unknown covariates, so it seems logical to combine both. 
If we take the heterogeneity to be discrete, but without the covariate dependence of class probabilities of (\ref{eq:latentclass}), if we assume the end-of-trial risk not to depend on the covariates, and if we choose the base hazard rates of all risks to be uniform in the cohort, 
 we obtain a model family in which $\MM(\bbeta_0,\ldots,\bbeta_R;\lambda_0,\ldots,\lambda_R|\bz)=\delta(\bbeta_0)\delta_{\rm F}[\lambda_0\!-\!\hat{\lambda}_0]\MM(\bbeta_1,\ldots,\bbeta_R;\lambda_1,\ldots,\lambda_R)$, with
\begin{eqnarray}
\MM(\bbeta_1,\ldots,\bbeta_R;\lambda_1,\ldots,\lambda_R)&=& \MM(\bbeta_1,\ldots,\bbeta_R)\prod_{r=1}^R
\delta_{\rm F}[\lambda_r\!-\!\hat{\lambda}_r]
\label{eq:our_model}
\\
\MM(\bbeta_1,\ldots,\bbeta_R)&=& 
 \sum_{\ell=1}^L w_\ell \prod_{r=1}^R
\delta(\bbeta_r\!-\!\hat{\bbeta}_r^{\ell})
\end{eqnarray}
Here $\hat{\bbeta}_r^{\ell}=(\hat{\beta}_r^{\ell 0},\ldots,\hat{\beta}_r^{\ell p})$.
See Figure \ref{fig:heterogeneity} for an illustration of what this parametrisation (\ref{eq:our_model}) means in terms of individual cause-specific hazard rates in our cohort. 
For any choice for the number $L$ of assumed latent classes, 
the remaining parameters to be estimated from the data are: the cause-specific hazard rates $\{\hat{\lambda}_r(t)\}$ of all risks, the $L$ class sizes $w_\ell\in[0,1]$ (subject to $\sum_{\ell=1}^L w_\ell=1$), the regression coeficients $\{\hat{\beta}_r^{\ell\mu}\}$ and  frailty parameters $\{\hat{\beta}_r^{\ell 0}\}$ of all risks $r=1\ldots R$ and all latent classes. The remaining parametrisation invariance is 
$\{\hat{\lambda}_r(t),\hat{\beta}_r^{\ell 0}\}\to \{\hat{\lambda}_r(t)\rme^{-\zeta_r},\hat{\beta}_r^{\ell 0}+\zeta_r\}$ for all $\ell$, which is removed by defining $\hat{\beta}_r^{10}=0$ for all $r$. Finding the optimal number $L$ of classes is in principle a simple Bayesian model selection problem.

The log-likelihood (\ref{eq:generic_loglikelihood}) of our survival data is at the core of all parameter estimation procedures. For the multi-risk parametrisation (\ref{eq:our_model})  it simplifies to the following expression, with our usual short-hand 
$\bbeta_r^{\ell}\cdot\bz=\beta_r^{\ell 0}+\sum_{\mu=1}^p\beta_r^{\ell \mu}z^\mu$ and with $\notdelta_{ab}=1-\delta_{ab}$:
\begin{eqnarray}
\LL(\MM)&=&
\sum_{i=1}^N 
\log   \hat{\lambda}_{r_i}(t_i)
+
\sum_{i=1}^N 
\log  \Big\{\sum_{\ell=1}^L w_\ell ~
\rme^{\hat{\bbeta}_{r_i}\cdot\bz_i-\sum_{r=0}^R
\hat{\Lambda}_{r}(t_i)\exp(\hat{\bbeta}_{r}^\ell\cdot\bz_i)}
\Big\}
\nonumber
\\
&=& \LL_0(\MM)+ \LL_{\rm risks}(\MM)
\end{eqnarray}
with a first term that includes the (mostly irrelevant) end-of-trial censoring information, and a second term that contains the quantities related to actual risks:
\begin{eqnarray}
 \LL_0(\MM) &=&
 \sum_{i=1}^N \delta_{0 r_i}
\log   \hat{\lambda}_{0}(t_i)
-
\sum_{i=1}^N \hat{\Lambda}_0(t_i)
\\
 \LL_{\rm risks}(\MM)
&=&
\sum_{i=1}^N \notdelta_{0 r_i}
\log   \hat{\lambda}_{r_i}(t_i)
+
\sum_{i=1}^N 
\log  \Big\{\sum_{\ell=1}^L w_\ell ~
\rme^{\notdelta_{0 r_i}\hat{\bbeta}_{r_i}\cdot\bz_i-\sum_{r=1}^R\!
\hat{\Lambda}_{r}(t_i)\exp(\hat{\bbeta}_{r}^\ell\cdot\bz_i)}
\Big\}~~~~
\label{eq:LL_our_model}
\end{eqnarray}
Inserting (\ref{eq:our_model}) into our formulae 
for the decontaminated cause-specific survival function and hazard rates of the true risks $r>0$ gives the relatively simple and intuitive expressions
\begin{eqnarray}
\tilde{S}_r(t|\bz)&=& \sum_{\ell=1}^L w_\ell  ~\rme^{- \exp(\hat{\bbeta}^\ell_{r}\cdot \bz)\hat{\Lambda}_{r}(t)}
\label{eq:StildeLC}
\\
\tilde{h}_r(t|\bz)&=&  \hat{\lambda}_r(t)~\frac{ \sum_{\ell=1}^L w_\ell  ~\rme^{\hat{\bbeta}^\ell_r\cdot \bz- \exp(\hat{\bbeta}^\ell_{r}\cdot \bz)\hat{\Lambda}_{r}(t)}}
{  \sum_{\ell=1}^L w_\ell  ~\rme^{- \exp(\hat{\bbeta}^\ell_{r}\cdot \bz)\hat{\Lambda}_{r}(t)}}
\end{eqnarray}
The crude hazard rate and the data probability become
\begin{eqnarray}
h_r(t|\bz)&=&  \hat{\lambda}_r(t)~\frac{\sum_{\ell=1}^L w_\ell ~
\rme^{\hat{\bbeta}^\ell_r\cdot \bz-\sum_{r^\prime=1}^R \exp(\hat{\bbeta}^\ell_{r^\prime}\cdot \bz)\hat{\Lambda}_{r^\prime}(t)}}{\sum_{\ell=1}^L w_\ell  ~
\rme^{-\sum_{r^\prime=1}^R \exp(\hat{\bbeta}^\ell_{r^\prime}\cdot \bz)\hat{\Lambda}_{r^\prime}(t)}}
\label{eq:cruderateLC}
\\
P(t,r|\bz)&=& \hat{\lambda}_r(t)\rme^{-\hat{\Lambda}_0(t)}\sum_{\ell=1}^L w_\ell ~\rme^{\hat{\bbeta}^\ell_r\cdot \bz-\sum_{r^\prime=1}^R \exp(\hat{\bbeta}^\ell_{r^\prime}\cdot \bz)\hat{\Lambda}_{r^\prime}(t)}
\end{eqnarray}
From the crude cause-specific hazard rates $h_r(t|\bz)$ follow as usual the crude cause-specific survival functions $S_r(t|\bz)$ for $r=1\ldots R$, via 
te relation $S_r(t|\bz)=\exp[-\int_0^t\!\rmd s~ h_r(s|\bz)]$. 
The cumulative cause-specific incidence functions corresponding to (\ref{eq:our_model}) for $r=1\ldots R$ are
\begin{eqnarray}
F_r(t|\bz)&=& \int_0^t\!\!\rmd t^\prime~\hat{\lambda}_r(t^\prime)\rme^{-\hat{\Lambda}_0(t^\prime)}~
\sum_{\ell=1}^L w_\ell ~
\rme^{\hat{\bbeta}^\ell_r\cdot \bz-\sum_{r^\prime=1}^R \exp(\hat{\bbeta}^\ell_{r^\prime}\cdot \bz)\hat{\Lambda}_{r^\prime}(t^\prime)}
\end{eqnarray}
The specific parametrisation (\ref{eq:our_model}) has two further useful features:
\begin{itemize}
\item
After the 
class sizes $(w_1,\ldots,w_L)$ have been inferred, for a chosen (or optimised) value of $L$, one obtains
 the {\em effective} number $L_{\rm eff}$ of classes via Shannon's information-theoretic entropy $S$ \citep{InfoTheory}, which takes into account any class size differences:
\begin{eqnarray}
L_{\rm eff}&=& \rme^{S},~~~~~~~~S=-\sum_{\ell=1}^L w_\ell \log w_\ell
\label{eq:Leff}
\end{eqnarray}
\item
Since our latent classes are defined in terms of the relation between covariates and risk, one cannot predict class membership for individuals on the basis of covariate information alone. However, after having identified the parameters of our cohort,  Bayesian arguments allow us to calculate retrospective class membership probabilities for any individual on which we have the covariates $\bz$ and survival information $(t,r)$. For each class label $\ell$, the model (\ref{eq:our_model}) gives
\begin{eqnarray}
P(t,r|\bz,\ell)&=& \hat{\lambda}_r(t)~\rme^{\hat{\bbeta}_r^{\ell}\cdot\bz-\hat{\Lambda}_0(t)-\sum_{r^\prime=1}^R \exp(\hat{\bbeta}_{r^\prime}^{\ell}\cdot\bz)
\hat{\Lambda}_{r^\prime}(t)}
\end{eqnarray}
Hence, using 
$P(t,r,\ell|\bz)= 
P(t,r|\bz,\ell)w_\ell$ and $P(t,r|\bz)= 
\sum_{\ell^\prime=1}^L P(t,r|\bz,\ell^\prime)w_{\ell^\prime}$, we obtain
\begin{eqnarray}
P(\ell|t,r,\bz)&=&  \frac{w_\ell P(t,r|\bz,\ell)}{\sum_{\ell^\prime=1}^L w_{\ell^\prime} P(t,r|\bz,\ell^\prime)}
\nonumber
\\
&=&  \frac{w_\ell ~\rme^{\hat{\bbeta}_r^{\ell}\cdot\bz-\sum_{r^\prime=1}^R \exp(\hat{\bbeta}_{r^\prime}^{\ell}\cdot\bz)
\hat{\Lambda}_{r^\prime}(t)}}{\sum_{\ell^\prime=1}^L w_{\ell^\prime}~ \rme^{\hat{\bbeta}_r^{\ell^\prime}\cdot\bz-\sum_{r^\prime=1}^R \exp(\hat{\bbeta}_{r^\prime}^{\ell^\prime}\cdot\bz)
\hat{\Lambda}_{r^\prime}(t)}}
\label{eq:find_class}
\end{eqnarray}
Formula (\ref{eq:find_class}) allows us to assign each individual in our cohort retrospectively to the identified latent classes: the probability that individual $i$ belongs to class $\ell$ is given by $P(\ell|t_i,r_i,\bz_i)$. 
\end{itemize}
The effective number of classes (\ref{eq:Leff}) can also be a practical tool for identifying the optimal value of $L$, complementary to Bayesian model selection. 
A useful application of (\ref{eq:find_class})
would be to aid the search for informative new covariates that could increase our ability to predict personalised risk  in heterogeneous cohorts. Such missing covariates 
 are expected to be features that impact upon risk and which patients in the same class tend to have in common.  

 Finally, instead of imposing by hand the independence of the end-of-trial risk on covariates (to reduce the number of model parameters), one could also treat the end-of-trial risk as any other risk. 
Any parameter estimation protocol should then report that $\hat{\beta}_0^{\ell\mu}=0$ for all $\ell$ and all $\mu=1\ldots p$, which gives a convenient sanity test of numerical imlementations. In addition one sometimes finds that this trivial enlargement of the search space reduces the impact of  spurious local minima.

\subsection{A unimodal parametrisation for heterogeneity-induced competing risks}

For unimodal distributions of individual  regression parameters 
a more appropriate parametrisation would be to replace in (\ref{eq:our_model}) the 
latent class distribution $\MM(\bbeta_1,\ldots,\bbeta_R)$ by a
Gaussian one:
\begin{eqnarray}
\MM(\bbeta_1,\ldots,\bbeta_R;\lambda_1,\ldots,\lambda_R)&=& \MM(\bbeta_1,\ldots,\bbeta_R)\prod_{r=1}^R
\delta_{\rm F}[\lambda_r\!-\!\hat{\lambda}_r]
\label{eq:gaussian_model}
\\
\MM(\bbeta_1,\ldots,\bbeta_R)&=& \frac{\rme^{-\frac{1}{2}
\left(\!\!\!\begin{array}{c}\bbeta_1 \!-\!\hat{\bbeta}_1\\[-1mm]
\vdots
\\[-1mm]
\bbeta_R\!-\!\hat{\bbeta}_R\end{array}\!\!\!\right)
\cdot \bC^{-1}
\left(\!\!\!\begin{array}{c}\bbeta_1 \!-\!\hat{\bbeta}_1\\[-1mm]
\vdots
\\[-1mm]
\bbeta_R\!-\!\hat{\bbeta}_R\end{array}\!\!\!\right)
}}{(2\pi)^{(p+1)R/2}~{\rm Det}^{\frac{1}{2}}\bC}
\label{eq:gaussian_measure}
\end{eqnarray} 
For this choice the parameters to be estimated are:  the base hazard rates $\{\hat{\lambda}_r(t)\}$ of all risks, the location  $\{\hat{\bbeta}_1,\ldots,\hat{\bbeta}_R\}$ of the centre of (\ref{eq:gaussian_measure}), and the entries of the $(p\!+\!1)R\!\times\! (p\!+\!1)R$ covariance matrix $\bC$. 
 The corresponding risk-dependent part of the data log-likelihood (\ref{eq:generic_loglikelihood}) is 
\begin{eqnarray}
\LL_{\rm risks}(\MM)&=&
\sum_{i=1}^N \notdelta_{0 r_i}
\log\hat{\lambda}_{r_i}(t_i)+
\sum_{i=1}^N 
\log\int\!\rmd\bbeta_1\ldots\rmd\bbeta_R~\MM(\bbeta_1,\ldots,\bbeta_R) 
\nonumber
\\
&&
\hspace*{30mm}
\times ~\rme^{\notdelta_{0 r_i}\bbeta_{r_i}\cdot\bz_i-\sum_{r=1}^R
\hat{\Lambda}_{r}(t_i)\exp(\bbeta_{r}\cdot\bz_i)
}
\label{eq:gauss_model_big}
\end{eqnarray}
For each risk $r$ and each $\bz$  the linear combination $\bbeta_r\cdot\bz$ 
will also be a Gaussian variable. This allows us to simplify the above $(p\!+\!1)R$-dimensional integral to a $R$-dimensional one, which involves the 
$R\!\times\!R$ matrix $\bK(\bz)$ with entries 
\begin{eqnarray}
K_{rr^\prime}(\bz)=\bz\cdot\bC^{rr^\prime}\bz,~~~~~~~~r,r^\prime=1\ldots R
\label{eq:Kaz}
\end{eqnarray}
Here $\bC^{rr^\prime}$ denotes the $(p\!+\!1)\!\times\!(p\!+\!1)$ sub-matrix of $\bC$ with entries 
$(\bC^{rr^\prime})_{\mu\mu^\prime}=\bra \beta_r^\mu\beta_{r^\prime}^{\mu^\prime}\ket_{\MM}-
\bra \beta_r^\mu\ket_{\MM}\bra\beta_{r^\prime}^{\mu^\prime}\ket_{\MM}$.  Averages refer to the dstribution (\ref{eq:gaussian_measure}). The result is (see Appendix \ref{app:GaussianIntegral} for details):
\begin{eqnarray}
\LL_{\rm risks}(\MM)&=&
\sum_{i=1}^N \notdelta_{0 r_i}
\log\hat{\lambda}_{r_i}(t_i)
+\sum_{r=1}^R
\sum_{i=1}^N\delta_{r r_i} \Big\{\hat{\bbeta}_{r}\cdot\bz_i+\frac{1}{2}\bz_i\cdot \bC^{rr}\bz_i 
\nonumber
\\
&&
+~
\log\! \int\!{\rm D}\by
 ~
\rme^{-\sum_{r^\prime=1}^R
\hat{\Lambda}_{r^\prime}(t_i)\exp[\hat{\bbeta}_{r^\prime}\cdot\bz_i
+\bz_i\cdot \bC^{r r^\prime}\!\bz_i+\sum_{r^\pprime=1}^R [\bK^{\frac{1}{2}}(\bz_i)]_{r^\prime r^\pprime}y_{r^\pprime}]
}
\Big\}
\label{eq:gauss_model_simpler}
\end{eqnarray}
Here $\by\in\R^R$ and $\bK^{\frac{1}{2}}(\bz)$ is the matrix defined by the property $[\bK^{\frac{1}{2}}(\bz)]^2=\bK(\bz)$. It is unique because $\bK(\bz)$ is non-negative definite and symmetric for any $\bz$.

Finally, Jenssen's inequality tells us that  $\int\! {\rm D}\by~\exp[u(\by)]\geq \exp \int\! {\rm D}\by~u(\by)$. This allows us after integration over $\by$ to obtain an explicit lower bound for $\LL_{\rm risks}(\MM)$,  which becomes an equality in the absence of heterogeneity (i.e. when $\bC\to\bnull$) and which is convenient in numerical calculations:
\begin{eqnarray}
\LL_{\rm risks}(\MM)&\geq &
\sum_{i=1}^N \notdelta_{0 r_i}
\log\hat{\lambda}_{r_i}(t_i)
+\sum_{r=1}^R
\sum_{i=1}^N\delta_{r r_i} \Big\{\hat{\bbeta}_{r}\cdot\bz_i+\frac{1}{2}\bz_i\cdot \bC^{rr}\bz_i 
\nonumber
\\
&&
-~\sum_{r^\prime=1}^R
\hat{\Lambda}_{r^\prime}(t_i)\rme^{\hat{\bbeta}_{r^\prime}\cdot\bz_i
+\bz_i\cdot \bC^{r r^\prime}\!\bz_i+ \frac{1}{2}\bz_i\cdot\bC^{r^\prime r^\prime}\bz_i}
\Big\}
\end{eqnarray}

\section{Application to synthetic survival data}

To test our regression method under controlled conditions we apply it first to synthetic data with heterogeneity-induced competing risks,  generated from populations of the type (\ref{eq:our_model}). 
Details of the numerical generation of these data are given in Appendix \ref{app:numerical}.
Our method is required to  uncover and map a cohort's risk and association substructure, if such substructure exists, i.e. report the number and sizes of sub-cohorts and their distinct regression parameters for all risks. It should then use this extracted information to generate correct decontaminated survival curves, and assign individuals retrospectively to their latent classes, with statistically significant accuracy.

\subsection{Cohort substructure and regression parameters}

We generated numerically event times and event types for three heterogeneous data sets A,B and C. Each set has $N=1600$ individuals from $L=2$ latent classes of equal size,  with at most two real risks, and with end-of-trial censoring at time $t=50$. 
Each indivdual $i$ has three  covariates $(z_i^1,z_i^2,z_i^3)$, drawn randomly and independently from $P(z)=(2\pi)^{-1/2}e^{-z^2/2}$. All frailty parameters $\beta_r^{\ell 0}$ are zero. 
The base hazard rates of the risks are time-independent: $\hat{\lambda}_1(t)=0.05$ (primary risk) and $\hat{\lambda}_2(t)=0.1$ (if the secondary risk is enabled). Table 1 shows the further specifications of the data sets, together with the results of performing proportional hazards regression \citep{Cox}, and our generic  heterogeneous regression according to the latent class log-likelihood   (\ref{eq:LL_our_model}) where the MAP protocol was complemented with Aikake's Information Criterion as described in  (\ref{eq:Psi_to_minimise}). 
The data sets were constructed such that they have fully {\em identical} primary risk characterics. In set A there is heterogeneity but no competing risk. In set B a secondary risk is introduced, which in one of the two classes targets individuals similar to those most sensitive to the primary risk (with respect to the first covariate); here one expects 
false protectivity effects. In set C a secondary risk is introduced, which in one of the two classes targets individuals similar to those least sensitive to the primary risk (with respect to the first covariate); here one expects 
false exposure effects.

\begin{figure}[t]
\vsp{\small
\hspace*{3mm}
\begin{tabular}{|r| lll|}
\hline
\smallroom								& {\sc class structure}~	& {\sc primary risk} & {\sc secondary risk}\\
\hline
\smallroom {\normalsize\sc data A~~}	 && 1194 events &\hspace*{10mm}--
\\
	\smallroom				& class 1: ~$w_1\!=\!0.5$		& $\bbeta_1^1\!=\!(~2, ~0,~0)$ 	& \hspace*{10mm}-- \\
					& class 2: ~$w_2\!=\!0.5$ 		& $\bbeta_1^2\!=\!(-2, 0,0)$ 	& \hspace*{10mm}--
\hspace*{24.3mm} \\[1mm]
\hline
\smallroom {\em Cox} 		& \hspace*{10mm} --   			& $\bbeta_1\!=\!(0.01, 0.05, -0.01)$ 	&   \hspace*{10mm}--\\[1mm]
\hline
\smallroom {\footnotesize\em generic}		& class 1: ~$w_1\!=\!0.51$		& $\bbeta_1^1\!=\!(~1.99, ~0.01, ~0.06)$ 	& \hspace*{10mm}-- \\
					& class 2: ~$w_2\!=\!0.49$ 		& $\bbeta_1^2\!=\!(-0.94, 0.03,\!-0.01)$\hspace*{2.1mm}	& \hspace*{10mm}-- \\[0.5mm]
\hline
\end{tabular}
\vspace*{5mm}

\hspace*{3mm}
\begin{tabular}{|r| lll|}
\hline
	\smallroom							& {\sc class structure}~	& {\sc primary risk} & {\sc secondary risk}\\
\hline
\smallroom{\normalsize\sc data B~~}	&& 512 events  & 943 events  
\\
\smallroom	& class 1:~ $w_1\!=\!0.5$		& $\bbeta_1^1\!=\!(~2, ~0,~0)$ 	& $\bbeta_2^1\!=\!(3, 0, 0)$ \\
					& class 2:~ $w_2\!=\!0.5$		& $\bbeta_1^2\!=\!(-2, 0, 0)$ & $\bbeta_2^2\!=\!(0, 0, 0)$ \\[1mm]
\hline
\smallroom {\em Cox} 		&  \hspace*{10mm}--   			& $\bbeta_1\!=\!(-0.13, 0.10, 0.00)$ 	&  \hspace*{10mm}--\\[1mm]
\hline
\smallroom {\footnotesize\em generic} 		& class 1: ~$w_1\!=\!0.49$ 		& $\bbeta_1^1\!=\!(-1.97, 0.18,~0.03)$\hspace*{2.2mm}	& $\bbeta_2^1\!=\!(-0.06,\!-0.07,\!-0.17)$ \\
					& class 2: ~$w_2\!=\!0.51$ 		& $\bbeta_1^2\!=\!(~1.98, ~0.01,\!-0.01)$\hspace*{3.5mm} 	& $\bbeta_2^2\!=\!(~3.01, ~0.00, ~0.07)$ 
\\[0.5mm]
\hline
\end{tabular}
\vspace*{5mm}

\hspace*{3mm}
\begin{tabular}{|r| lll|}
\hline
	\smallroom							& {\sc class structure}~	& {\sc primary risk} & {\sc secondary risk}\\
\hline
\smallroom {\normalsize\sc data C~~}	 &&  682 events & 914 events 
\\
\smallroom	& class 1: ~$w_1\!=\!0.5$		& $\bbeta_1^1\!=\!(~2,~0,~0)$ 		& $\bbeta_2^1\!=\!(-3, 0, 0)$ \\
					& class 2: ~$w_2\!=\!0.5$\ 		& $\bbeta_1^2\!=\!(-2, 0,0)$ 	& $\bbeta_2^2\!=\!(~0,~0,0)$ \\[1mm]
\hline
\smallroom {\em Cox} 		&  \hspace*{10mm}--   			& $\bbeta_1\!=\!(-0.34, 0.05, 0.06)$ 	&   \hspace*{10mm}--\\[1mm]
\hline
\smallroom {\footnotesize\em generic} 			& class 1: ~$w_1\!=\!0.49$		& $\bbeta_1^1\!=\!(~1.82,~0.05,~0.12)$ 	& $\bbeta_2^1\!=\!(-3.04,\!-0.09,0.06)$~~~ \\
					& class 2: ~$w_2\!=\!0.51$ 		& $\bbeta_1^2\!=\!(-2.01,\!-0.04,\!-0.02)$ 	& $\bbeta_2^2\!=\!(~0.05,\!-0.05,\!-0.05)$ \\[0.5mm]
\hline
\end{tabular}}
\vspace*{8mm}
\label{table1}

\noindent
{\bf Table 1. }Characteristics of three synthetic data sets A, B and C, all of the form 
(\ref{eq:our_model}) with two equally large latent classes and $N=1600$ individuals. All three have {\em identical} primary risk parameters; they differ only in characteristics of the secondary risk. We also show the results for each set of Cox's propertional hazards regression, and of our generic regression method 
based on  (\ref{eq:LL_our_model}) and (\ref{eq:Psi_to_minimise}), with parameters extimated via Maximum A Posteriori likelihood augmented with Aikake's Information Criterion. 
Error bars in regression parameters are of the order of the last specified decimal. Cox regression cannot cope with heterogeneity and reports non-informative parameters, whereas
the generic heterogeneous regression method is able to identify reasonably accurately the cohort's substructure (number and sizes of the classes) and its class-specific regression parameters.
\end{figure}

As expected, the proportional hazards regression method \citep{Cox} fails to report meaningful results, since it aims to describe the relation between covariates and the primary risk in each data set with a single regression vector $(\beta_1^1,\beta_1^2,\beta_1^3)$.  The heterogeneous regression based on  (\ref{eq:LL_our_model},\ref{eq:Psi_to_minimise}) always reports the correct number of classes ($L=2$), and the correct class-specific parameters (within accuracy limits determined by numerical search accuracy and finite sample size). Note that the assigment of class labels to identified classes is in principle arbitrary; see e.g. the  regression results for data set B, where the class labelled $\ell=2$ is labelled $\ell=1$ in the data definition. 

\subsection{Decontaminated survival functions}

The second test of the regression method and its numerical implementation is to verify that for all three data sets A, B and C described in table \ref{table1} it can extract the correct decontaminated covariate-conditioned survival curve $\tilde{S}_1(t|\bz)$ for the primary risk, from the survival data alone. The result 
\clearpage

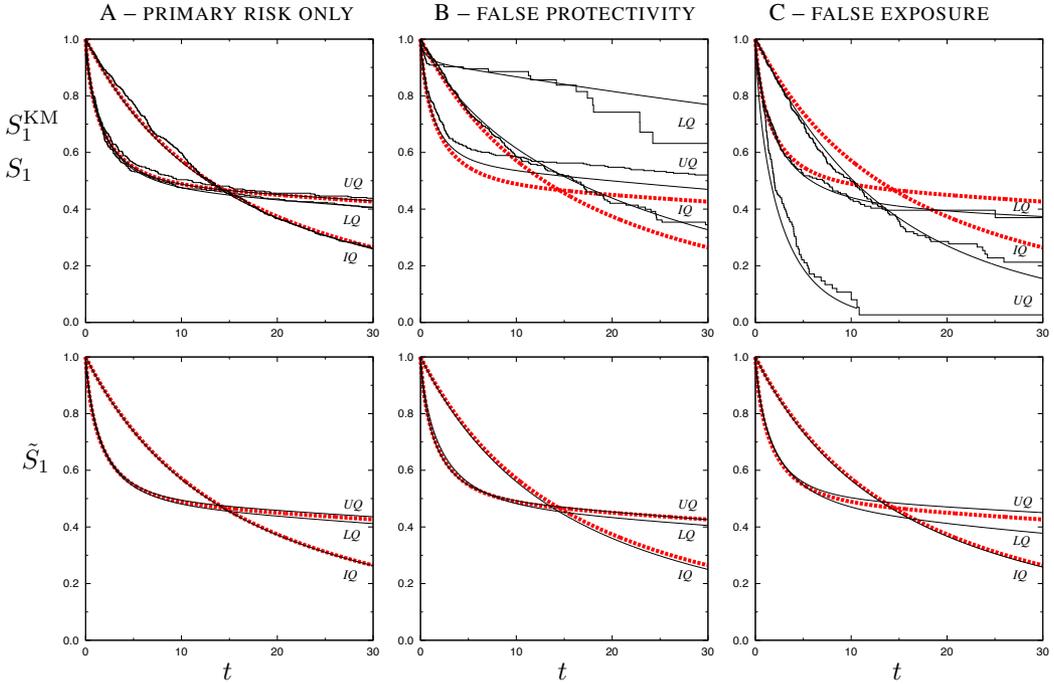
\begin{figure}[t]
\unitlength=0.44mm
\vsp

\hspace*{-0mm}
\begin{picture}(300,200)
\put(-0,163){$S_1^{\tiny\rm KM}$}
\put(-0,150){$S_1$}
\put(0,100){\includegraphics[width=138\unitlength,height=95\unitlength]{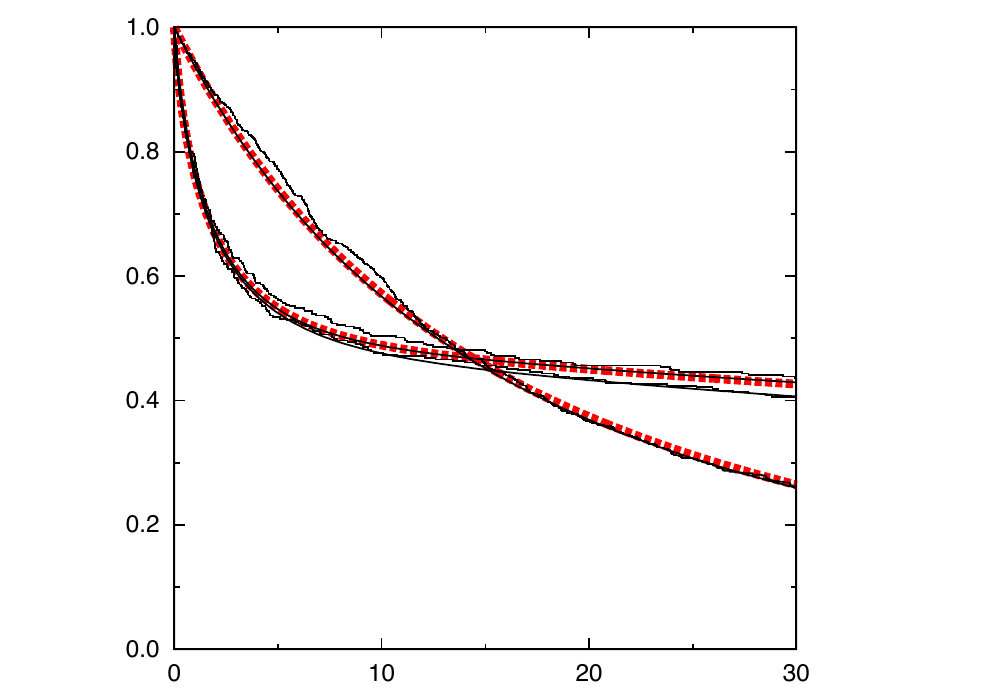}}
\put(100,100){\includegraphics[width=138\unitlength,height=95\unitlength]{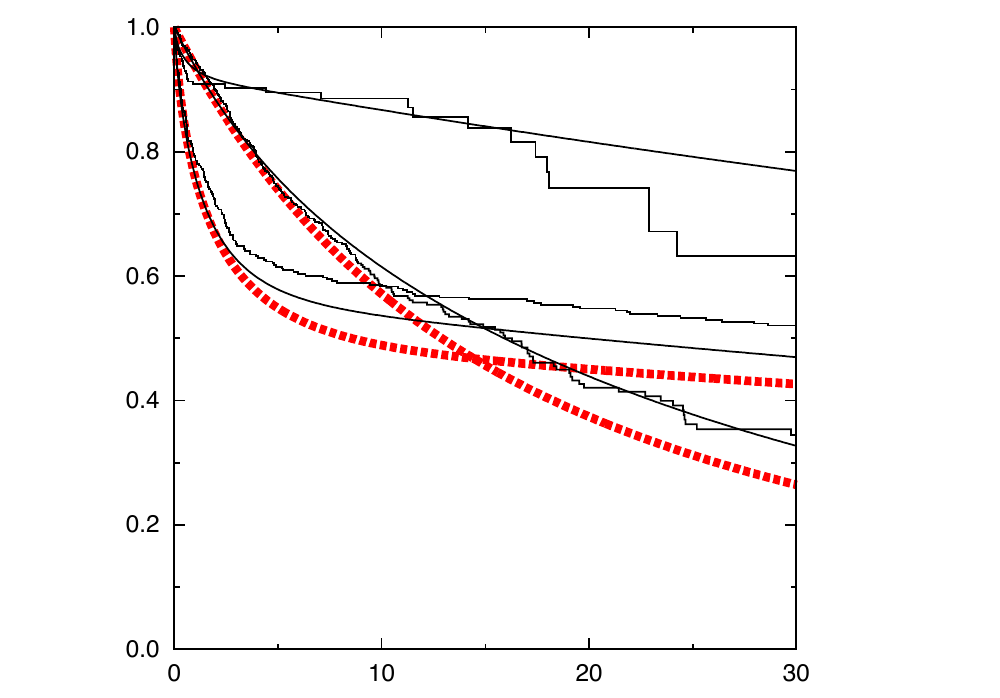}}
\put(200,100){\includegraphics[width=138\unitlength,height=95\unitlength]{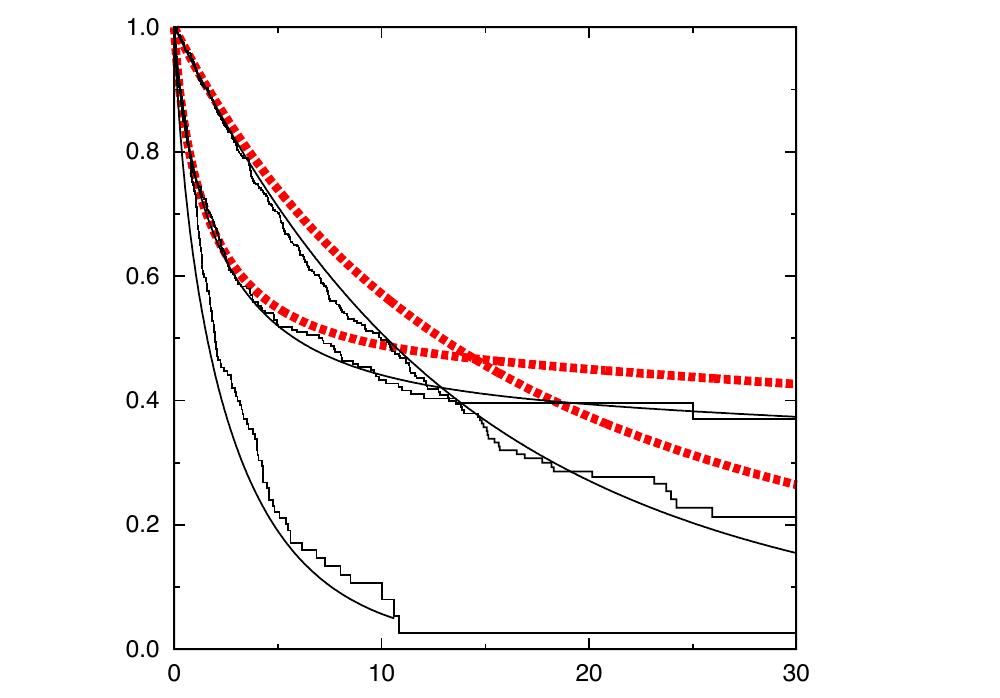}}
%\put(65,95){$t$}\put(165,95){$t$}\put(265,95){$t$}
\put(28,197){\small\sc A -- primary risk only}
\put(128,197){\small\sc B -- false protectivity}
\put(228,197){\small\sc C -- false exposure}

\put(101,147){\tiny\em UQ} \put(101,136){\tiny\em LQ} \put(101,125){\tiny\em IQ}
\put(201,165){\tiny\em LQ} \put(201,153){\tiny\em UQ} \put(201,139){\tiny\em IQ}
\put(301,140){\tiny\em LQ} \put(301,112){\tiny\em UQ} \put(301,127){\tiny\em IQ}

\put(5,63){$\tilde{S}_1$}
\put(0,5){\includegraphics[width=138\unitlength,height=95\unitlength]{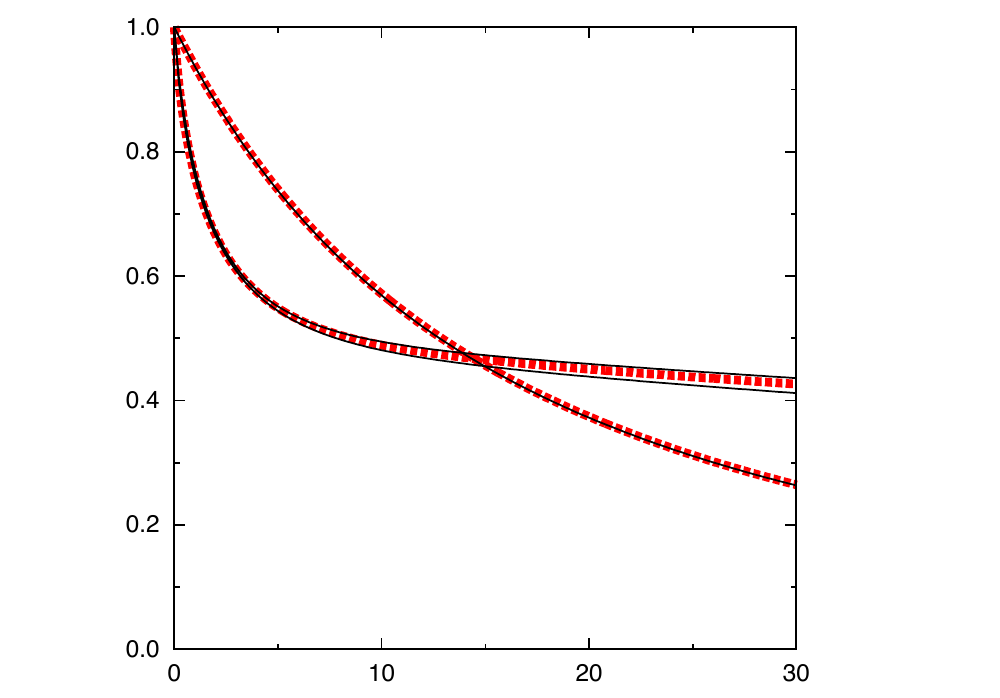}}
\put(100,5){\includegraphics[width=138\unitlength,height=95\unitlength]{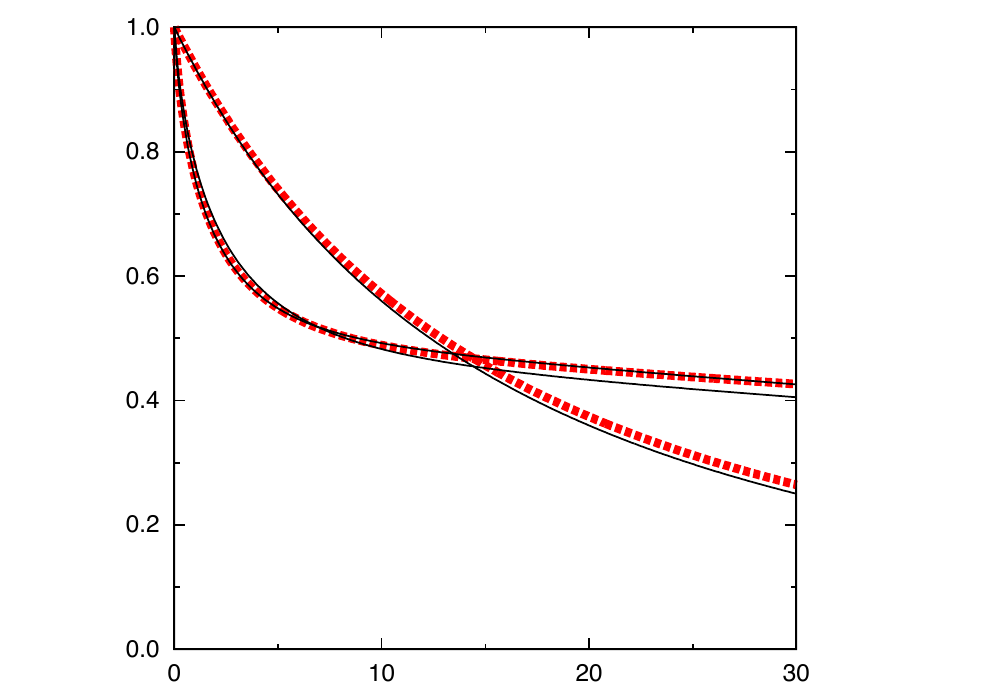}}
\put(200,5){\includegraphics[width=138\unitlength,height=95\unitlength]{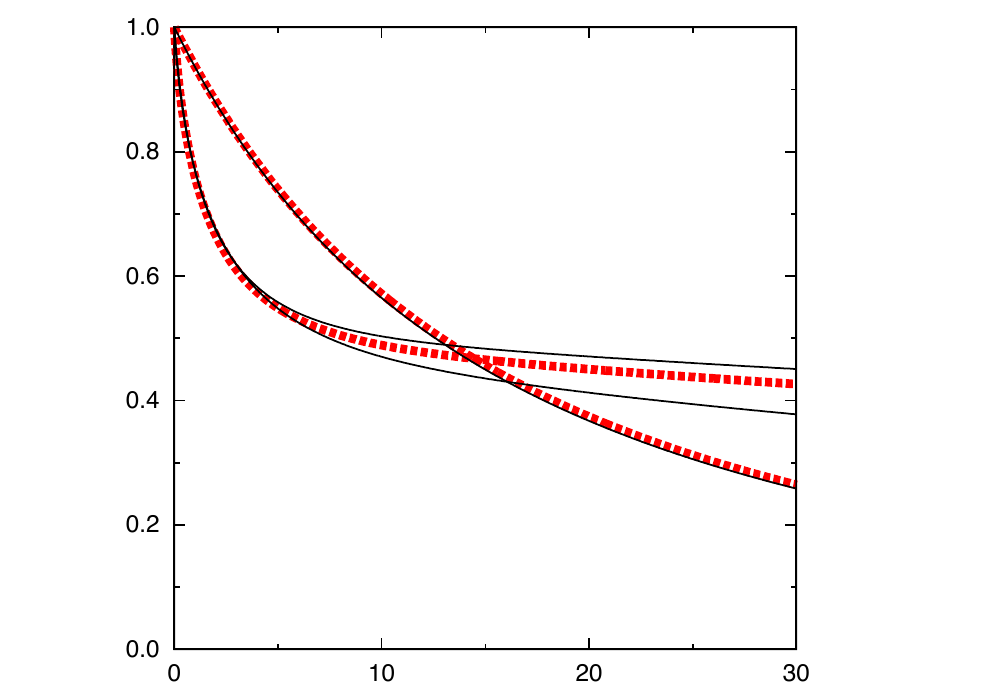}}
\put(65,0){$t$}\put(165,0){$t$}\put(265,0){$t$}

\put(101,51){\tiny\em UQ} \put(101,42){\tiny\em LQ} \put(101,30){\tiny\em IQ}
\put(201,51){\tiny\em UQ} \put(201,42){\tiny\em LQ} \put(201,30){\tiny\em IQ}
\put(301,52){\tiny\em UQ} \put(301,40){\tiny\em LQ} \put(301,30){\tiny\em IQ}

\end{picture}\vspace*{1mm}

\caption{Top row: Kaplan-Meier curves $S_1^{\rm KM}(t)$  (jagged solid curves) and crude survival functions $S_1(t)$ (smooth solid curves) of the primary risk,  
calculated for the upper and lower quartiles (UQ, LQ), and the inter-quartile range (IQ) of covariate 1, for  the synthetic data of Table \ref{table1}.  Dotted  red curve: the true primary risk survival functions (\ref{eq:LQUQ},\ref{eq:IQ}) of the upper/lower quartiles (which are identical) and of the inter-quartile range. 
The primary risk characteristics of all three data sets are {\em identical}. 
In set A there is only the primary risk; although the cohort is heterogeneous, the estimators quantify the  primary risk correctly. In set B the secondary risk is seen to cause `false protectivity' for the primary risk. In set C it causes `false exposure'. Risk correlations in data sets B and C clearly invalidate the use of Kaplan-Meier estimators and crude survival functions.
Bottom row: decontaminated survival curves  for the primary risk (solid black), for the same data and covariate ranges. The decontaminated curves $\tilde{S}_1$ are seen to be reliable estimates of the  quantitative characteristics  of the primary risk.
 }
\label{fig:synthetic}
\end{figure}

\noindent
should be {\em identical} in all three cases, since the data sets differ only in the interference effects of a secondary risk. 
For the primary risk in table \ref{table1}, the correct expression (\ref{eq:StildeLC}) simplifies to
\begin{eqnarray}
\tilde{S}_1(t|z_1)&=& \frac{1}{2}e^{-\frac{t}{20}\exp(2z_1)}+ \frac{1}{2}e^{-\frac{t}{20}\exp(-2z_1)}
\label{eq:syntheticStilde}
\end{eqnarray}
From this we can calculate the true primary risk survival curves for the upper and lower quartiles (UQ, LQ) and for the inter-quartile range (IQ).  For the present Gaussian-distributed covariates, with zero average and unit variance, 
the upper and lower quartile survival curves are identical, due to the symmetry $\tilde{S}_1(t|-z_1)=\tilde{S}_1(t|z_1)$. 
With the usual short-hand $Dz=(2\pi)^{-1/2}e^{-z^2/2}dz$, and with
  the quartile point $z_Q$ defined via $\int_{z_Q}^\infty\!Dz=\frac{1}{4}$, giving $z_Q\!\approx\! 0.67449$, we obtain from (\ref{eq:syntheticStilde}) the following (exact) formulae:
\begin{eqnarray}
{\rm LQ,UQ:} &&
 \tilde{S}_1(t|z_1\!\in\![z_Q,\infty))= ~~~2\int_{z_Q}^{\infty}\!\!Dz~\Big(
e^{-\frac{t}{20}\exp(2z)}+e^{-\frac{t}{20}\exp(-2z)}\Big)
\label{eq:LQUQ}
\\
{\rm IQ:} &&
\tilde{S}_1(t|z_1\!\in\![-\!z_Q,z_Q])= 2\int_{0}^{z_Q}\!\!Dz~\Big(e^{-\frac{t}{20}\exp(2z)}+e^{-\frac{t}{20}\exp(-2z)}
\Big)
\label{eq:IQ}
\end{eqnarray}
Figure \ref{fig:synthetic} shows the true LQ, UQ and IQ survival curves (\ref{eq:LQUQ},\ref{eq:IQ})   for the data sets A, B and C of table \ref{table1}, together with the 
decontaminated curves $\tilde{S}_1$ in (\ref{eq:StildeLC}), as calculated from application of our 
heterogeneous 
regression method (\ref{eq:LL_our_model},\ref{eq:Psi_to_minimise}), see  bottom row. We show also for comparison the Kaplan-Meier estimators $S_1^{\rm KM}$ \citep{KM} of the primary risk survival function and  the crude primary risk survival functions for the same covariate subsets (LQ,UQ,IQ), see top row. The crude survival functions  are calculated via  $S_1(t|\bz)=\exp[-\int_0^t\!\rmd s~ h_r(s|\bz)]$, using the crude hazard rate  (\ref{eq:cruderateLC}) with parameters as estimated from our heterogeneous regression. 

As expected, the Kaplan-Meier estimators $S_1^{\rm KM}$ are estimators of the {\em crude} survival functions $S_1$. However, both $S_1^{\rm KM}$ and $S_1$ are, in turn,  only estimators of the {\em true} survival functions $\tilde{S}_1$ of risk 1, i.e. of  (\ref{eq:LQUQ},\ref{eq:IQ}), if the different risks are uncorrelated. Here the risks are uncorrelated only for data set A (where there is no secondary risk). As soon as the risks are correlated,  in data sets B and C, we see in Figure \ref{fig:synthetic} that the Kaplan-Meier estimators and the crude survival functions no longer  predict the true (red) curves. As anticipated from Table \ref{table1}, they both underestimate grossly the primary risk in data set B (where the competing risk filters out high-primary-risk individuals)    and overestimate the primary risk in data set C (where the competing risk filters out low-primary-risk individuals). In fact, plotting only the upper and lower quartile curves for $S^{\rm KM}_1$ or $S_1$ would suggest a strong overall impact of covariate 1 on the primary risk in data sets B and C, where in reality there is none. 
 In contrast, the decontaminated curves $\tilde{S}_1$ calculated from our generic heterogeneous regression protocol
(lower three panels of Figure \ref{fig:synthetic}) 
do capture and predict the true survival functions of the primary risk from survival data alone, in spite of the presence of the competing risk.

\subsection{Retrospective class identification}

\begin{figure}[t]

\unitlength=0.35mm
\hspace*{-4mm}
\begin{picture}(400,120)
\put(0,0){\includegraphics[height=100\unitlength]{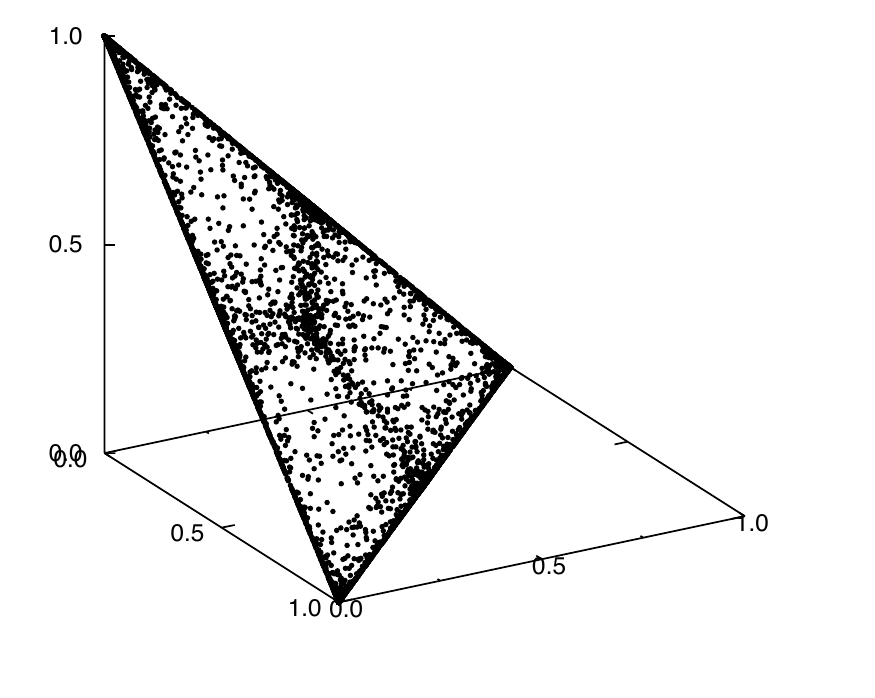}} 
\put(100,0){\includegraphics[height=100\unitlength]{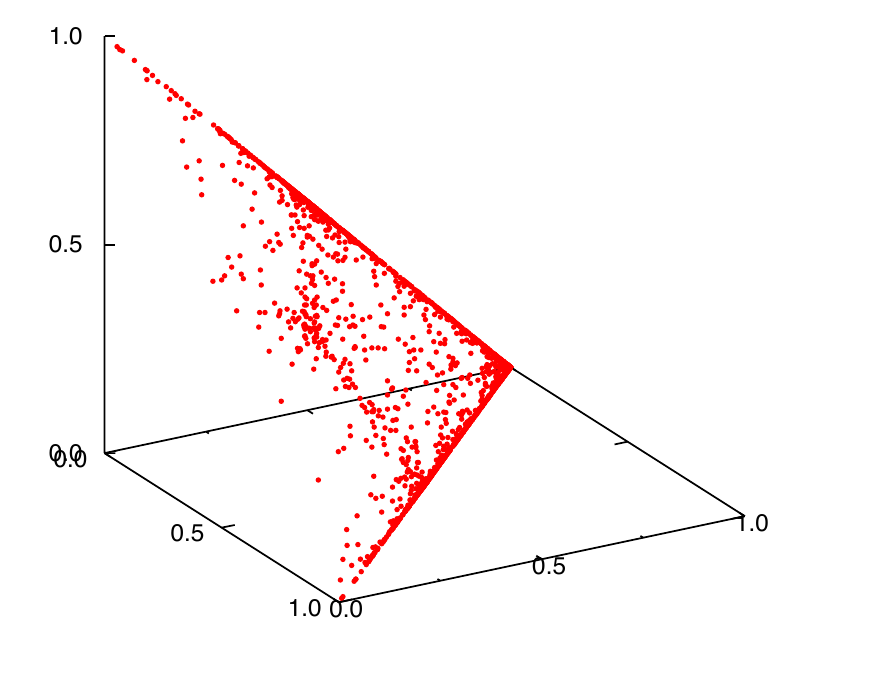}}
\put(200,0){\includegraphics[height=100\unitlength]{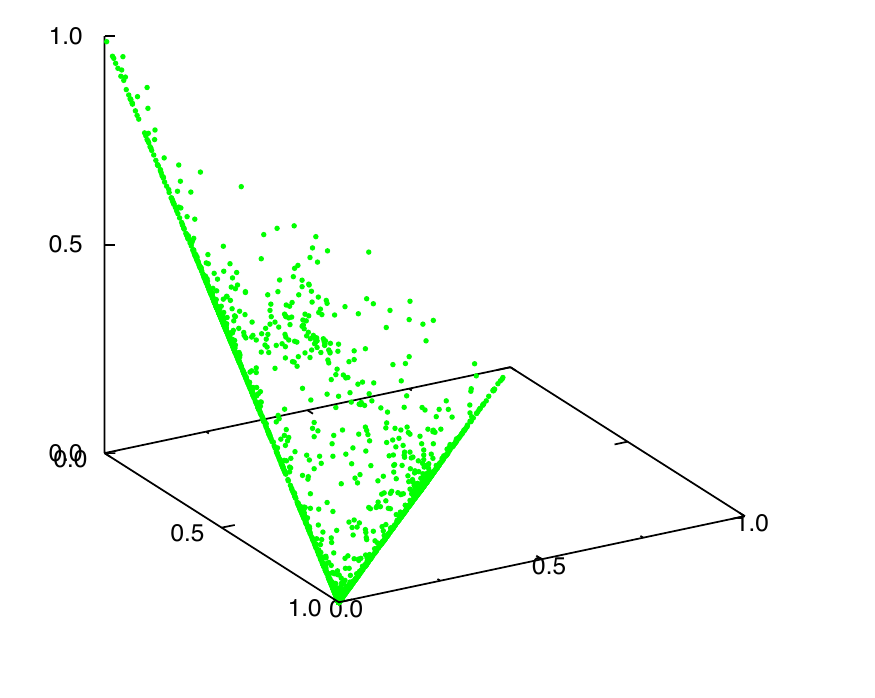}}
\put(300,0){\includegraphics[height=100\unitlength]{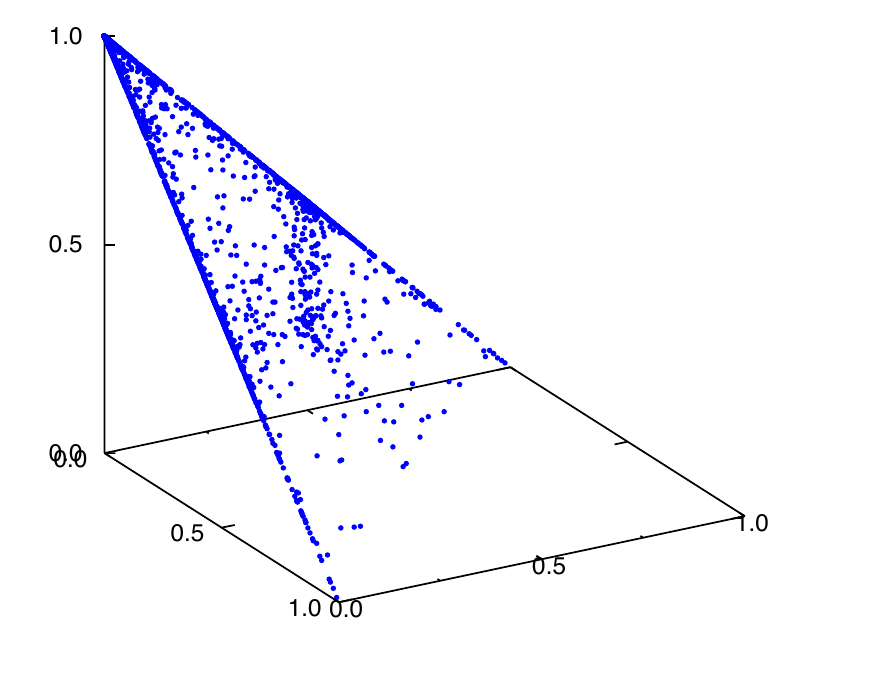}}

\put(40,100){\small\sc all}\put(140,100){\small\sc class 1}\put(240,100){\small\sc class 2}\put(340,100){\small\sc class 3}
\end{picture}
\vsp

\hspace*{30mm}
\unitlength=0.27mm
\begin{picture}(300,320)

\put(0,200){\includegraphics[height=100\unitlength]{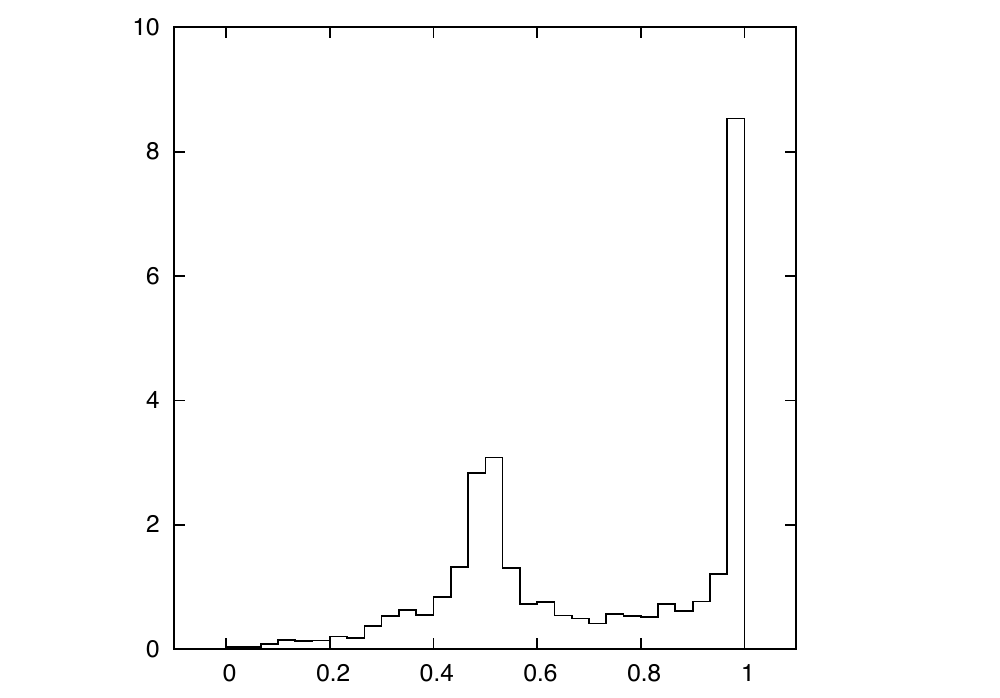}} 
\put(100,200){\includegraphics[height=100\unitlength]{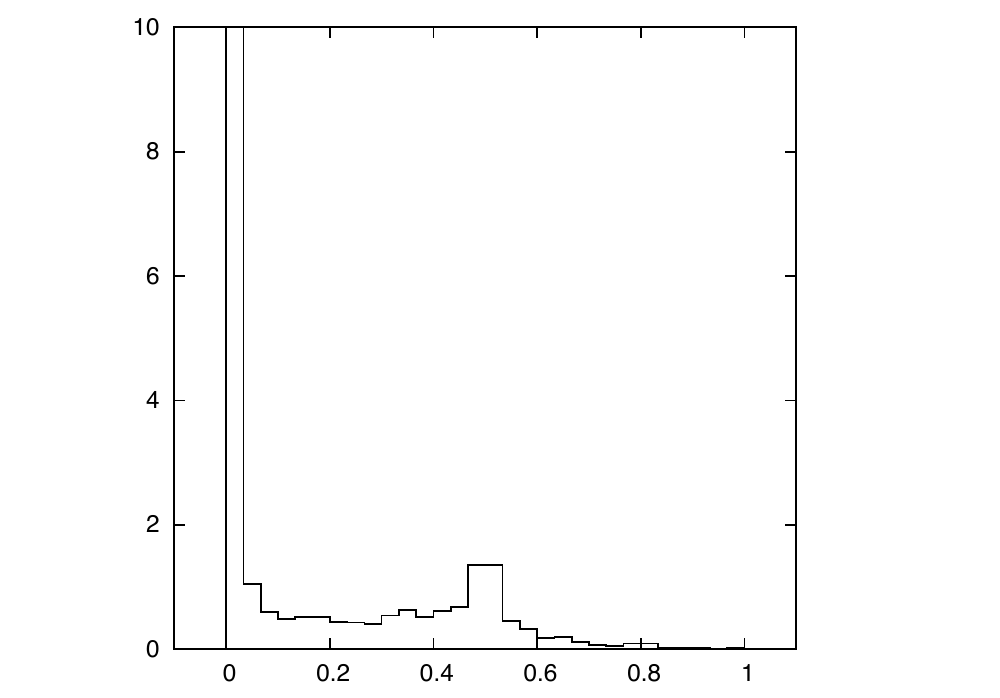}}
\put(200,200){\includegraphics[height=100\unitlength]{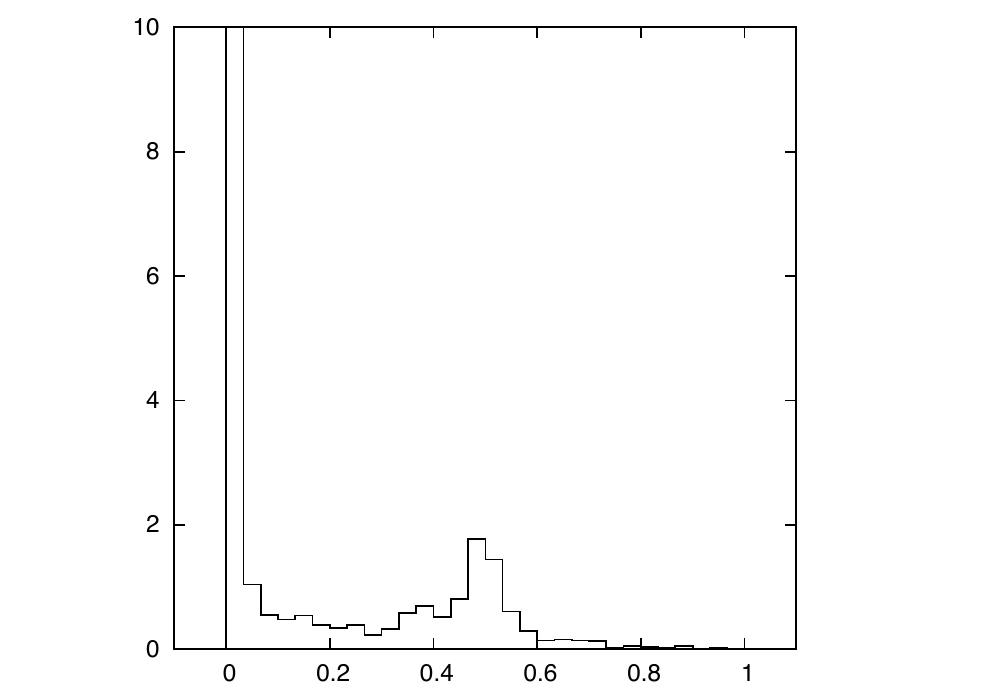}}

\put(0,100){\includegraphics[height=100\unitlength]{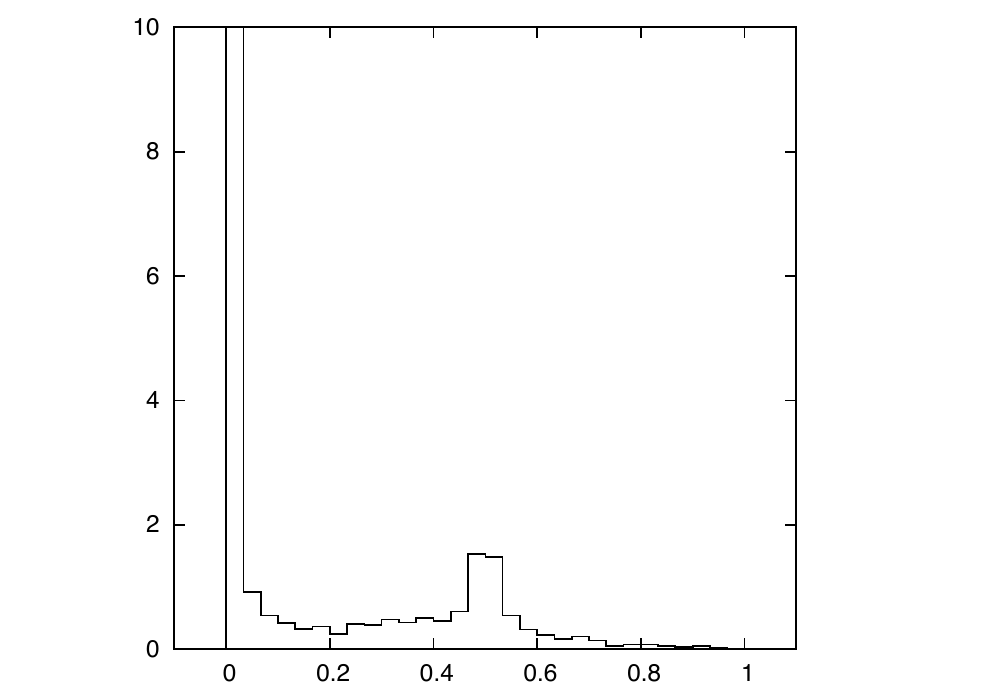}} 
\put(100,100){\includegraphics[height=100\unitlength]{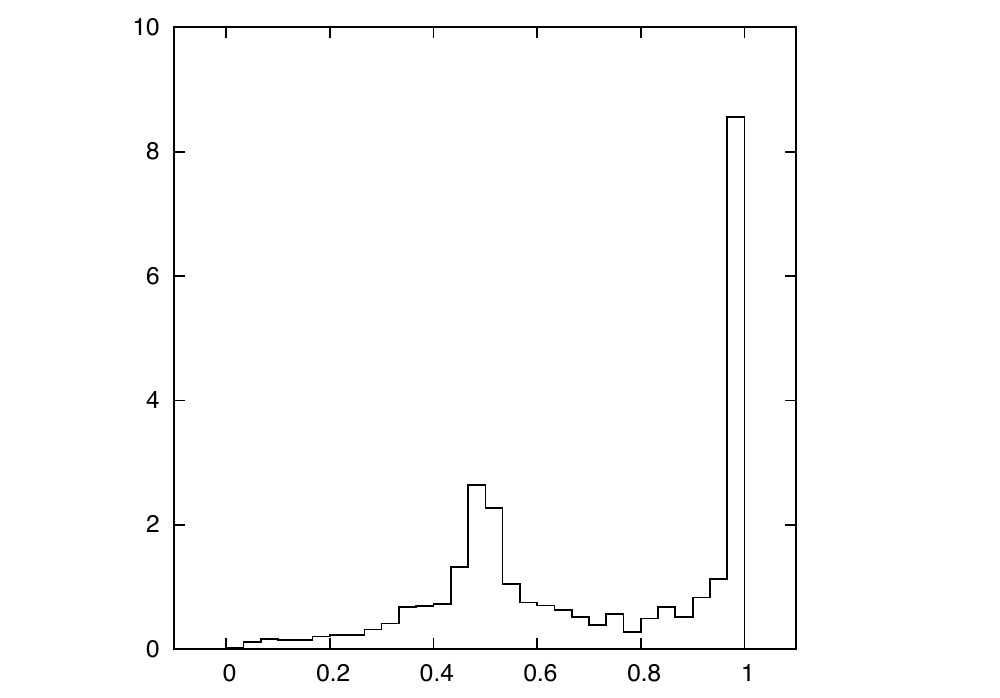}}
\put(200,100){\includegraphics[height=100\unitlength]{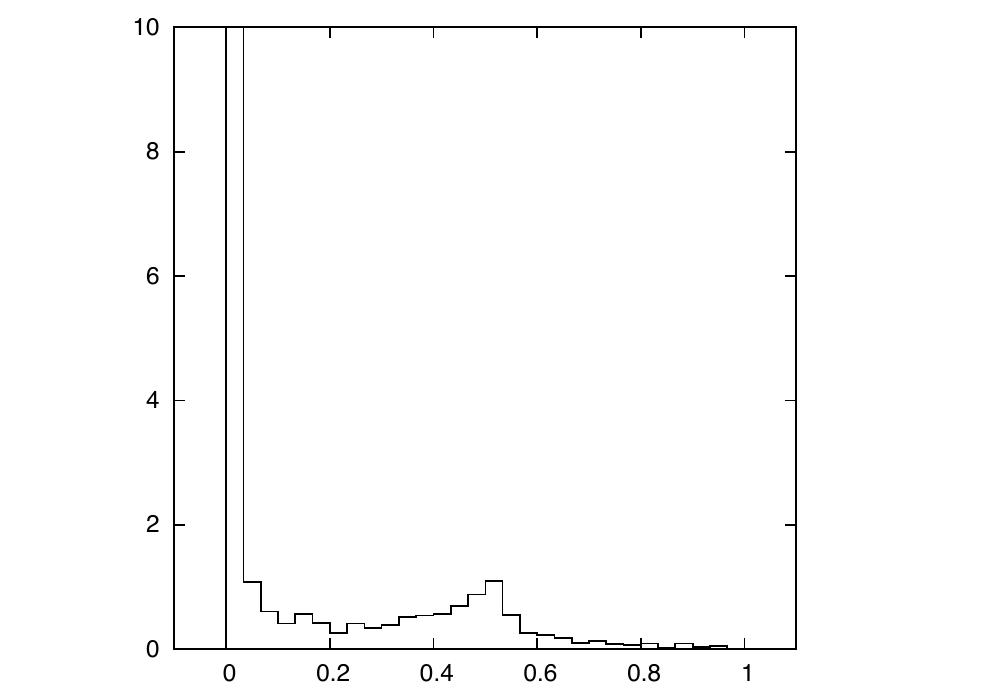}}

\put(0,0){\includegraphics[height=100\unitlength]{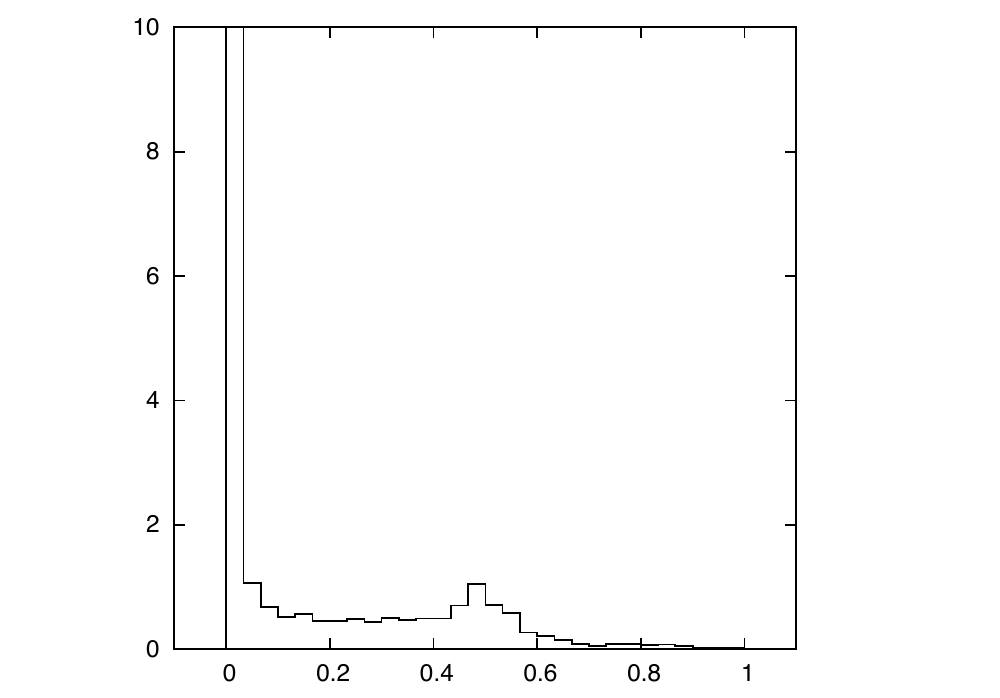}} 
\put(100,0){\includegraphics[height=100\unitlength]{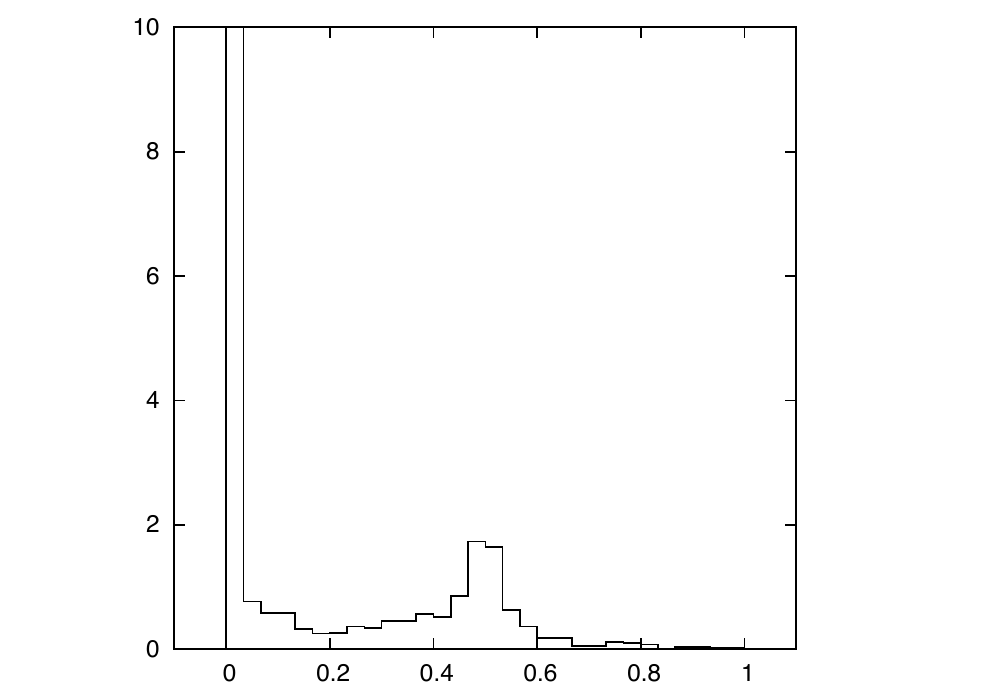}}
\put(200,0){\includegraphics[height=100\unitlength]{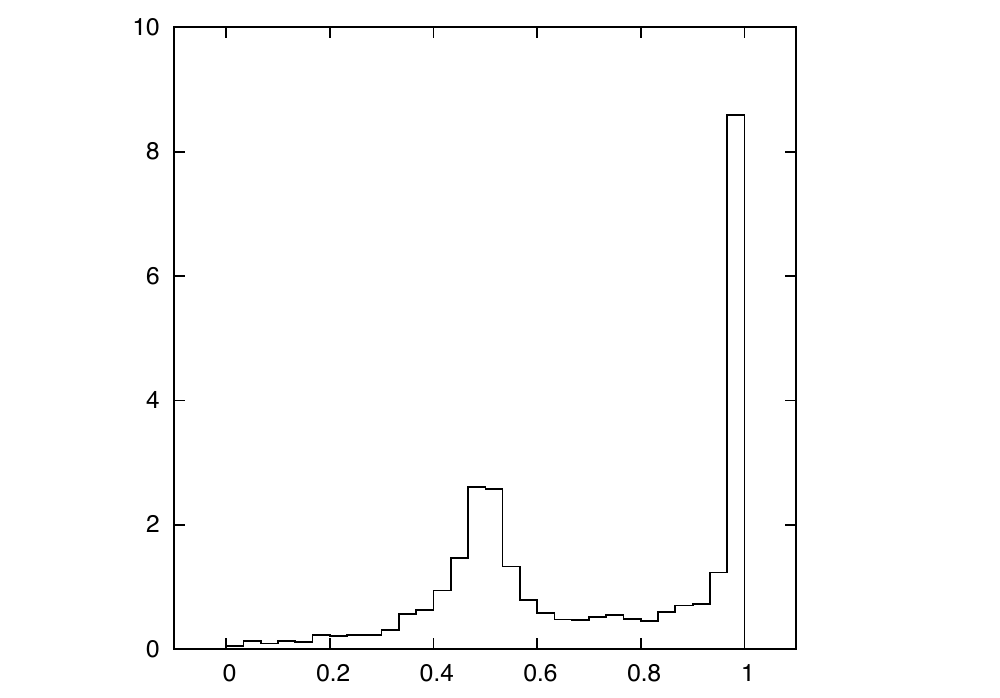}}

\put(-30,50){\small\sc class 3}\put(-30,150){\small\sc class 2}\put(-30,250){\small\sc class 1}
\put(55,307){$\{p_{i1}\}$}\put(155,307){$\{p_{i2}\}$}\put(255,307){$\{p_{i3}\}$}
\end{picture}
\vsp

\caption{Retrospective assigment to classes of individuals in a heterogeneous synthetic cohort of the type (\ref{eq:synthetic2}), with $N=9600$ and $\varrho=4$. Of the 9600 events, 4968 were primary. 
Top row: posterior joint probabilities $\bp_i=(p_{i1},p_{i2},p_{i3})$ for each individual to belong to the three classes,  calculated from the survival data alone, drawn as points in $\R^3$.
Black: joint likelihoods $\bp_i$  for all $i$.  Red,  green and blue: joint likelihoods $\bp_i$ of the three subsets of 3200 individuals that were generated respectively from the classes $\ell=1,2,3$. 
The points of each class indeed tend to be positioned close to the corners $(1,0,0)$, $(0,1,0)$ and $(0,0,1)$ that would correspond  to perfect allocation. Histograms (below): distributions of $\{p_{i1}\}$, $\{p_{i2}\}$ and $\{p_{i3}\}$, calculated for members of the three classes separately. Perfect assignment corresponds to finding most $\{p_{i\ell}\}$ for each $\ell$ close to the value $1$ only for  the row of class $\ell$, with values close to zero for the other rows. In this example the correctly assigned fraction is $f\approx 0.758$  }
\label{fig:class_assignment}
\end{figure}
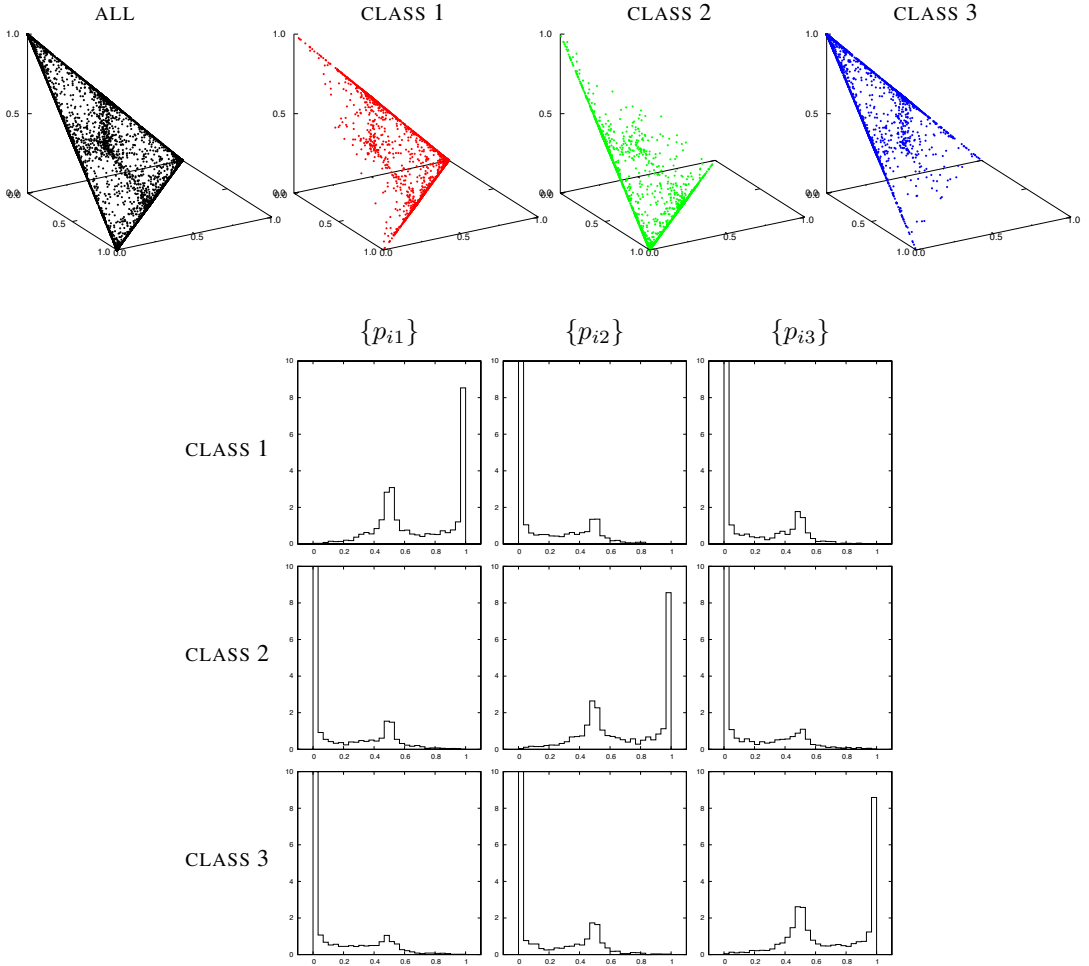

\noindent
Finally we illustrate with synthetic data the ability of our methodology to identify the classes of the individuals 
in a given data set, retrospectively, via (\ref{eq:find_class}), after having estimated the generating cohort's structure and  parameters.  
We generate heterogeneous data sets with three risks,  zero frailty parameters,  and three independently generated (zero average and unit variance) Gaussian covariates. Each set has $L=3$ classes, with $w_1\!=\!w_2\!=\!w_3\!=\!\frac{1}{3}$, and the following  regression parameter vectors:
\begin{eqnarray}
\bbeta_1^1&=&(\frac{1}{2},\frac{1}{2},\frac{1}{2})+\varrho(1,0,1) \nonumber
\\
\bbeta_1^2&=& (\frac{1}{2},\frac{1}{2},\frac{1}{2})+\varrho(-1,-1,0)
\label{eq:synthetic2}
\\
\bbeta_1^3&=& (\frac{1}{2},\frac{1}{2},\frac{1}{2})+\varrho(0,1,-1)
\nonumber
\end{eqnarray}
For the two non-primary risks $r=2,3$ we set $\bbeta_r^{\ell}=\bnull$ for all $\ell$. The base hazard rates are $\hat{\lambda}_1(t)=1/10$, $\hat{\lambda}_2(t)=1/20$ and $\hat{\lambda}_3(t)=1/30$. 
For any value of the parameter $\varrho$ the average primary risk regression vector over the cohort is $\frac{1}{3}(\bbeta^1\!+\!\bbeta^2\!+\!\bbeta^3)=(\frac{1}{2},\frac{1}{2},\frac{1}{2})$, with 
the cohort becoming homogeneous for $\varrho=0$, and increasingly heterogeneous (i.e. separable) as $\varrho$ increases.

Formula (\ref{eq:find_class}) gives for each individual $i$, with covariates $\bz_i$ and outcome values $(t_i,r_i)$,  the probabilities  $p_{i\ell}=P(\ell|t_i,r_i,\bz_i)$ for $i$ to belong to each of the classes $\ell=1\ldots L$. 
For $L=3$ one
\clearpage

\noindent
can visualise the overall class assigment  by showing the triplets $\bp_i=(p_{i1},p_{i2},p_{i3})$ for all $i$ as points in $\R^3$, on the simplex defined by the conditions $p_1\!+\!p_2\!+\!p_3\!=\!1$ and  $p_1,p_2,p_3\!\in\![0,1]$. The result is shown in Fig. \ref{fig:class_assignment} for the synthetic data (\ref{eq:synthetic2}), with $N=9600$ and $\lambda=4$. Those individuals $i$ that in reality originate from class 1 are indeed seen to be assigned probability vectors $\bp_i$ close to the point $(1,0,0)$, those from class 2 tend to have  $\bp_i$ close to the point $(0,1,0)$, and those from class 3 
have $\bp_i$ close to $(0,0,1)$. The three corner points correspond to fully confident assignment. 

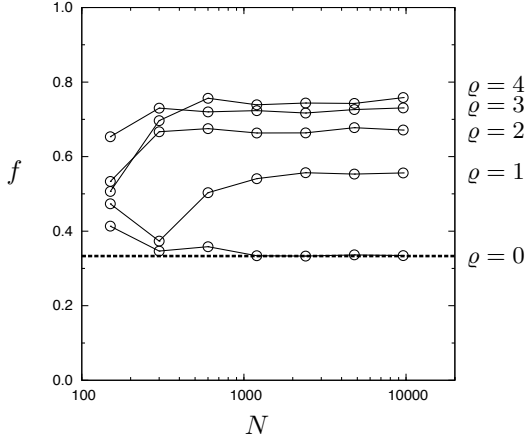
\begin{figure}[t]
\unitlength=0.55mm
\hspace*{30mm}\begin{picture}(100,100)
\put(0,0){\includegraphics[height=100\unitlength]{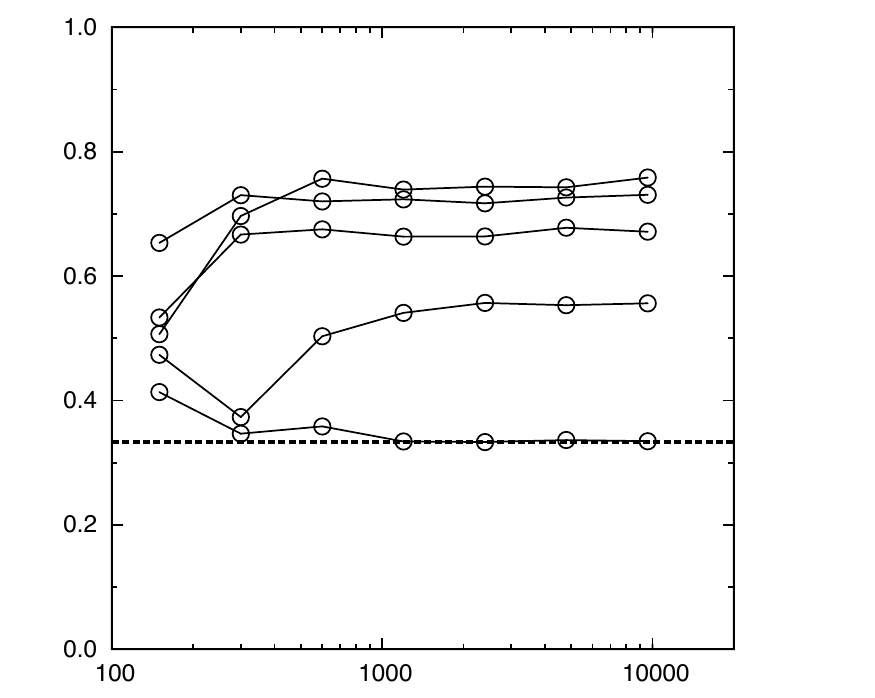}} 
\put(55,-6){$N$}
\put(-2,55){$f$}
\put(108,35){\small $\varrho=0$}
\put(108,55){\small $\varrho=1$}
\put(108,65){\small $\varrho=2$}
\put(108,71){\small $\varrho=3$}
\put(108,76){\small $\varrho=4$}
\end{picture}\vsp

\caption{Connected markers: classification quality $f$ (the fraction of correctly classified individuals) for synthetic data sets of the type (\ref{eq:synthetic2}) as a function of the data set size $N$, for different values of the degree $\varrho$ of heterogeneity.
Dashed: the value $f=1/3$ corresponding to random class assigment. In all sets there are three (zero-average and unit-variance random Gaussian) covariates and three risks. The number of primary events in each set is roughly $N/2$. }
\label{fig:classification_vs_size}
\end{figure}

To quantify the quality of the retrospective class identification (\ref{eq:find_class}) we can assign each $i$ to its most probable class ${\rm arg max}_{\ell=1\ldots L} ~p_{i\ell}$, and define the fraction 
$f$ of correctly assigned individuals: 
\begin{eqnarray}
f&=&\frac{1}{N}\sum_{i=1}^N \delta_{\ell_i,{\rm arg max}_{\ell=1\ldots L} ~p_{i\ell}}
\end{eqnarray}
where $\ell_i$ is the true class label of $i$. Even if the cohort's parameters were known perfectly, due to the intrinsic stochasticity of event times  there will always be a fraction of unlikely events 
and $f$ will always be less than 1. For the data of Fig. \ref{fig:class_assignment}  one finds the value $f\approx 0.758$, to be compared to the benchmark value  $f=\frac{1}{3}$ that would be obtained for random assignment to the three classes. 
One must expect the assigment quality $f$ for the synthetic data (\ref{eq:synthetic2}) to increase monotonically with both the degree of class dissimilarity (as measured by $\varrho$)  and the size $N$ of the data set (due to the improved recovery of the true model parameters). This is borne out by simulation experiments with different choices for $(\varrho,N)$, the results of which are shown in Fig. \ref{fig:classification_vs_size}. On average the number of primary (i.e. informative) events in these sets is about $N/2$, so for the present example one requires
 data sets with about 250 primary events or more to have good retrospective overall class allocation. From  Fig. \ref{fig:class_assignment} one can also conclude that the classification reliability  can be increased further by limiting oneself to the subset of patients $i$ that have joint probability vectors $\bp_i$ close to one of the corners $(1,0,0)$, $(0,1,0)$ or  $(0,0,1)$ of the simplex (with obvious generalisations to $L\neq 3$).

\section{Applications to prostate cancer data}

\begin{figure}[t]
\unitlength=0.55mm
\hspace*{30mm}\begin{picture}(200,102)
\put(0,0){\includegraphics[height=100\unitlength]{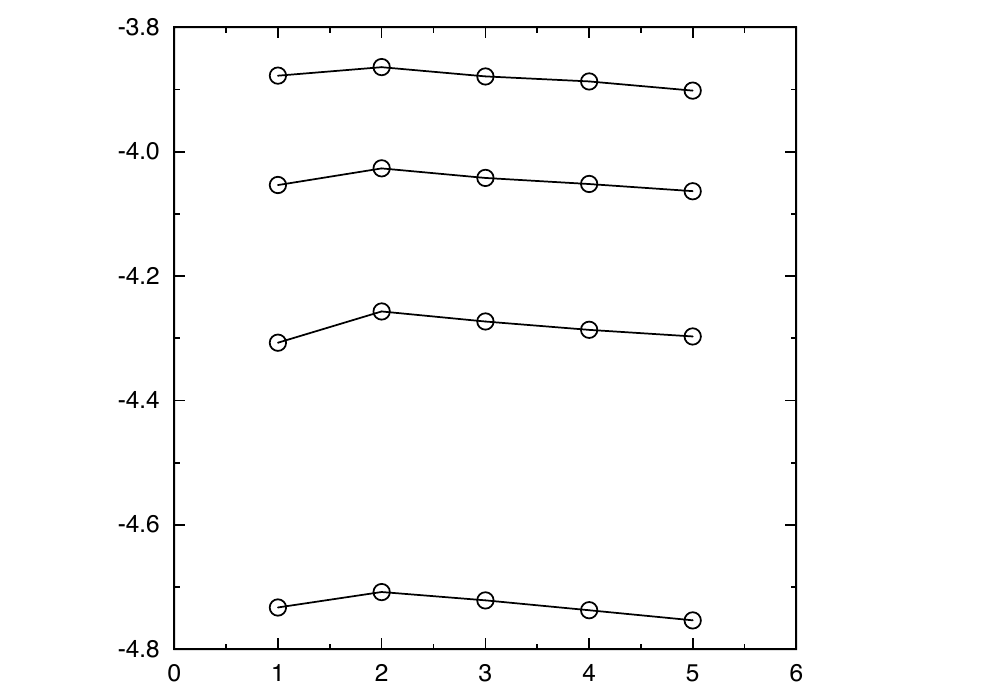}} 
\put(67,-6){$L$}
\put(-15,65){$-\Psi_{L,K}/N$}
\put(97,15){\small $K\!=\!1$}
\put(97,43){\small $K\!=\!2$}
\put(97,65){\small $K\!=\!3$}
\put(97,79){\small $K\!=\!4$}
\end{picture}\vsp

\caption{Rescaled Aikake-score as a function of the number of classes $L$, for $K\in\{1,2,3,4 \}$ 
 for the ULSAM prostate cancer data set.  This set contains survival data on $N=2047$ individuals, of whom 208 reported the primary event (prostate cancer), 910 reported a secondary event (death of other causes), and 929 were consored due to end of trial.  We observe that for the above values of $K$, which controls the number of interpolation points in the base hazard rates, the most probable model to explain the ULSAM data would have two classes.}
\label{fig:ULSAMscores}
\end{figure}
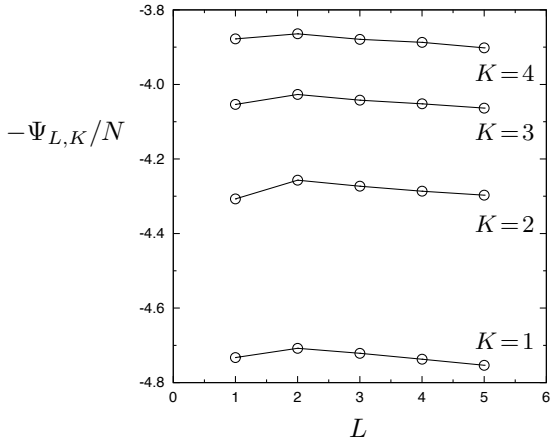

\begin{figure}[t]
\vsp{\footnotesize
\hspace*{-1mm}
\begin{tabular}{|r| cll|}
\hline
\smallroom								& {\sc structure}~	& {\sc primary risk} & {\sc secondary risk}\\
\hline 
\smallroom                                            & & 208 events & 910 events \\
\smallroom{\em covariates}                    &  & \!{\em BMI~~~selen~~phys1~~phys2~~smok}	& \!{\em BMI~~~selen~~phys1~~phys2~~smok}\\
\hline
\room {\em Cox regression} 		& --   			& 0.14~~ -0.15~~ 0.20~ -0.09~~ -0.08 	&   \\[1mm]
\room {\footnotesize\em $L\!=\!1$: $~~K\!=\!3$}		& -- 	& 0.15 ~ -0.15 ~ 0.20~ -0.09 ~ -0.08 	&  0.21 ~ -0.06 ~ -0.11 ~ 0.01 ~ 0.34 \\[0mm]
\smallroom {\footnotesize\em $~~~~~~K\!=\!4$}		& -- 	& 0.14 ~ -0.15 ~ 0.20~ -0.09 ~ -0.08 	& 0.21 ~ -0.06 ~ -0.11 ~ 0.01 ~ 0.34 \\[1mm]
\hline
\room {\footnotesize\em 
$L\!=\!2$: $~~K\!=\!3$}	 & class 1: ~$w_1\!=\!0.51$	& 1.31 ~ -0.39 ~ 0.79 ~~ 0.03 ~ 1.40	&  0.88 ~ -0.44 ~ -0.30 ~ \!-0.16 ~ \!1.39 \\
				 	 & class 2: ~$w_2\!=\!0.49$ 	& \!-0.06 ~ \!-0.16 ~ 0.20 ~ \!-0.10 ~ \!-0.27	&  0.09 ~ -0.06 ~ -0.07 ~ 0.04 ~ 0.17 \\[0.5mm]
\smallroom                      & relative frailties:                    & $\beta_{10}^1\!-\!\beta_{10}^2\!=\! -4.84$ (HR 0.008) & $\beta_{20}^1\!-\!\beta_{20}^2\!=\! -4.10$ (HR 0.017)
\\[2mm]
\smallroom {\footnotesize\em 
$~~~~~~K\!=\!4$}		& class 1: ~$w_1\!=\!0.51$	& 1.22 ~ -0.41 ~ 0.73 ~ \!-0.01 ~ 1.43 	&  0.82 ~ -0.42 ~ -0.31 ~ -0.14 ~1.35 \\
					& class 2: ~$w_2\!=\!0.49$ 	& \!-0.07 ~ \!-0.16 ~ 0.19 ~ \!-0.10 ~-0.27 &  0.10 ~ -0.07 ~ -0.07 ~ 0.04 ~ 0.18 \\[0.5mm]
\smallroom                      & relative frailties:                    & $\beta_{10}^1\!-\!\beta_{10}^2\!=\! -4.61$ (HR 0.010)& $\beta_{20}^1\!-\!\beta_{20}^2\!=\! -4.06$ (HR 0.017)
\\[2mm]
\hline
\end{tabular}
\vspace*{5mm}
}
\vspace*{1mm}

\noindent
{\bf Table 2. }
Regression results for the ULSAM prostate cancer (PC) data set, corresponding to some of the $(L,K)$ combinations in Figure \ref{fig:ULSAMscores}. We included five covariates: body mass index ({\em BMI}, real-valued), serum selenium level ({\em selen}, integer-valued), leisure time physical activity ({\em phys1}, discrete levels 0/1/2), work physical activity ({\em phys2}, discrete levels 0/1/2), and smoking ({\em smok}, discrete levels 0/1/2). We report the results of Cox's propertional hazards method, and of our generic regression method 
based on  (\ref{eq:LL_our_model},\ref{eq:Psi_to_minimise}), with parameters extimated via Maximum A Posteriori likelihood augmented with Aikake's Information Criterion. 
We limit ourselves to $K=3$ and $K=4$. For $K\leq 2$ the time dependence of the base hazard rates cannot be captured, as emphasised by nonnegligible regression coefficients for the end-of-trial risk (not shown here). For $K\geq 5$  the 
growing parameter complexity (relative to the number of data points) pushes us increasingly towards the trivial $L=1$ option. 
Error bars in regression parameters are of the order of the last specified decimal. See the main text for discussion and interpretation of the above data. 
\label{table2}
\end{figure}

Prostate cancer (PC) data are notorious for exhibiting significant competing risk effects. The main reason for this is the fact that the disease tends to occur late in life, 
when there are an increased number of non-primary events whose incidence could correlate with prostate cancer. Here we analyse data from the ULSAM study \citep{ULSAM}, with $N=2047$ individuals of which 208 reported PC as the first event. In this study we limited ourselves to a subset of five typical covariates (the ULSAM data set contains more), in view of CPU constraints; this is not a fundamental limitation, and we expect in the near future to have computer code upgrades that will allow us to include  more. We refer again to Appendix \ref{app:numerical} 
for numerical details. 

\subsection{Cohort substructure and regression parameters}

\begin{figure}[t]
\unitlength=0.49mm
\vspace*{1mm}

\hspace*{23mm}
\begin{picture}(200,100)

\put(114,38){$S_1^{\small\rm KM}$}\put(114,52){$S_1$}\put(114,24){$\tilde{S}_1$}
\put(0,5){\includegraphics[width=138\unitlength,height=95\unitlength]{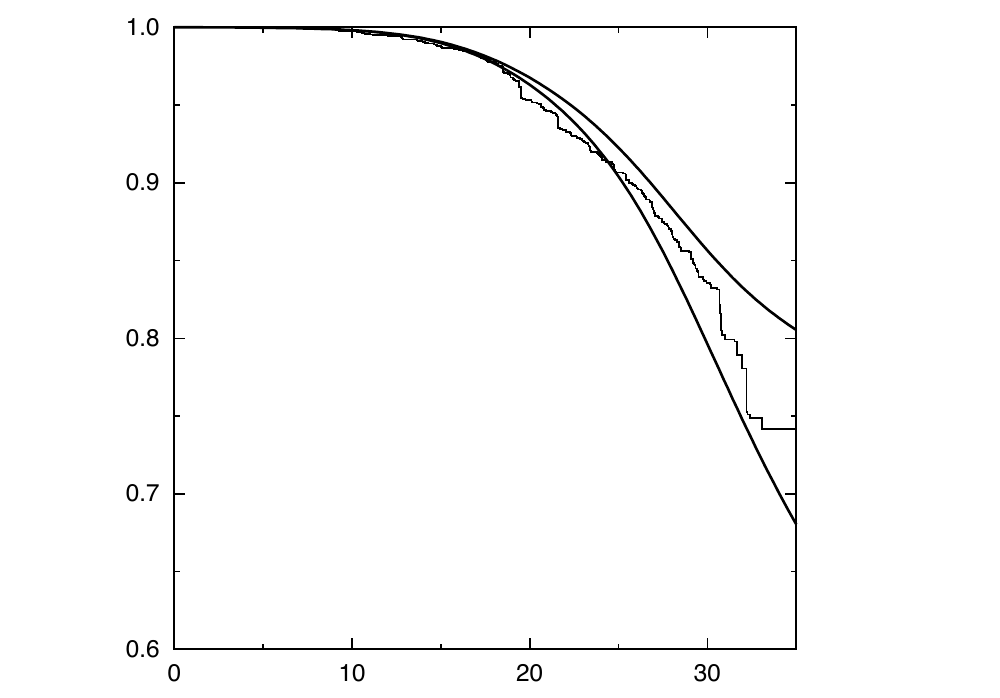}}
\put(65,0){$t$}

\end{picture}\vspace*{1mm}

\caption{Kaplan-Meier curves $S_1^{\rm KM}(t)$  (jagged solid curve), crude survival functions $S_1(t)$ (upper smooth solid curve)  and decontaminated survival functions $\tilde{S}_1(t)$ (lower smooth solid curve) of the primary risk.  The crude and decontaminated survival curves are calculated from the model $(L,K)=(2,4)$; see Table 2.  
Since the crude and decontaminated curves are different, there is informative censoring in the ULSAM cohort. The true PC risk is predicted to be  higher than that which would be found upon assuming risk independence,  so here the competing risks act to give false protectivity. 
 }
\label{fig:ULSAM_Scurves1}
\end{figure}

\begin{figure}[t]
\unitlength=0.49mm
\vspace*{1mm}

\hspace*{3mm}
\begin{picture}(200,100)

\put(2,85){$S_1^{\rm KM}$}\put(5,72){$S_1$}
\put(0,5){\includegraphics[width=138\unitlength,height=95\unitlength]{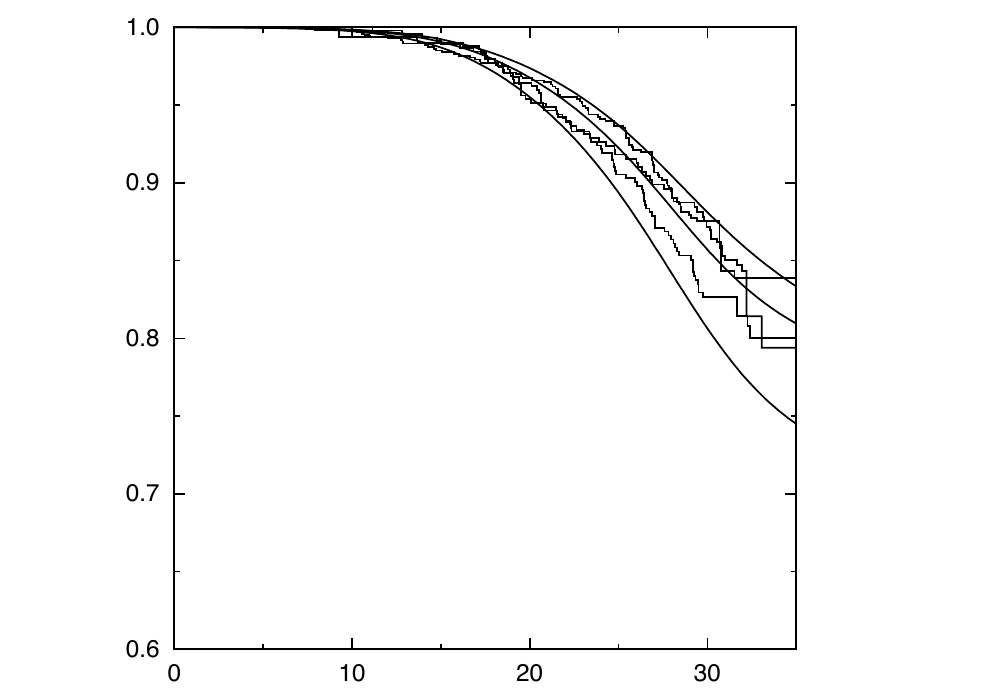}}

\put(142,85){$S_1^{\rm KM}$}\put(145,72){$\tilde{S}_1$}
\put(140,5){\includegraphics[width=138\unitlength,height=95\unitlength]{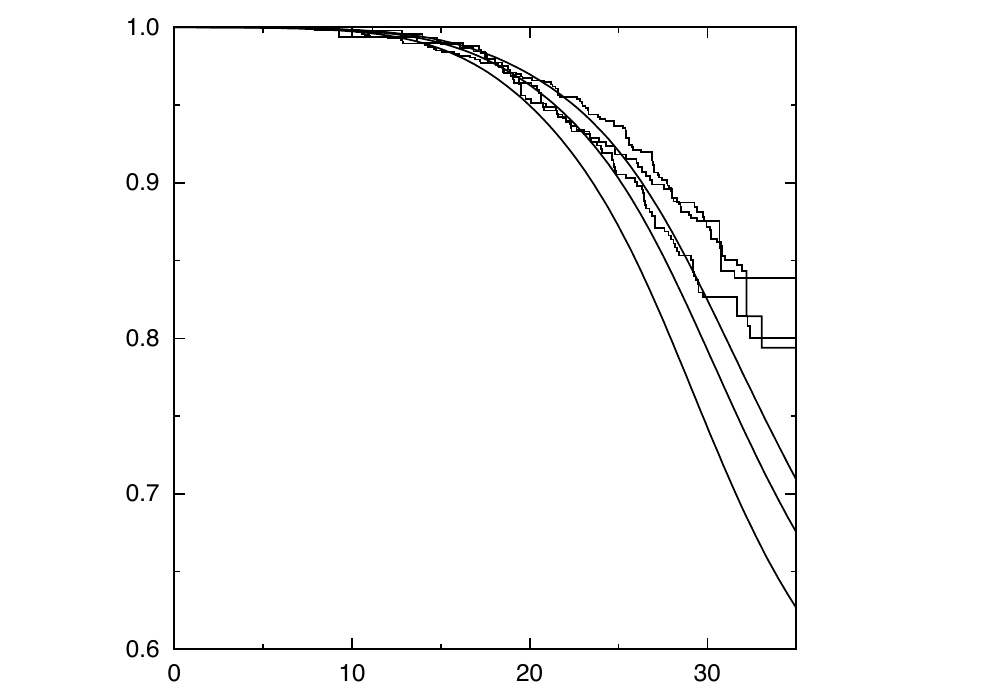}}
\put(65,0){$t$}\put(205,0){$t$}

\put(113,59){\footnotesize $z_5\!=\!2$} \put(113,51){\footnotesize $z_5\!=\!1$} \put(113,39){\footnotesize $z_5\!=\!0$}
\put(253,32){\footnotesize $z_5\!=\!2$} \put(253,25){\footnotesize $z_5\!=\!1$} \put(253,15){\footnotesize $z_5\!=\!0$}

\end{picture}\vspace*{1mm}

\caption{Left: Kaplan-Meier curves $S_1^{\rm KM}(t)$  (jagged solid curve) and crude survival functions $S_1(t)$ (smooth solid curve) of the primary risk,  
calculated for the three covariate subgroups $z_5=0$ (non-smokers), $z_5=1$ (ex-smokers), and $z_5=2$ (smokers).  
Right: Kaplan-Meier curves $S_1^{\rm KM}(t)$  (jagged solid curve) and {\em decontaminated} survival functions $\tilde{S}_1(t)$ (smooth solid curve) of the primary risk,  
calculated for the same three covariate subgroups. The crude and decontaminated survival curves are calculated from the model $(L,K)=(2,4)$; see Table 2.  
 }
\label{fig:ULSAM_Scurves2}
\end{figure}

We compare the outcomes of Cox's propertional hazards regression \citep{Cox} and our generic heterogeneous regression method, 
based on  (\ref{eq:LL_our_model}) and (\ref{eq:Psi_to_minimise}), with parameters extimated via Maximum A Posteriori likelihood augmented with Aikake's Information Criterion. 
In Figure \ref{fig:ULSAMscores} we show the rescaled Aikake scores for different combinations of the discrete parameters $L$ (number of classes in the cohort) and $K$ (which controls the time resolution in the parametrised base hazard rates of all risks). Each score is the result of optimising numerically (via the MAP protocol)  the regression and frailty coefficients and base hazard rates, for each risk and all classes,  as well as the relative class sizes.  
For $K=1,2,3,4$ the most probable explanation of the ULSAM data involves two distinct classes, marking either residual disease or host heterogeneity not captured by covariates.  
For larger $K$ values the Aikake score will increase further, but mainly due to the model's increasing ability to capture the base rate of the end-of-trial censoring events, which in the ULSAM data set is sharply peaked in time. Moreover, the score differences for different $L$ values will be  increasingly dominated by the Aikake-term, which acts to suppress models with many parameters and eventually forces us to accept only the trivial homogeneous explanation $L=1$. Note that the score differences in Figure \ref{fig:ULSAMscores} are nonneglible: 
since the posterior likelihood of models relates to the Aikake score approximately via ${\rm Prob}(L|D)\propto\exp[\Psi(L|D)]$, a difference of e.g. $\Delta\Psi/N=0.001$  implies 
a model likelihood ratio of ${\rm Prob}(L|D)/{\rm Prob}(L^\prime|D)\approx 11.8$. 

In Table 2 we show the regression outcomes for some of the $(L,K)$ combinations in Figure \ref{fig:ULSAMscores} in more detail, together with the corresponding results from proportional hazards regression \citep{Cox}. As expected, Cox regression and the $L=1$ explanation (which for the primary risk differs from the Cox protocol in the definition of the base hazard rate and a simple Bayesian prior for regression coefficients) give the same results. They both report weak effects of the five covariates, some of which are slightly unexpected (e.g. increased PC risk due to leisure time physical activity,  weak protective effect of smoking). 
The apparently more probable $L=2$ explanation generated by our generic heterogeneous regression method points consistently to a very different  picture. It suggests that the ULSAM cohort should be viewed as consisting of two distinct classes of similar size: one class $\ell=1$ with relatively healthy individuals (in terms of both primary and secondary risk), and one class $\ell=2$ with rather frail individuals.  This overall frailty difference follows from the significantly and consistently different frailty parameters of the two classes, and corresponds to hazard ratios (HR) of class membership in the range $0.01-0.02$ for primary and secondary risks.  The similar sizes of the two classes is consistent with the fact that about half of the ULSAM individuals are censored due to end of trial. 
In the relatively healthy class, the regression coefficients are now much more pronounced, and especially BMI and smoking are recognised as serious PC risk factors. 
In the frail class the regression coeficients are weak, and one expects the negative coefficients for e.g. BMI and smoking to reflect reverse causal effects: within this group, having a higher BMI and still being {\em able} to smoke may well be an indicator of {\em relative} health. 

One should not make the mistake of concluding that the above two-class explanation of the ULSAM data is for certain the best. There are multiple alternative 
ways to carry out the regression, e.g. by combining all non-primary risks (including the end-of-trial risk) or by setting the regression coefficients of the end-of-trial risk to zero 
(instead of leaving them to be determined by optimisation, as a test). In addition one could focus on larger $K$ values, where the resulting complexity increase forces us to reject $L>1$ solutions. The only safe conclusion to be drawn is that the new two-class explanation of the ULSAM data is both probabilistically and intuitively  plausible. 

\subsection{Decontaminated survival curves}

To assess whether the two-class explanation for the ULSAM data in Table 2 leads to different survival predictions, compared to standard methods,  as a consequence of informative censoring by the non-primary risks, we 
calculate the crude and decontaminated primary risk survival functions, see  (\ref{eq:StildeLC},\ref{eq:cruderateLC}),  with parameters as estimated from our heterogeneous regression. 
We show also for comparison the Kaplan-Meier estimators $S_1^{\rm KM}$ \citep{KM} of the primary risk; note, however, that in the presence of competing risks these KM curves can no longer be trusted to estimate real survival curves (as emphasised by our previous synthetic data). The results are shown in Figures \ref{fig:ULSAM_Scurves1}, where we plotted the survival function estimators for all members of the cohort, and \ref{fig:ULSAM_Scurves2}, where we plotted the survival function estimators conditioned on the value of the fifth covariate (smoking, for which there aqre three values, viz. 0,1,2). We see a clear difference between the crude and the decontaminated primary risk survival curves, i.e. there are clear competing risk effects.  The  decontaminated curves are significantly lower, implying a false protectivity effect of the competing risk. 
This makes sense in view of the date in Table 2, where we can see quite clearly that the regression coefficients of primary and secondary risks are indeed correlated; hence the secondary risk tends to remove from the cohort those individuals that are also more likely to have the primary (PC) events. However, the survival functions still show that smoking is associated with slightly {\em decreased} PC risk. The explanation is that the type-1 events (PC)  occur prdominantly in the {\em frail} sub-class of individuals.

\section{Discussion}

When censoring by non-primary risks is informative, i.e. when non-primary events tend to occur at times that are not statistically independent of the  primary event times, most of the commonly used survival analysis methods involving risk-specific quantities are no longer applicable, as they tend to be based on the assumption of risk independence. 
 The observed (crude) cause specific hazard rates are no longer estimators of what these hazard rates would have been if all non-primary risks were disabled. Finding the latter `decontamined' hazard rates and their associated `decontaminated' cause-specific survival functions from observed survival data is called the competing risk problem.  

In this study we have developed a generic statistical description of survival analysis with competing risks that unifies the main schools of thought, such as frailty and random effect models, latent class models, and the Fine and Gray approach.  We introduced the concept of heterogeneity-induced competing risks, i.e. informative censoring caused at cohort level, by residual (disease- or patient-) heterogeneity that is not captured by covariates, in populations where only at the level of individuals the different risks are independent. 
This differentiation of types of competing risks  leads in a natural way to a classification of survival cohorts from the viewpoint of informative censoring, into four distinct complexity levels.  Assuming heterogeneity-induced competing risks is much weaker than assuming risk independence, yet we demonstrated that it still imposes sufficient constraints to solve the competing risk problem. 
The canonical statistical description of cohorts with heterogeneity-induced competing risks  is in terms of its covariate-constrained functional distribution of individual hazard rates of all risks.  We derive exact formulae for decontaminated primary risk hazard rates and cause-specific survival functions, expressed in terms of this distribution. 

Translating the above formalism into a practical epidemiological tool requires constructing rational and efficient parametrisations of the covariate-conditioned functional distribution of hazard rates. 
We showed that assuming proportional hazards at the level of {\em individuals} leads to a natural family of such  parametrisations,  from which Cox regression, frailty and random effects models, and latent class models can all be recovered in special limits, and which also generates parametrised cumulative incidence functions (the language of Fine and Gray). 
Applications of the formalism include a better understanding of the nature and sub-structure of patient cohorts, tools for detecting and quantifying informative censoring, and improved outcome prediction from covariates (via decontaminated survival curves). In addition it may aid the search for informative  biomarkers, via retrospective class assignment. 
We used a simple numerical implementation of the method  to analyse synthetic data, which revealed how the generic method can uncover and map a cohort's substructure, if such substructure exists, and can indeed remove heterogeneity-induced false protectivity and false exposure effects. 
We showed that application to real survival data from the ULSAM study, with prostate cancer (PC) as the primary risk, leads to plausible alternative explanations for previous counter-intuitive inferences (such as a weak protective effect on PC of smoking), in terms of distinct sub-groups of patients with distinct risk factors and overall frailties.

Although the statistical formalism in the first part of this study is for now completed, we regard the second part, where we construct parametrisations and numerical implementations, only as a first step. In future studies we will implement full Bayesian sampling (instead of the present MAP+AIC parameter estimation protocol), introduce cause specific base hazard rates that may depend on the class label (which was not implemented here to reduce model complexity), and investigate Gaussian mixture models (which can be seen as the integration of the presently proposed latent class and unimodel Gaussian parametrisations).

\section*{Acknowledgements}

We are grateful for fruitful discussions with  Shola Agbaje, Salma Ayis, Maria D'Iorio, Niels Keiding, Katherine Lawler and Tony Ng, and for financial support from Prostate Cancer UK and the European Union's FP-7 Programme.

\vspace*{\fill}\clearpage

\appendix
\section{Connection between cohort level and individual level cause-specific hazard rates}
\label{app:link_for_rates}

Here we give the derivation of identities (\ref{eq:link1}) and (\ref{eq:link2}). Starting from (\ref{eq:hazardrates}), we multiply both sides by $S(t)$ and insert $\Prob(t_0,\ldots,t_R)=\frac{1}{N}\sum_{i=1}^{N}\Prob_i(t_0,\ldots,t_R)$. This gives, via (\ref{eq:rates_fori}):
\begin{eqnarray}
h_r(t)S(t)&=&\frac{1}{N}\sum_{i=1}^{N}\int_0^\infty\!\!\!\!\ldots\!\int_0^\infty\!\dt_0\ldots \dt_R~\Prob_i(t_0,\ldots,t_R)\delta(t-t_r)\prod_{r^\prime\neq r}^R\theta(t_{r^\prime}-t)
\nonumber
\\
&=&
\frac{1}{N}\sum_{i=1}^{N}S_i(t)h_r^i(t)
\end{eqnarray}
Insertion of the identity $S(t)=\frac{1}{N}\sum_i S_i(t)$ and our formula
$S_i(t)=\exp[-\sum_{r=0}^R \int_0^t\ds~h^i_r(s)]$ for the individual survival functions  then leads to the claimed identity (\ref{eq:link1}):
\begin{eqnarray}
h_r(t)&=& \frac{\sum_{i=1}^{N}S_i(t)h_r^i(t)}{\sum_{i=1}^N S_i(t)}
~=~ \frac{\sum_{i=1}^N h_r^i(t)\rme^{-\sum_{r^\prime=0}^R \int_0^t\ds~h^i_{r^\prime}(s)}}
{\sum_{i=1}^N \rme^{-\sum_{r^\prime=0}^R \int_0^t\ds~h^i_{r^\prime}(s)}}
\end{eqnarray}
In the case of covariate conditioning we repeat the above steps, but now we start from (\ref{eq:covariates_h}) and we use the sub-cohort distribution $\Prob(t_0,\ldots,t_R|\bZ)=[\sum_{i,~\bz_i=\bz}\Prob_i(t_0,\ldots,t_R)]/[\sum_{i,~\bz_i=\bz}1]$ instead of 
$\Prob(t_0,\ldots,t_R)$. We then obtain:
\begin{eqnarray}
h_r(t|\bz)S(t|\bz)&=&\frac{\sum_{i,~\bz_i=\bz}\int_0^\infty\!\!\!\!\ldots\!\int_0^\infty\!\dt_0\ldots \dt_R~\Prob_i(t_0,\ldots,t_R)\delta(t-t_r)\prod_{r^\prime\neq r}^R\theta(t_{r^\prime}-t)}{\sum_{i,~\bz_i=\bz}1}
\nonumber
\\&=& \frac{\sum_{i,~\bz_i=\bz} S_i(t)h_r^i(t)}{\sum_{i,~\bz_i=\bz}1}
\end{eqnarray}
Insertion of the identity $S(t|\bz)=[\sum_{i,~\bz_i=\bz} S_i(t)]/[\sum_{i,~\bz_i=\bz}1]$ and our formula for 
$S_i(t)$  now leads to the claimed result (\ref{eq:link2}) for the covariate-conditioned case:
\begin{eqnarray}
h_r(t|\bz)&=& \frac{\sum_{i,~\bz_i=\bz} h_r^i(t)\rme^{-\sum_{r^\prime=0}^R \int_0^t\ds~h^i_{r^\prime}(s)}}{\sum_{i,~\bz_i=\bz} \rme^{-\sum_{r^\prime=0}^R \int_0^t\ds~h^i_{r^\prime}(s)}}
\end{eqnarray}
This completes the proof of identities (\ref{eq:link1}) and (\ref{eq:link2}).

\section{Equivalence of formulae for data likelihood in terms of $W[h_0,\ldots,h_R|\bz]$}
\label{app:twoformulas}

In section \ref{sec:HICR} we have derived two routes for expressing the covariate-conditioned data likelihood $P(t,r|\bz)$ in terms of $\WW[h_0,\ldots,h_R|\bz]$. The first (direct) route is via (\ref{eq:dataprob_inW}):
\begin{eqnarray}
P_A(t,r|\bz)=\int\!\{\rmd h_0\ldots\rmd h_R\}~\WW[h_0,\ldots,h_R|\bz] ~h_r(t)\rme^{-\sum_{r^\prime=0}^R \int_0^t\ds~h_{r^\prime}(s)}
\end{eqnarray}
The second route is to combine (\ref{eq:conditioned_dataprob}) with formula (\ref{eq:falseh_inW})
for the crude covariate-conditioned cause-specific hazard rates, i.e. use the pair
\begin{eqnarray}
P_B(t,r|\bz)&=&h_r(t|\bz)\rme^{-\sum_{r^\prime=0}^R\int_0^t\ds~h_{r^\prime}(s|\bz)}
\\
h_r(t|\bz)&=& \frac{\int\!\{\rmd h_0\ldots\rmd h_R\}~\WW[h_0,\ldots,h_R|\bz]~
h_r(t)\rme^{-\sum_{r^\prime=0}^R\int_0^t\ds~h_{r^\prime}(s)}}{\int\!\{\rmd h_0\ldots\rmd h_R\}~\WW[h_0,\ldots,h_R|\bz]~
\rme^{-\sum_{r^\prime=0}^R\int_0^t\ds~h_{r^\prime}(s)}}
\label{eq:part2}
\end{eqnarray}
Although at first sight the two recipes for $P(t,r|\bz)$ may appear to be different, one can show that they are identical (as they should be). 
We first note that at $t=0$ both expressions agree, since
\begin{eqnarray}
P_A(0,r|\bz)&=&\int\!\{\rmd h_0\ldots\rmd h_R\}~\WW[h_0,\ldots,h_R|\bz] ~h_r(0)\\
P_B(0,r|\bz)&=&h_r(0|\bz)~=~\frac{\int\{\rmd h_0\ldots\rmd h_R\}~\WW[h_0,\ldots,h_R|\bz]~
h_r(0)}{\int\{\rmd h_0,\ldots\rmd h_R\}~\WW[h_0,\ldots,h_R|\bz]}
\nonumber
\\
&=&\int\!\{\rmd h_0\ldots\rmd h_R\}~\WW[h_0,\ldots,h_R|\bz]~
h_r(0)
\end{eqnarray}
Next we show that the ratio of $P_A(t,r|\bz)$ and $P_B(t,r|\bz)$ is time-independent. 
We note that the numerator of (\ref{eq:part2}) is identical to $P_A(t,r|\bz)$, so we may write
\begin{eqnarray}
\frac{\rmd}{\dt}\frac{P_B(t,r|\bz)}{P_A(t,r|\bz)}
&=& 
\frac{\rmd}{\dt}\Big\{
\frac{\rme^{-\sum_{r^\prime=0}^R\int_0^t\ds~h_{r^\prime}(s|\bz)}}
{\int\!\{\rmd h_0\ldots\rmd h_R\}~\WW[h_0,\ldots,h_R|\bz]~
\rme^{-\sum_{r^\prime=0}^R\int_0^t\ds~h_{r^\prime}(s)}}
\Big\}
\nonumber
\\
&&\hspace*{-10mm} =~ 
\Big\{\int\!\{\rmd h_0\ldots\rmd h_R\}~\WW[h_0,\ldots,h_R|\bz]~
\rme^{-\sum_{r^\prime=0}^R\int_0^t\ds~h_{r^\prime}(s)}\Big\}^{-1}
\nonumber
\\
&&
\hspace*{-5mm}
\times
\left\{
\frac{\rmd}{\dt}
\rme^{-\sum_{r^\prime=0}^R\int_0^t\ds~h_{r^\prime}(s|\bz)}
\nonumber
\right.
\\
&&
\left.
\hspace*{-8mm}
-
\frac{\rme^{-\sum_{r^\prime=0}^R\int_0^t\ds~h_{r^\prime}(s|\bz)}
\int\!\{\rmd h_0\ldots\rmd h_R\}~\WW[h_0,\ldots,h_R|\bz]~\frac{\rmd}{\dt}
\rme^{-\sum_{r^\prime=0}^R\int_0^t\ds~h_{r^\prime}(s)}}{\int\!\{\rmd h_0\ldots\rmd h_R\}~\WW[h_0,\ldots,h_R|\bz]~
\rme^{-\sum_{r^\prime=0}^R\int_0^t\ds~h_{r^\prime}(s)}}
\right\}
\nonumber
\\
&&\hspace*{-10mm} =~ 
\frac{\rme^{-\sum_{r^\prime=0}^R\int_0^t\ds~h_{r^\prime}(s|\bz)}}{\int\!\{\rmd h_0 \ldots\rmd h_R\}~\WW[h_0,\ldots,h_R|\bz]~
\rme^{-\sum_{r^\prime=0}^R\int_0^t\ds~h_{r^\prime}(s)}}
\nonumber
\\
&&\hspace*{-8mm}
\times\sum_{r^\pprime=0}^R 
\Big\{\frac{
\int\!\{\rmd h_0\ldots \rmd h_R\}~\WW[h_0,\ldots,h_R|\bz]~h_{r^\pprime}(t)
\rme^{-\sum_{r^\prime=0}^R\int_0^t\ds~h_{r^\prime}(s)}}{\int\!\{\rmd h_0\ldots\rmd h_R\}~\WW[h_0,\ldots,h_R|\bz]~
\rme^{-\sum_{r^\prime=0}^R\int_0^t\ds~h_{r^\prime}(s)}}
-
h_{r^\pprime}(t|\bz)
\Big\}
\nonumber
\\
&&\hspace*{-10mm} =~  0
\end{eqnarray}
with the last line following directly from (\ref{eq:part2}). Since we know that $P_B(t,r|\bz)/P_A(t,r|\bz)=1$ at $t=0$, we have now established that 
$P_B(t,r|\bz)/P_A(t,r|\bz)=1$ for all $t\geq 0$. 

\section{Connection with standard regression methods}
\label{app:links}

Here we show how the equations of some conventional regression methods, that focus only on the hazard rates of the primary risk and assume primary and non-primary risks to be independent, are recovered from our generic formulae. 
 For such models one has
$\MM(\bbeta_0,\ldots,\bbeta_R;\lambda_0,\ldots,\lambda_R|\bz) =\MM(\bbeta_1;\lambda_1|\bz)\MM(\bbeta_0,\bbeta_2,\ldots,\bbeta_R;\lambda_0,\lambda_2,\ldots, \lambda_R|\bz)$. Inserting this into (\ref{eq:generic_loglikelihood})  gives
\begin{eqnarray}
\LL(\MM)&=& \LL_1(\MM)~+~{\rm terms~independent~of}~\MM(\bbeta_1;\lambda_1|\bz)
\end{eqnarray}
with, after some simple manipulations, 
\begin{eqnarray}
\LL_1(\MM)&=&
\sum_{i=1}^N \delta_{r_i,1}
\log\Big\{ \frac{\int\!\rmd\bbeta_1\{\rmd\lambda_1\}~\MM(\bbeta_1;\lambda_1|\bz_i)
~\lambda_{1}(t_i)~\rme^{\bbeta_{1}\cdot\bz_i-
\Lambda_{1}(t_i)\exp(\bbeta_{1}\cdot\bz_i)
}}
{ \int\!\rmd\bbeta_1\{\rmd\lambda_1\}~\MM(\bbeta_1;\lambda_1|\bz_i)
~\rme^{-
\Lambda_{1}(t_i)\exp(\bbeta_{1}\cdot\bz_i)
}}
\Big\}
\nonumber
\\
&&+
\sum_{i=1}^N 
\log \int\!\rmd\bbeta_1\{\rmd\lambda_1\}~\MM(\bbeta_1;\lambda_1|\bz_i)
~\rme^{-
\Lambda_{1}(t_i)\exp(\bbeta_{1}\cdot\bz_i)
}
\label{eq:simple_loglikelihood}
\end{eqnarray}
The independence of primary and non-primary risks causes (\ref{eq:h_generic}) to simplify to $h_1(t|\bz)=\tilde{h}_1(t|\bz)$, and (\ref{eq:Stilde_generic},\ref{eq:htilde_generic}) give the following decontaminated primary  risk survival function and hazard rate: 
\begin{eqnarray}
\tilde{S}_1(t|\bz)&=&  \int\!\rmd\bbeta_1\{\rmd\lambda_1\}~\MM(\bbeta_1;\lambda_1|\bz) ~\rme^{- \exp(\bbeta_{1}\cdot \bz)\Lambda_{1}(t)}
\label{eq:Sapp}
\\
\tilde{h}_1(t|\bz)&=&  \frac{ \int\!\rmd\bbeta_1\{\rmd\lambda_1\}~\MM(\bbeta_1;\lambda_1|\bz) ~\lambda_1(t)\rme^{\bbeta_1\cdot \bz- \exp(\bbeta_{1}\cdot \bz)\Lambda_{1}(t)}}
{  \int\!\rmd\bbeta_1\{\rmd\lambda_1\}~\MM(\bbeta_1;\lambda_1|\bz) ~\rme^{- \exp(\bbeta_{1}\cdot \bz)\Lambda_{1}(t)}}
\label{eq:happ}
\end{eqnarray}
Let us work out such formulae for the two most popular methods:

\begin{itemize}
\item {\em Cox's proportional hazards regression}
\\[2mm]
Cox's  regression method implies choosing (\ref{eq:Cox}) for the distribution of primary risk parameters, viz. $\MM(\bbeta_1;\lambda_1|\bz)= \delta_{\rm F}[\lambda_1\!-\!\hat{\lambda}]~\delta(\beta_1^0)\prod_{\mu=1}^p\delta(\beta_1^\mu\!-\!\hat{\beta}^\mu)$. Insertion into our formulae for the decontaminated survival function and hazard rate of the primary risk gives Cox's recipes
\begin{eqnarray}
\tilde{S}_1(t|\bz)=  \rme^{- \exp(\sum_{\mu=1}^p\hat{\beta}^\mu z^\mu)\hat{\Lambda}(t)},~~~~~~~
\tilde{h}_1(t|\bz)= \hat{\lambda}(t) \rme^{\sum_{\mu=1}^p\hat{\beta}^\mu z^\mu}
\end{eqnarray}
The contribution to  (\ref{eq:simple_loglikelihood}) that contains primary risk parameters  is then found to be
\begin{eqnarray}
\LL_1(\MM)=
\sum_{i=1}^N \delta_{r_i,1}\log \hat{\lambda}(t_i)
+\sum_{\mu=1}^p\hat{\beta}^\mu\mu\sum_{i=1}^N \delta_{ r_i,1} z_i^\mu-\sum_{i=1}^N \rme^{\sum_{\mu=1}^p\hat{\beta}_\mu z_i^\mu}\hat{\Lambda}(t_i)
\label{eq:CoxFullL}
\end{eqnarray}
If we calculate from this the maximum likelihood estimate for the base hazard rate, via functional differentiation of (\ref{eq:CoxFullL}) with respect to $\hat{\lambda}(t)$, we recover Breslow's formula 
\begin{eqnarray}
\hat{\lambda}(t)&=& \frac{\sum_{j=1}^N \delta_{r_j,1}\delta(t-t_j)}{\sum_{j=1}^N \rme^{\sum_{\mu=1}^p\hat{\beta}^\mu z^\mu_j}\theta(t_j-t)}
\end{eqnarray}
Insertion into (\ref{eq:CoxFullL})  then gives, after some rewriting, the standard formula for the log-likelihood in terms of the remaining primary risk regression parameters $(\hat{\beta}^1,\ldots,\hat{\beta}^p)$ in Cox regression:
\begin{eqnarray}
\LL_1(\hat{\beta}^1,\ldots,\hat{\beta}^p)= \LL_1(\bnull)+\sum_{i=1}^N \Big[\sum_{\mu=1}^p\hat{\beta}^\mu z_i^\mu+\log \Big(\frac{\sum_{j=1}^N\theta(t_j-t_i)}{\sum_{j=1}^N\rme^{\sum_{\mu=1}^p \hat{\beta}^\mu z_j^\mu}\theta(t_j-t_i)}\Big)\Big]
\end{eqnarray}

\item {\em Frailty models}
\\[2mm]
The simple frailty models correspond to $\MM(\bbeta_1;\lambda_1|\bz)= \delta_{\rm F}[\lambda_1\!-\!\hat{\lambda}]~g(\beta_1^0)\prod_{\mu=1}^p\delta(\beta_1^\mu\!-\!\hat{\beta}^\mu)$
Inserting this into (\ref{eq:simple_loglikelihood}) leads to
\begin{eqnarray}
\LL_1(\MM)&=&
\sum_{i=1}^N \delta_{r_i,1}
\log\hat{\lambda}(t_i)
+\sum_{\mu=1}^p \hat{\beta}^\mu
\sum_{i=1}^N \delta_{r_i,1}z^\mu_i
\nonumber
\\&&
+
\sum_{i=1}^N \delta_{r_i,1}
\log\Big(
\frac{\int\!\rmd\beta^0_1~g(\beta_1^0)~\rme^{\beta_0^1-
\hat{\Lambda}(t_i)\exp(\beta^0_1+\sum_{\mu=1}^p \hat{\beta}^\mu z^\mu_i)
}}
{ \int\!\rmd\beta^0_1~g(\beta_1^0)
~\rme^{-
\hat{\Lambda}(t_i)\exp(\beta^0_1+\sum_{\mu=1}^p \hat{\beta}^\mu z^\mu_i)
}}\Big)
\nonumber
\\
&&+
\sum_{i=1}^N 
\log \int\!\rmd\beta^0_1~g(\beta_1^0)
~\rme^{-
\hat{\Lambda}(t_i)\exp(\beta^0_1+\sum_{\mu=1}^p \hat{\beta}^\mu z^\mu_i)
}
\end{eqnarray}
If for the frailty distribution $g(\beta_1^0)$ one chooses
\begin{eqnarray}
g(\beta_1^0)&=& \frac{\alpha^{\alpha}}{\Gamma(\alpha)}\int_0^\infty\!\rmd\theta~\theta^{\alpha-1}\rme^{-\alpha\theta}
\delta(\beta_1^0\!-\!\log\theta)
\end{eqnarray}
 with $\Gamma(z)=\int_0^\infty\!\rmd x~x^{z-1}\rme^{-x}$, 
we will have the normalisation $\int\!\rmd\beta_1^0~\rme^{\beta_1^0}=1$ (which removes the parametrisation redundancy), and
one can do the relevant integrals  analytically using
\begin{eqnarray}
\int\!\rmd\beta^0_1~g(\beta_1^0)~\rme^{-y\exp(\beta_1^0)}&=& 
 \Big(\frac{\alpha}{\alpha\!+\!y}\Big)^{\!\alpha}
\\
\int\!\rmd\beta^0_1~g(\beta_1^0)~\rme^{\beta_1^0-y\exp(\beta_1^0)}&=& 
\Big(\frac{\alpha}{\alpha\!+\!y}\Big)^{\!\alpha+1}
\end{eqnarray}
Here we used the standard identity $\Gamma(\alpha\!+\!1)=\alpha\Gamma(\alpha)$ for the gamma function. 
The formulae (\ref{eq:Sapp},\ref{eq:happ}) now give
\begin{eqnarray}
\tilde{S}_1(t|\bz)&=&  
\Big(
1\!+\!\frac{\hat{\Lambda}(t_i)}{\alpha}\rme^{\sum_{\mu=1}^p\hat{\beta}^\mu z_i^\mu}
\Big)^{-\alpha}
\\
\tilde{h}_1(t|\bz)&=& 
\hat{\lambda}(t)\rme^{\sum_{\mu=1}^p\hat{\beta}^\mu z_i^\mu}
\Big(
1\!+\!\frac{\hat{\Lambda}(t_i)}{\alpha}\rme^{\sum_{\mu=1}^p\hat{\beta}^\mu z_i^\mu}
\Big)^{-1}
\end{eqnarray}
For the primary contribution to the data likelihood we obtain
\begin{eqnarray}
\LL_1(\MM)&=&
\sum_{i=1}^N \delta_{r_i,1}
\log\hat{\lambda}(t_i)
+\sum_{\mu=1}^p \hat{\beta}^\mu
\sum_{i=1}^N \delta_{r_i,1}z^\mu_i
-\alpha
\sum_{i=1}^N 
\log
\Big(
1\!+\!\frac{\hat{\Lambda}(t_i)}{\alpha}\rme^{\sum_{\mu=1}^p\hat{\beta}^\mu z_i^\mu}
\Big)
\nonumber
\\&&
-
\sum_{i=1}^N \delta_{r_i,1}
\log\Big(
1\!+\!\frac{\hat{\Lambda}(t_i)}{\alpha}\rme^{\sum_{\mu=1}^p\hat{\beta}^\mu z_i^\mu}
\Big)
\end{eqnarray}
It is no longer possible to find an analytical expression for the base hazard rate in terms of the other parameters, so here one is limited to either choosing convenient parametrised forms or to numerical maximisation.  
Since $\int\!\rmd\beta_1^0~\rme^{2\beta_1^0}=1\!+\!\frac{1}{\alpha}$ we have $\lim_{\alpha\to\infty}g(\beta_1^0)=\delta(\beta_1^0\!-\!1)$, so for $\alpha\to\infty$ all the above equations must reduce to those corresponding to Cox regression, which indeed they do. 
The parameter $\alpha$ can also be estimated from the data via Bayesian methods
\citep{Ibrahim}.

\end{itemize}
For random effects  and latent class models one can recover from our generic formulation the 
various published results in a similar way. Since these models are more involved, fewer  steps can typically be taken analytically, and their authors have to turn to numerical determination of parameters sooner.

\section{The Gaussian integral over regression parameters}
\label{app:GaussianIntegral}

Here we derive a simplified expression for the log-likelihood (\ref{eq:gauss_model_big}), by manipulation of the $(p\!+\!1)\!\times\!R$-dimensional Gaussian integral 
\begin{eqnarray}
I(t,r,\bz)&=& 
\int\!\rmd\bbeta_1\ldots\rmd\bbeta_R~\MM(\bbeta_1,\ldots,\bbeta_R) ~
\rme^{\notdelta_{0r}\bbeta_{r}\cdot\bz-\sum_{r^\prime=1}^R
\hat{\Lambda}_{r^\prime}(t)\exp(\bbeta_{r^\prime}\cdot\bz)
}
\end{eqnarray}
We first define for $r=1\ldots R$ new Gaussian variables $x_r=(\bbeta_{r}-\hat{\bbeta}_r)\cdot\bz$. According to the distribution (\ref{eq:gaussian_measure}) they have zero average and covariance matrix entries 
$\bra x_r x_{r^\prime}\ket= K_{rr^\prime}(\bz)$, with the matrix $\bK(\bz)$ as defined in (\ref{eq:Kaz}). Hence we may write, with $\bx=(x_1,\ldots,x_R)$, 
\begin{eqnarray}
I(t,r,\bz)&=& \rme^{\notdelta_{0r}\hat{\bbeta}_{r}\cdot\bz}
\int\!\frac{\rmd\bx~\rme^{-\frac{1}{2}\bx\cdot\bK^{-1}(\bz)\bx}}{
\sqrt{(2\pi)^R{\rm Det}\bK(\bz)}}
 ~
\rme^{\notdelta_{0r} x_r-\sum_{r^\prime=1}^R
\hat{\Lambda}_{r^\prime}(t)\exp(\hat{\bbeta}_{r^\prime}\cdot\bz+x_{r^\prime})
}
\end{eqnarray}
A final transformation $x_s=\sum_{r^\prime=1}^R[\bK^{1/2}(\bz)]_{sr^\prime}y_{r^\prime}+\notdelta_{0r}K_{sr}(\bz)$  then brings us after some simple manipulations to
the following expression, with the short-hand ${\rm D}\by=\prod_{r=1}^R[ (2\pi)^{-1/2}\rme^{-y_r^2/2}\rmd y_r]$:
\begin{eqnarray}
I(t,r,\bz)&=& \rme^{\notdelta_{0r}[\hat{\bbeta}_{r}\cdot\bz+\frac{1}{2}K_{rr}(\bz)]}
\nonumber
\\
&&\times 
\int\!{\rm D}\by
 ~
\rme^{-\sum_{r^\prime=1}^R
\hat{\Lambda}_{r^\prime}(t)\exp[\hat{\bbeta}_{r^\prime}\cdot\bz
+\notdelta_{0r}K_{rr^\prime}(\bz)]\exp[\sum_{r^\pprime=1}^R [\bK^{1/2}(\bz)]_{r^\prime r^\pprime}y_{r^\pprime}]
}
\nonumber
\\[-1mm]
&&
\end{eqnarray}
With this result, and recalling the definition (\ref{eq:Kaz}),  we may write the data log-likelihood (\ref{eq:gauss_model_big}) as
\begin{eqnarray}
\LL_{\rm risks}(\MM)&=&
\sum_{i=1}^N \notdelta_{0r_i}
\log\hat{\lambda}_{r_i}(t_i)+
\sum_{i=1}^N 
\log I(t_i,r_i,\bz_i)
\nonumber
\\
&=& 
\sum_{i=1}^N \notdelta_{0r_i}
\log\hat{\lambda}_{r_i}(t_i)
+
\sum_{i=1}^N  \notdelta_{0r_i}\Big\{\hat{\bbeta}_{r_i}\cdot\bz_i+\frac{1}{2}\bz_i\cdot \bC^{r_i r_i}\bz_i \Big\}
\label{eq:gauss_model_small}
\\
&&
+
\sum_{i=1}^N 
\log \int\!{\rm D}\by
 ~
\rme^{-\sum_{r^\prime=1}^R
\hat{\Lambda}_{r^\prime}(t_i)\exp[\hat{\bbeta}_{r^\prime}\cdot\bz_i
+\bz_i\cdot\bC^{r_i r^\prime}\bz_i+\sum_{r^\pprime=1}^R [\bK^{1/2}(\bz_i)]_{r^\prime r^\pprime}y_{r^\pprime}]
}\nonumber
\end{eqnarray}
As a simple test one confrms that in the limit $\bC\to\bnull$, i.e. in the case of vanishing heterogeneity,  one recovers from this expression the corresponding log-likelihood formula of the Cox model. 

\section{Numerical details}
\label{app:numerical}

Here we give details of a number of numerical procedures that were used in the applications of our method to synthetic and real survival data, to facilitate the reproduction of our results.

\begin{itemize}
\item {\em Synthetic data}
\\[2mm]
All synthetic data used in this paper were generated as follows. For each individual $i$  and each risk $r=1\ldots R$ we generate numerically a uniformly distributed random variable $u_{ir}\in[0,1]$, and define a latent event time $t_{ir}=-\tau_{ir}\log u_{ir}$. The survival data for $i$ are then given by
\begin{eqnarray}
t_i={\rm min}_{r\in\{1,\ldots,R\}} t_{ir},~~~~~~
r_i={\rm argmin}_{r\in\{1,\ldots,R\}} t_{ir}
\end{eqnarray}
These synthetic data will correspond to risks that are  independent at the level of individuals:
\begin{eqnarray}
\Prob_i(t_1,\ldots,t_R)= \prod_{r=1}^R \Big(
 \tau^{-1}_{ir}\rme^{-t_r/\tau_{ir}}\Big)
\end{eqnarray}
with average latent times $\tau_{ir}=\int_0^\infty\!\dt_r~t_r\Prob_i(t_r)$. Let $\ell$ be the latent class in our cohort to which $i$ belongs. We define the $\tau_{ir}$ in terms of the individuals' covariate vector $\bz_i\in\R^p$:
\begin{eqnarray}
\tau_{ir}&=& \lambda^{-1}_r~ \rme^{-\sum_{\mu=1}^p \hat{\beta}_r^{\ell \mu}z_i^\mu}
\end{eqnarray}
The individual cause-specific hazard rates of our synthetic data will then be as in the model (\ref{eq:our_model}) (see Figure \ref{fig:heterogeneity}), but with time-independent base hazard rates and without frailty terms:
\begin{eqnarray}
h_r^i(t)&=& \lambda_r~\rme^{\sum_{\mu=1}^p \hat{\beta}_r^{\ell \mu}z_i^\mu}
\end{eqnarray}

\begin{figure}[t]
\unitlength=0.338mm

\begin{picture}(400,120)
\put(-2,70){$\lambda_r(t)$}
\put(0,0){\includegraphics[width=138\unitlength,height=115\unitlength]{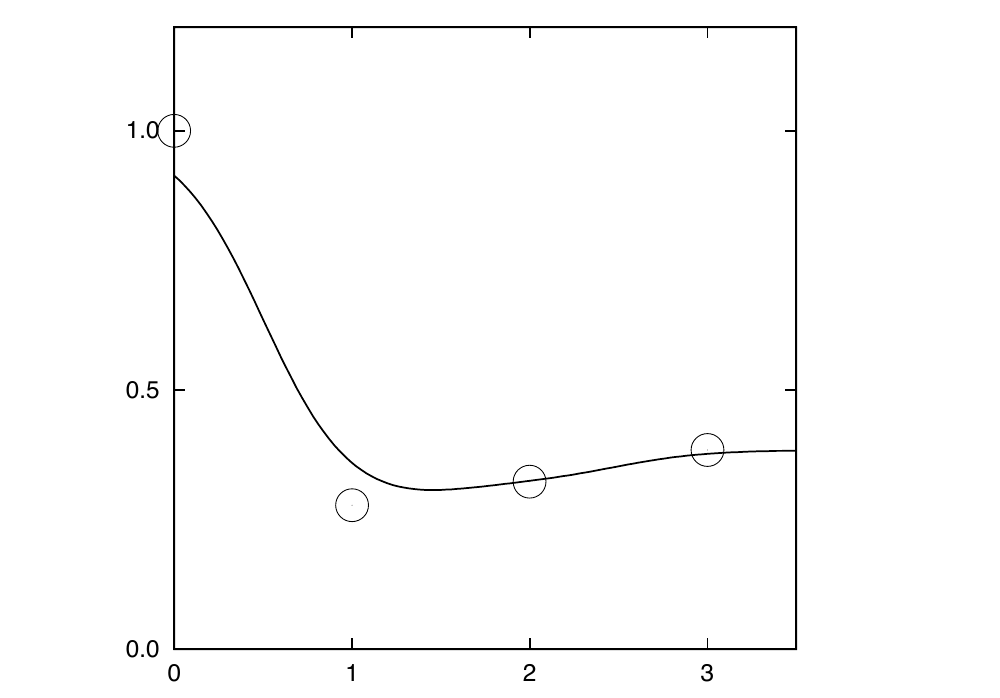}}
\put(100,0){\includegraphics[width=138\unitlength,height=115\unitlength]{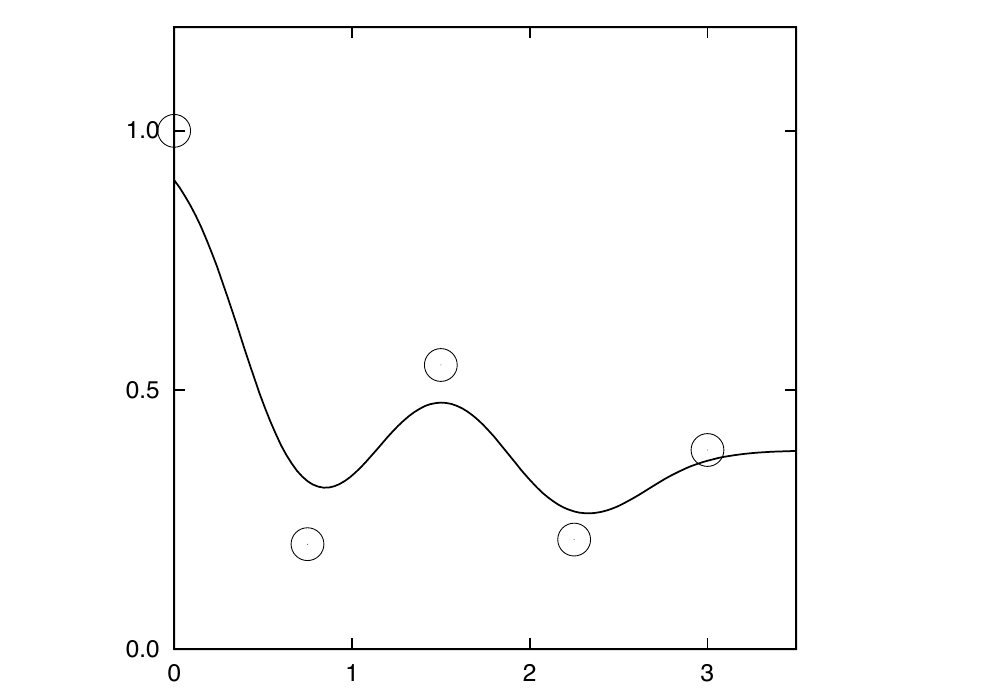}}
\put(200,0){\includegraphics[width=138\unitlength,height=115\unitlength]{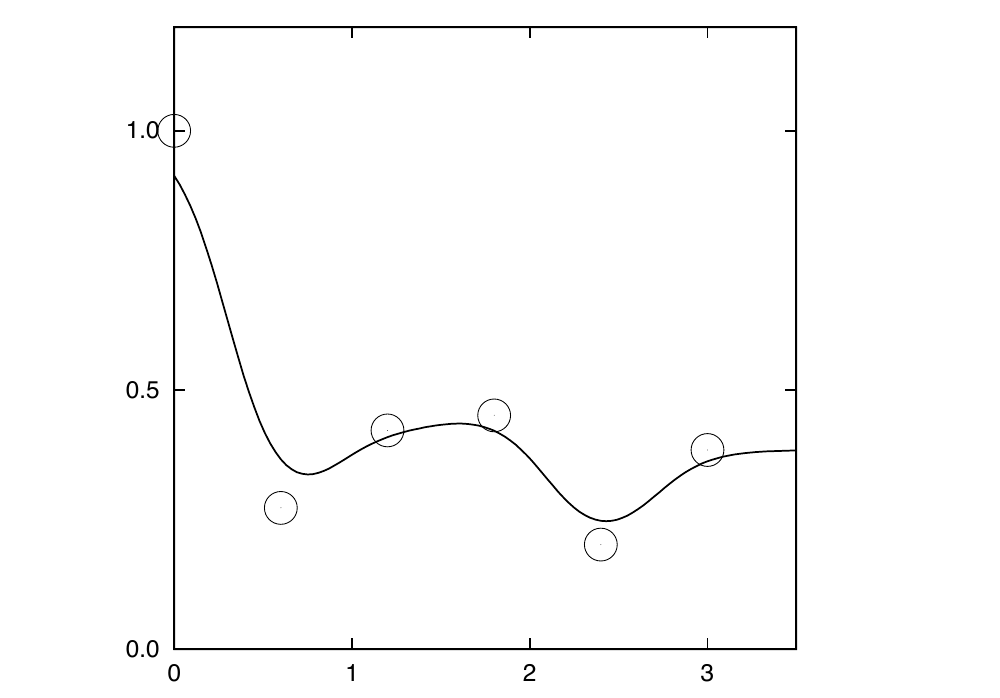}}
\put(300,0){\includegraphics[width=138\unitlength,height=115\unitlength]{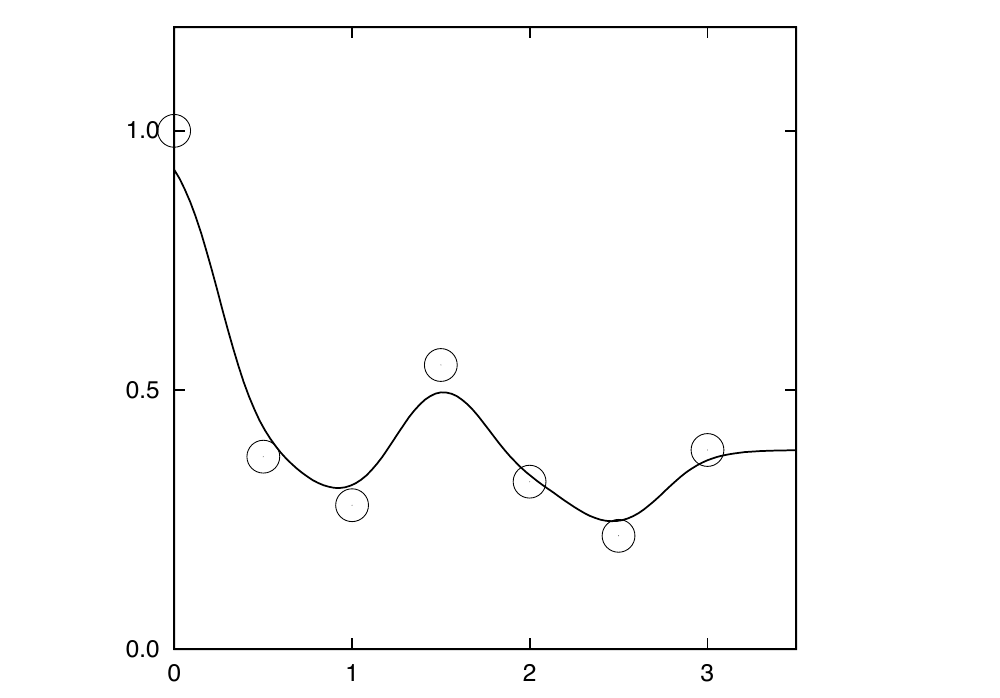}}
\put(65,-7){$t$}\put(165,-7){$t$}\put(265,-7){$t$}\put(365,-7){$t$}
\put(70,95){$K\!=\!3$}\put(170,95){$K=4$}\put(270,95){$K=5$}\put(370,95){$K=6$}
\end{picture}\vsp

\caption{Illustration of the base hazard rate parametrisation via Gaussian interpolation. Circles: the base points $(\tilde{t}_k,\xi_{kr})$, with $k=0\ldots K$. The equi-distant times $\tilde{t}_k$ are fixed and given by (\ref{eq:spline_times});  in this example $t_{\rm min}=0$ and $t_{\rm max}=3$.  The $K\!+\!1$ values $\xi_{kr}$ are optimised via MAP (maximum a posteriori data likelihood). Solid lines: the corresponding smooth parametrised hazard rates $\lambda(t|\bxi)$ defined in (\ref{eq:spline_values}).  This parametrisation procedure is applied to the base hazard rates of all  risks $r=0\ldots R$. }
\label{fig:splines}
\end{figure}
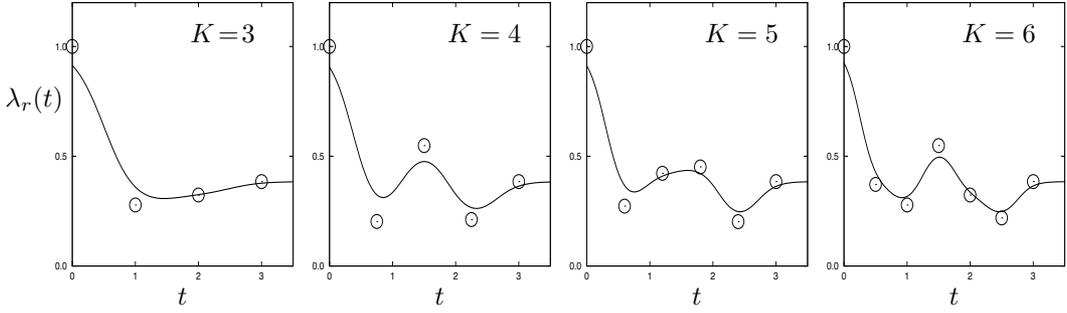

\item {\em Parametrisation of base hazard rates}
\\[2mm]
In our applications of the models (\ref{eq:our_model},\ref{eq:gaussian_model})  to synthetic and real data the base hazard rates $\lambda_r(t)$ and the parameters $\{w_\ell,\hat{\beta}_r^{\ell\mu}\}$ are estimated by a version of the MAP method (\ref{eq:MAP}), i.e. by numerical maximization of the posterior probability $P(D|\btheta)P(\btheta)$. This requires the base rates to be parametrised, for which we chose an interpolation method. Given the survival data $D=\{(t_1,r_1),\ldots,(t_N,r_N)\}$, which cover a time interval  from $t_{\rm min}={\rm min}_{i\in\{1,\ldots,N\}}t_i$ to $t_{\rm max}={\rm max}_{i\in\{1,\ldots,N\}}t_i$, we define 
$K\!+\!1$ equidistant time points $\tilde{t}_k$, with $k=0\ldots K$,
\begin{eqnarray}
\tilde{t}_k&=& t_{\rm min}+\frac{k}{K}(t_{\rm max}-t_{\rm min})
\label{eq:spline_times}
\end{eqnarray}
To each $\tilde{t}_k$  we assign  associated base rate parameters $(\xi_{k0},\ldots,\xi_{kR})$ for all risks $r=0\ldots R$, which allows us to define the $R\!+\!1$ base hazard rates $\lambda_r(t)$ for each $r$ and each $t\geq 0$ as smooth Gaussian convolutions, with uniform variation time scale $\sigma=\frac{1}{2}(\tilde{t}_{k+1}-\tilde{t}_k)$:
\begin{eqnarray}
\lambda_r(t|\bxi)= \frac{\sum_{k=0}^K \xi_{kr} ~\rme^{-\frac{1}{2}(t-\tilde{t}_k)^2/\sigma^2}}
{\sum_{k=0}^K \rme^{-\frac{1}{2}(t-\tilde{t}_k)^2/\sigma^2}},~~~~~~~~\sigma=\frac{t_{\rm max}-t_{\rm min}}{2K}
\label{eq:spline_values}
\end{eqnarray}
See also Figure \ref{fig:splines}. 
The number $K$ is either chosen or determined by Bayesian model selection. Increasing the value of  $K$ allows more irregular functions $\lambda_r(t)$ to be modelled, but it also increases considerably the number of parameters, which slows down the parameter estimation and adds to the danger of overfitting. The integrated rates $\Lambda_r(t|\bxi)\!=\!\int_0^t\!\rmd s~\lambda_r(s|\bxi)$ are obtained numerically from (\ref{eq:spline_values}) via 11-point Gaussian Quadrature, see e.g. \citep{NR}, applied separately to the $n$ intervals $[jt/n,(j\!+\!1)t/n]$ with $j=0\ldots n\!-\!1$ and $n=20$.

\item {\em Numerical parameter optimisation}
\\[2mm]
The parameter optimisations used in the applications of our method to data all follow the MAP (Maximum A Posteriori Probability) protocol described in subsection \ref{sub:estimation}, complemented by Aikake's Information Criterion (AIC) to limit overfitting; see e.g. \citep{MacKay}. In combination this implies that 
the optimal parameters of each of the models (\ref{eq:our_model}) and (\ref{eq:gaussian_model})
 are given in terms of their data log-likelihood $\LL(\btheta)=\log P(D|\btheta)$ and  parameter prior $P(\btheta)$ by 
\begin{eqnarray}
\hat{\btheta}&=& {\rm argmin}_{\btheta}~\Psi(\btheta)\\
\Psi(\btheta)&=& n_{\rm par}-\LL(\btheta)-\log P(\btheta)
\label{eq:Psi_to_minimise}
\end{eqnarray}
Here $n_{\rm par}={\rm dim}(\btheta)$ is the number of parameters of each parametrisation, i.e. 
\begin{eqnarray}
{\rm parametrisation~(\ref{eq:our_model}) }:&\!\!\!& n_{\rm par}=RL(p\!+\!1)+KR+L-1
\\
{\rm parametrisation~(\ref{eq:gaussian_model}) }:&\!\!\!& n_{\rm par}=\frac{1}{2}(p\!+\!1)^2
R^2+\frac{3}{2}(p\!+\!1)R+KR~~~~~~
\end{eqnarray}
We choose flat priors for the parameters $\{\xi_r\}$ of the base hazard rates (since the chosen parametrisation already ensures smoothness), flat priors for the weights $W_\ell$ in (\ref{eq:our_model})  (i.e. the maximum entropy measure), and unit-variance zero-average Gaussian priors for all other parameters (which for the regression coefficients is justified by our decision to normalise all covariates by linear rescaling to zero average and unit variance over the cohort). \\
\hspace*{5mm}
The optimum $\hat{\btheta}$ is determined numerically via a simple adaptation of the Nelder-Mead or downhill simplex method, see e.g. \cite{NR}, in which we complement simplex iterations  by repeated randomizations of decreasing amplitude.
Setting $\xi_{kr}=\xi_r$ for all $k$ in (\ref{eq:spline_values})  gives the stationary hazard rates $\lambda_r(t|\bxi)=\xi_r$ for all $t\geq 0$,  so a rational initialisation of our search algorithms is to set our parameters such that 
\begin{eqnarray}
\MM(\bbeta_1,\ldots,\bbeta_R) = \prod_{r=1}^R\delta(\bbeta_r),~~~~~~\hat{\lambda}_r(t)=\xi_r~~(\forall t\geq 0)
\label{eq:initialisation}
\end{eqnarray}
in which the $\{\xi_r\}$ are chosen such as to maximise the data log-likelihoods (\ref{eq:LL_our_model}) and (\ref{eq:gauss_model_big}). For the choice (\ref{eq:initialisation}) the latter formulae both reduce to 
$\LL_{\rm risks}=\sum_{r=1}^R\big[
\log (\xi_{r})
\sum_{i=1}^N \delta_{r_i r}
-\xi_r
\sum_{i=1}^N t_i\big]$, from which by simple differentiation we obtain 
\begin{eqnarray}
\xi_r&=& \Big(\sum_{i=1}^N\delta_{r_i r}\Big)/\Big(\sum_{i=1}^N t_i\Big)
\end{eqnarray}
To break possible symmetries we assign small (Gaussian) random values to the regression coefficients, instead of the strictly zero values in (\ref{eq:initialisation}), and we repeat the numerical searches for multiple random initialisations to reduce the impact of sub-optimal local minima. \\
\hspace*{5mm}
This leaves the model selection question, i.e. the selection of the parameters $K$, which controls the complexity of the base hazard rates, and $L$ (the number of allowed distinct
sub-cohorts in parametrisation (\ref{eq:our_model})). 
For parametrisation (\ref{eq:our_model}) we determine
for each choice of $(L,K)$ numerically the value $\Psi_{L,K}\!=\! \Psi(\hat{\btheta})$, with  $\hat{\btheta}\!=\!{\rm argmin}_{\btheta}\Psi(\btheta)|_{L,K}$, via the above procedure. The most probable model then corresponds to $(L,K)\!=\!{\rm argmin}_{L,K>0}\Psi_{L,K}$. For  (\ref{eq:gaussian_model}) we determine
for each $K\!>\!0$ numerically the value $\Psi_{K}\!=\! \Psi(\hat{\btheta})$, with  $\hat{\btheta}\!=\!{\rm argmin}_{\btheta}\Psi(\btheta)|_{K}$. Here the most probable model corresponds to $K={\rm argmin}_{K>0}\Psi_{K}$.

\item {\em Calculation of parameter error bars via numerical estimation of curvature}
\\[2mm]
The error bar $\sigma_i$ associated with each estimate $\hat{\theta}_i$  is the standard deviation of the corresponding marginal of the posterior 
$P(\btheta|D)$. The latter can be written in terms of the log-likelihood $\LL(\btheta)$ and the function $\Psi(\btheta)$ (\ref{eq:Psi_to_minimise})
that is minimised in the MAP method:
\begin{eqnarray}
\sigma^2_i&=& \int\!\rmd\btheta~P(\btheta|D)\theta_i^2-\Big(\int\!\rmd\btheta~P(\btheta|D)\theta_i\Big)^2
\nonumber
\\
&=& \frac{\int\!\rmd\btheta~\rme^{-\Psi(\btheta)}\theta_i^2}{\int\!\rmd\btheta~\rme^{-\Psi(\btheta)}}-\Big(\frac{\int\!\rmd\btheta~\rme^{-\Psi(\btheta)}\theta_i}{\int\!\rmd\btheta~\rme^{-\Psi(\btheta)}}\Big)^2
\end{eqnarray}
Close to the most probable point $\hat{\btheta}$ we may expand $\Psi(\btheta)$ 
up to quadratic order in the deviation $\btheta\!-\!\hat{\btheta}$, which implies approximating $P(\btheta|D)$ by a Gaussian distribution around $\hat{\btheta}$, i.e.  
\begin{eqnarray}
\Psi(\btheta)&=& \Psi(\hat{\btheta})+\frac{1}{2}(\btheta-\hat{\btheta})\cdot \bC^{-1}(\btheta-\hat{\btheta})+\ldots
\end{eqnarray}
and find upon neglecting cubic and higher orders that $\sigma_i\approx C_{ii}$. \\
\hspace*{5mm} The curvature matrix of the function $\Psi(\btheta)$ at $\hat{\btheta}$ can in principle be calculated analytically. Alternatively, it can be obtained via simple numerical probes of $\Psi$  close to the minimum $\hat{\btheta}$, which is the method used in this paper.
We note that for a truly quadratic surface $\Psi(\btheta)$:
\begin{eqnarray}
\Psi(\hat{\btheta}\!+\!\Delta\btheta)-\Psi(\hat{\btheta})&=& \frac{1}{2}\Delta\btheta\cdot\bC^{-1}\Delta\btheta
\label{eq:quad_approx}
\end{eqnarray}
We now define specific probes close to $\hat{\btheta}$ and the corresponding responses of the function $\Psi$:
\begin{eqnarray}
\Delta\theta_i=\epsilon \delta_{ik}:&& \Delta\Psi_k = \Psi(\hat{\btheta}\!+\!\Delta\btheta)-\Psi(\hat{\btheta})
\\
\Delta\theta_i=\epsilon (\delta_{ik}\!+\!\delta_{i\ell})~~(k\neq \ell):&& \Delta\Psi_{k\ell} = \Psi(\hat{\btheta}\!+\!\Delta\btheta)-\Psi(\hat{\btheta})
\end{eqnarray}
For $\epsilon$ sufficiently small to ensure that the quadratic approximation (\ref{eq:quad_approx}) is good one then finds, after some rearranging of terms, the following simple formulae for the entries of $\bC^{-1}$:
\begin{eqnarray}
(\bC^{-1})_{kk}=2\epsilon^{-2}\Delta\Psi_k,~~~~~~~~k\neq \ell:~~~(\bC^{-1})_{k\ell}=\epsilon^{-2}(\Delta\Psi_{k\ell}\!-\!\Delta\Psi_k\!-\!\Delta\Psi_\ell) 
\label{eq:Cinverse}
\end{eqnarray}
Numerical matrix inversion then leads us to $\bC$ and the error bars $\sigma_i$.  We averaged (\ref{eq:Cinverse}) prior to inversion over 10 different choices for $\epsilon$, namely $\epsilon_\lambda=10^{-3}(\frac{1}{2})^{\lambda-1}$ with $\lambda=1\ldots 10$, to obtain robust estimates. 
Our covariate normalisation ensures that all relevant parameters are of order one, so for uni-modal posterior distributions $P(\btheta|D)$  the quadratic appoximation should be acceptable for the chosen $\epsilon$ values. For multi-modal distributions the error bars $\sigma_i$ will quantify only the local parameter uncertainty associated with the most probable point $\hat{\btheta}$. 

\end{itemize}
\end{document}

%% file: commands.tex
\newcommand{\rmd}{{\rm d}}
\newcommand{\rme}{{\rm e}}

\newcommand{\pprime}{{\prime\prime}}

\newcommand{\bra}{\langle}
\newcommand{\ket}{\rangle}

\newcommand{\bnull}{{\mbox{\boldmath $0$}}}
\newcommand{\be}{\begin{equation}}
\newcommand{\ee}{\end{equation}}
\newcommand{\bd}{\begin{displaymath}}
\newcommand{\ed}{\end{displaymath}}
\newcommand{\vsp}{\vspace*{3mm}}
\newcommand{\room}{\rule[-0.1cm]{0cm}{0.6cm}}

\newcommand{\smallroom}{\rule[-0.05cm]{0cm}{0.4cm}}

\newcommand{\R}{{\rm I\!R}}

\renewcommand{\be}{\ensuremath{\mathbf{e}}}

\newcommand{\bp}{\ensuremath{\mathbf{p}}}

\newcommand{\bx}{\ensuremath{\mathbf{x}}}
\newcommand{\by}{\ensuremath{\mathbf{y}}}
\newcommand{\bz}{\ensuremath{\mathbf{z}}}

\newcommand{\bC}{\ensuremath{\mathbf{C}}}

\newcommand{\bK}{\ensuremath{\mathbf{K}}}

\newcommand{\bZ}{\ensuremath{\mathbf{Z}}}

\newcommand{\bbeta}{{\mbox{\boldmath $\beta$}}}

\newcommand{\bxi}{{\mbox{\boldmath $\xi$}}}

\newcommand{\btheta}{{\mbox{\boldmath $\theta$}}}

\newcommand{\here}{\makebox(0,0)}

\newcommand{\notdelta}{\overline{\delta}}